\documentclass[twocolumn,superscriptaddress,floatfix,preprintnumbers, nofootinbib]{revtex4-2} 
\pdfoutput=1
\usepackage[colorlinks=true,breaklinks=true]{hyperref}
\usepackage[normalem]{ulem}
\usepackage[utf8]{inputenc}
\hypersetup{allcolors=[rgb]{0.0 0.0 0.6},linkcolor=[rgb]{0.75 0.05 0.05}}
\usepackage{amssymb,amsmath,latexsym,mathrsfs, bm}
\usepackage{physics}
\usepackage{mathtools}
\usepackage{graphicx}   
\usepackage{slashed}       
\usepackage{tikz}
\usepackage{url}
\usepackage{color}
\usepackage{multirow}
\usepackage{comment}
\usepackage{makecell}
\usepackage{enumitem}
\usepackage{dashrule}
\usepackage{xspace}
\usepackage{booktabs}
\setitemize{itemindent=25pt}

\hypersetup{
colorlinks,
    linkcolor={blue!50!black},
    citecolor={blue!50!black},
    urlcolor={blue!80!black}
}

\definecolor{forestgreen}{rgb}{0.13, 0.545, 0.13}

\newcommand{\dnds}{$\mathrm{d}N/\mathrm{d}S$\xspace}
\newcommand{\Fermi}{\textit{Fermi}}

\allowdisplaybreaks

\bibpunct{[}{]}{,}{n}{}{,}

\usepackage{colortbl}

\definecolor{GCcolor}{RGB}{230, 159, 0}      
\definecolor{ReLUcolor}{RGB}{178, 34, 34}    
\definecolor{BNcolor}{RGB}{0, 158, 115}      
\definecolor{Unpoolcolor}{RGB}{127, 0, 255}  
\definecolor{MaxPoolcolor}{RGB}{204, 121, 167}

\definecolor{Layer0}{RGB}{255, 255, 255}
\definecolor{Layer1}{RGB}{235, 245, 255} 
\definecolor{Layer2}{RGB}{200, 225, 255}
\definecolor{Layer3}{RGB}{150, 200, 255}
\definecolor{Layer4}{RGB}{100, 180, 255}
\definecolor{Layer5}{RGB}{50, 150, 255}  
\definecolor{Layer6}{RGB}{30, 120, 200}  
\definecolor{Layer7}{RGB}{15, 90, 180}   

\usepackage{lineno}

\begin{document}

\preprint{LAPTH-016/25}

\title{A robust neural determination of the source-count distribution of the \textit{Fermi}-LAT sky at high latitudes}

\author{Christopher Eckner}
\email{ceckner@ung.si}
\affiliation{Center for Astrophysics and Cosmology, University of Nova Gorica, Vipavska 11c, 5270 Ajdov\v{s}\v{c}ina, Slovenia}
\affiliation{LAPTh, CNRS, F-74000 Annecy, France}

\author{Noemi Anau Montel}
\email{noemiam@mpa-garching.mpg.de}
\affiliation{GRAPPA (Gravitation Astroparticle Physics Amsterdam),
University of Amsterdam, Science Park 904, 1098 XH Amsterdam, The Netherlands}
\affiliation{Max-Planck-Institut für Astrophysik, Karl-Schwarzschild-Str.\ 1, 85748 Garching, Germany}

\author{Florian List}
\email{florian.list@univie.ac.at}
\affiliation{Department of Astrophysics, University of Vienna, Türkenschanzstraße 17, 1180 Vienna, Austria}

\author{Francesca Calore}
\email{calore@lapth.cnrs.fr}
\affiliation{LAPTh, CNRS, F-74000 Annecy, France}

\author{Christoph Weniger}
\email{c.weniger@uva.nl}
\affiliation{GRAPPA (Gravitation Astroparticle Physics Amsterdam),
University of Amsterdam, Science Park 904, 1098 XH Amsterdam, The Netherlands}

\smallskip

\begin{abstract}
Over the past 16 years, the \textit{Fermi} Large Area Telescope (LAT) has significantly advanced our view of the GeV gamma-ray sky, yet several key questions remain -- such as the composition of the isotropic gamma-ray background, the origin of the \textit{Fermi} Bubbles or the potential presence of signatures from exotic physics like dark matter. Addressing these challenges requires sophisticated astrophysical modeling and robust statistical methods capable of handling high-dimensional parameter spaces.
In this work, we analyze 14 years of high-latitude ($|b|\geq30^{\circ}$) \textit{Fermi}-LAT data in the range from 1 to 10 GeV using simulation-based inference (SBI) via neural ratio estimation. This approach allows us to detect individual gamma-ray sources and derive a list of significant gamma-ray emitters containing more than 98\% of all sources listed in the \textit{Fermi}-LAT Fourth Source Catalog (4FGL) with a flux $S>3\times10^{-10}\;\mathrm{cm}^{-2}\,\mathrm{s}^{-1}$ (about a factor of three larger than the flux above which 4FGL is nearly complete), without any non-4FGL source detected in that flux range. Additionally, we reconstruct the source-count distribution in both parametric and non-parametric forms, achieving large agreement with previous literature results as well as those sources detected by our SBI pipeline. We also quantitatively validate our gamma-ray emission simulator via an anomaly detection technique, demonstrating that the synthetic data closely reproduces the complexity of the real observations.
\end{abstract}

\maketitle

\section{Introduction}
\label{sec:introduction}

A major milestone of the Large Area Telescope aboard NASA's \Fermi~satellite (\Fermi-LAT) 
has been the measurement of the extragalactic gamma-ray background (EGB)~\cite{2015PhR...598....1F} as residual emission after subtracting the diffuse gamma-ray foreground of the Milky Way (MW). The EGB is the combined emission of all discrete extragalactic gamma-ray sources and other, truly diffuse Galactic and extragalactic gamma-ray contributions. It is measured at high Galactic latitudes $|b| > 20^\circ$ from 100 MeV to several hundreds of GeV. The EGB can be decomposed, on one hand, into a part associated with resolved discrete sources. On the other hand, it features a diffuse component, the so-called isotropic gamma-ray background (IGRB)~\cite{Fermi-LAT:2016hfi}. Understanding the composition of the IGRB remains a central challenge in gamma-ray astrophysics. It receives contributions from the superposition of multiple, faint and unresolved gamma-ray source populations, (e.g., blazars, misaligned active galactic nuclei, and star-forming galaxies), characterized on a statistical level by their source-count distributions -- \dnds~for short. The \dnds~describes how many astrophysical (point-like) sources emit gamma rays at different brightness levels. In other words, it tells us how common faint sources are compared to bright ones. Additionally, other, genuinely diffuse, Poissonian processes, e.g. from annihilation or decay of dark matter particles, may contribute to the IGRB composition~\cite{2015PhR...598....1F}.

Since intensity measurements alone provide limited insight into the nature of the underlying sources,  complementary observables -- such as anisotropies~\cite{2012PhRvD..85h3007A,Fornasa:2016ohl,2017PhRvD..95l3006A, Fermi-LAT:2018udj}\footnote{Due to the existence of anisotropies in the IGRB, it is sometimes referred to as diffuse gamma-ray background, DGRB.} and cross-correlations with multi-wavelength tracers~\cite{Xia:2011ax, Xia:2015wka, Branchini:2016glc, Cuoco:2017bpv, Ammazzalorso:2018evf} -- are routinely employed to disentangle different  source populations. Crucially, the distinction between resolved and unresolved components is not absolute, 
but it depends on the instrument's sensitivity: pushing source-detection thresholds enables better constraints
on the gamma-ray luminosity functions of different source classes, and informs models of the IGRB's
unresolved contributions.
The current \Fermi-LAT catalog, 4FGL-DR4~\cite{Ballet:2023qzs}, includes more than 7000  sources detected in 14 years of cumulated gamma-ray observations, and it is complete above flux values of about $1 \times 10^{-10}$~cm$^{-2}$~s$^{-1}$
at latitudes above 30$^\circ$ in the energy range from 1 to 10 GeV (see Fig.~\ref{fig:detection_histogram}). 
Below this flux threshold, the detection efficiency drops significantly, and the \Fermi-LAT catalog becomes
incomplete. 

To probe this faint regime, statistical techniques such as photon-count statistics have been developed. 
In particular, the one-point photon-count probability distribution function (1p-PDF)~\cite{Zechlin:2015wdz, Zechlin:2016pme} and the non-Poissonian template fitting (NPTF)~\cite{Lee:2014mza} (both building on \cite{Malyshev2011StatisticsLimit}) have proven effective in extending measurements of the 
\dnds, to lower fluxes than the \Fermi-LAT catalog limit. Both the 1p-PDF \cite{Zechlin:2015wdz, Manconi:2019ynl} and the NPTF method \cite{Lisanti:2016jub, Somalwar:2020awt} have been tested on high Galactic latitudes. However, due to differences in the data selection criteria, the NPTF results are not directly comparable to ours, and we will instead often take the 1p-PDF results by~\cite{Zechlin:2015wdz} as a baseline here.

These results have led to important advances, including the study of the
\dnds~energy dependence \cite{Lisanti:2016jub}, refined estimates of the IGRB composition~\cite{Manconi:2019ynl} also in light of anisotropy measurements, 
and improved constraints on dark matter models~\cite{Zechlin:2017uzo}.
Despite the progress these techniques permitted, they are intrinsically limited to yield a mere statistical description of the
source-count distribution: While they can provide an accurate characterization of this quantity, they are by definition not capable of deriving exact sky realizations (in space, and flux) of the individual point-like sources, both, in the dim and bright regime.
Moreover, the extension to multiple energy bins -- which is important to study the energy evolution of the \dnds~and the contribution of spectrally different source populations -- is, in principle, possible but technically challenging, and it quickly hits computational limitations. 
Finally, background mis-modeling -- notably of the Galactic diffuse emission, albeit mild at high latitudes -- severely impacts the reconstruction of the \dnds~in more systematic-dominated regions such as the inner Galaxy~\cite{Leane:2019xiy, Chang:2019ars, Buschmann:2020adf, Leane:2020pfc}. Strategies have been developed to mitigate this systematic uncertainty in the inner Galaxy, 
combining data-driven background optimization and photon-count statistical methods, e.g.~\cite{Calore:2021jvg, Manconi:2024tgh}.
These techniques are, nonetheless, hybrid approaches where absolute care should be put into avoiding inconsistencies in the data treatment.

These more conventional inference techniques above fundamentally rely on explicitly defined likelihood functions based on the approximation that all pixels are statistically independent (effects of the finite angular resolution of the instrument are, however, accounted for in each pixel). While powerful in principle, such likelihood functions become increasingly intractable as model complexity and data dimensionality grow. This limitation is especially critical in scenarios involving many free parameters, intricate instrumental effects, stochastic physical processes that are difficult to model analytically, or nuisance parameters introduced to account for known model uncertainties. As a result, classical methods such as the 1p-PDF and NPTF approaches necessitate handcrafted summary statistics or reduced data representations to remain computationally feasible and to define an analytically tractable likelihood function in the first place. This can limit the scientific insight gained from the analysis. 

In contrast, simulation-based inference (SBI) methods -- often referred to as likelihood-free or implicit inference \cite{Cranmer:2019eaq} -- offer a more flexible and scalable alternative. SBI leverages forward simulations to learn the complex parameter-to-observation mapping directly, bypassing the need for a tractable likelihood (which is learned implicitly). Combined with neural networks, SBI can also derive a suitable summary statistic tailored to the problem at hand. This allows one to incorporate high-fidelity models, observational effects, and latent stochasticity without compromising the integrity of the inference. Crucially, SBI scales more favorably with the dimensionality of both the parameter (thanks to its ability to directly marginalize over parameters not of interest) and data spaces (thanks to recent advances in machine learning that enable efficient high-dimensional data processing), making it particularly well-suited for complex statistical problems in high-energy astrophysics and beyond.

A first step towards a \textit{neural}, i.e.~invoking machine learning, simulation-based reconstruction of the \dnds~at high latitudes has been reported in \cite{Amerio:2023uet}. The authors leverage convolutional neural networks trained on patches of simulated sky maps to derive the source-count distribution and its intrinsic uncertainty at high Galactic latitudes from \textit{Fermi}-LAT data. While the authors of this work resorted to a likelihood-based loss function to train the network, our approach presented here follows the spirit of SBI, and does not make any analytic assumptions regarding the likelihood. We defer a more detailed comparison between our analysis and \cite{Amerio:2023uet} to the concluding part of our study.

On the other hand, SBI has been applied to LAT data of the inner Galaxy using normalizing flows \cite{Mishra-Sharma:2021oxe} and quantile-based posterior estimation \cite{List:2021aer} (and in~\cite{Mishra-Sharma:2020kjb, Christy:2024gsl} on simulated data only). These studies seek to decompose the gamma-ray emission towards the central region of the MW. In particular, their objective is related to unraveling the nature of the GeV gamma-ray excess in the MW's center observed by the \textit{Fermi} LAT \cite{Goodenough:2009gk, Hooper:2010mq}. To this day, the nature of this excess remains an unsolved problem of gamma-ray astronomy. A key limitation of current SBI applications towards the inner Galaxy lies in the \emph{reality gap} -- the mismatch between simulated data
used for training and the complexities of the real sky~\cite{Caron:2022akb}. The authors of \cite{Mishra-Sharma:2021oxe} already showed how effective methods like Gaussian processes can be combined with SBI to partially mitigate the impact of model mis-specification. However, since the central part of the MW hosts a plethora of distinct gamma-ray emission components which all come with their known (and, most crucially, unknown) uncertainties, we argue that high Galactic latitudes offer a much cleaner environment to develop and validate robust SBI frameworks.

In this work, we explore the application of SBI to high-latitude \Fermi-LAT data with the goal of building 
a reliable inference pipeline based on neural ratio estimation (NRE) \cite{Hermans:2019ioj, Miller:2020hua}. By using realistic sky models and accounting for instrumental 
and astrophysical uncertainties, we aim to demonstrate the potential of SBI to probe the unresolved gamma-ray sky through the determination of the \dnds~of faint sources.
\medskip

Given the highly technical nature of this manuscript -- necessary to rigorously demonstrate the validity of our approach -- we find it useful to first provide an overview of the main results.
\begin{itemize}[leftmargin=-0.02in]
\item \textit{SBI framework:} We developed a complete SBI pipeline tailored to gamma-ray astrophysics. 
It comprises: \emph{(i)} a fast and realistic simulator for the high-latitude GeV gamma-ray sky (cf.~Sec.~\ref{sec:model}), 
incorporating both point-like sources and extended diffuse backgrounds, 
including variations in Galactic diffuse emission via Gaussian random fields (GRFs), 
and \emph{(ii)} a neural network-based inference scheme, relying on NRE and 
spherical convolutional architectures, designed to operate directly on photon-count sky maps (cf. Sec.~\ref{sec:method}). 

\item \textit{Source detection performance:} Applying our detection network to real \Fermi-LAT data 
at $|b|>30^{\circ}$ over the energy range 1--10 GeV: \emph{(i)} we recover up to 98\% of unflagged 4FGL-DR4 
sources above a flux threshold of $S = 3 \times 10^{-10}\mathrm{cm}^{-2}\mathrm{s}^{-1}$ (cf.~Sec.~\ref{sec:results_det}), 
\emph{(ii)} we identify potential source candidates below the 4FGL completeness limit (cf.~Fig.~\ref{fig:detection_histogram} and Fig.~\ref{fig:lat-detection-wGRF}), 
and \emph{(iii)} we demonstrate that training with GRF-enhanced simulations improves both completeness and precision by reducing spurious detections by almost a factor of 2 (cf.~App.~\ref{app:lat-detection-wGRF}).

\item \textit{Reconstruction of the source-count distribution (\dnds):} We infer the high-latitude \dnds~through two complementary strategies: \emph{(i)} explicitly from the set of detected point-like sources,
and \emph{(ii)} from full-sky binned photon-count maps using a parametric and a non-parametric SBI inference network.  
The \dnds~results from both approaches are consistent with one another at the $1\sigma$ level, agree with the distribution implied by the 4FGL-DR4 catalog, and reproduce earlier literature results based on traditional 
likelihood analyses (cf.~Sec.~\ref{sec:results_dndS_woGRF}).

\item \textit{Performance and calibration:} We validate our inference framework on simulations, showing: \emph{(i)} robust recovery of the \dnds~even in the low-flux regime, where sources contribute fewer than 10 photons on average (cf.~Sec.~\ref{app:simulation-1pPDF-parametric}), 
and \emph{(ii)} well-calibrated posterior distributions, confirmed through Bayesian coverage tests (see Appendix \ref{app:coverage}). 

\item \textit{Robustness of the simulator:} We show that: \emph{(i)} incorporating GRF-based distortions of the diffuse emission enhances the realism of the simulated maps without biasing the inferred \dnds, \emph{(ii)} the inclusion of GRFs leads to improved agreement with catalog-based and literature results (for both points see Sec.~\ref{sec:results_dndS_wGRF}), and \emph{(iii)} using a neural-network-based anomaly detection test, we verify that the observed LAT sky lies within the data space spanned by our simulations, both with and without GRFs (see Appendix \ref{app:svdd}). 
\end{itemize}

\medskip
The outline of this paper reflects our guiding principle: to emphasize the results obtained with our SBI approach on the \textit{Fermi}-LAT high-latitude sky. Accordingly, the main body of this work focuses exclusively on findings based on real gamma-ray data. The performance of our method on simulated datasets is presented in a series of appendices. Specifically, the structure of this study is the following one:
In Sec.~\ref{sec:data}, we detail the selection, cleaning, and preparation of the \textit{Fermi}-LAT dataset used in our analysis. Sec.~\ref{sec:model} outlines the technical implementation of our gamma-ray emission forward model, which generates synthetic all-sky maps. This includes the modeling of the LAT’s instrument response functions, point-like source populations, and astrophysical background components. In Sec.~\ref{sec:method}, we describe the SBI inference strategies, the associated neural network architectures, and relevant technical aspects. The main results derived from the \textit{Fermi}-LAT high-latitude dataset are presented in Sec.~\ref{sec:results}, where we highlight the findings from two complementary approaches to reconstruct the source-count distribution: one based on point-source detection and the other via direct inference from the data. These results are interpreted and discussed in Sec.~\ref{sec:discussion}, followed by our conclusions in Sec.~\ref{sec:conclusion}.
The appendices provide supplementary material, validation results, and sanity checks of our SBI framework. In App.~\ref{sec:data-4fgl}, we explain how we utilize the \textit{Fermi}-LAT collaboration’s most recent gamma-ray source catalog, 4FGL-DR4. App.~\ref{app:NN-architectures} documents the neural network architectures and training strategies used throughout this study. Additional results supporting our inference on real LAT data are collected in App.~\ref{app:realData-material}, while App.~\ref{app:simulations} demonstrates the performance of our framework on simulated datasets. In App.~\ref{app:gof}, we assess the calibration of the inferred posterior distributions through Bayesian coverage tests. We also evaluate the realism of our gamma-ray emission simulator using neural network-based anomaly detection. Finally, in App.~\ref{app:mismodeling}, we test the robustness of our framework by applying it to synthetic data intentionally designed to exhibit specific mis-modeling in the Galactic diffuse emission and source-count distribution.

\section{\textit{Fermi}-LAT gamma-ray data preparation}
\label{sec:data}

We analyze the cumulative all-sky LAT Pass 8 data of 14 years in the energy range from 1 GeV to 10 GeV collected from the 4th of August 2008 to the 2nd of August 2022. From the available events, we select those that fall within the \texttt{SOURCEVETO} event class and are classified as \texttt{FRONT}-type events. We apply the event quality cut (\texttt{DATA\_QUAL>0 \&\& LAT\_CONFIG==1}). We also require zenith angles of less than $90^{\circ}$ to reduce the contamination by photons from the Earth's limb. These data selection and cleaning steps, as well as all tasks requiring LAT data products, are performed with the \textit{Fermi} Science Tools\footnote{\url{https://github.com/fermi-lat/Fermitools-conda}} (version 2.2.0) \cite{2019ascl.soft05011F}. We note that up to 10 GeV, the \texttt{SOURCEVETO} event class utilizes the same selection criteria as the \texttt{SOURCE} event class -- the recommended class for analyses examining the properties of point-like or slightly extended sources. Beyond 10 GeV, the \texttt{SOURCEVETO} event class is almost identical to the \texttt{ULTRACLEANVETO} class. The adopted data selection specifications allow us to compare the results of our work with some previous assessments of the high-latitude source-count distribution reported in the literature \cite{Zechlin:2015wdz, Amerio:2023uet}. 

\begin{figure}[h]
    \includegraphics[width=\columnwidth]{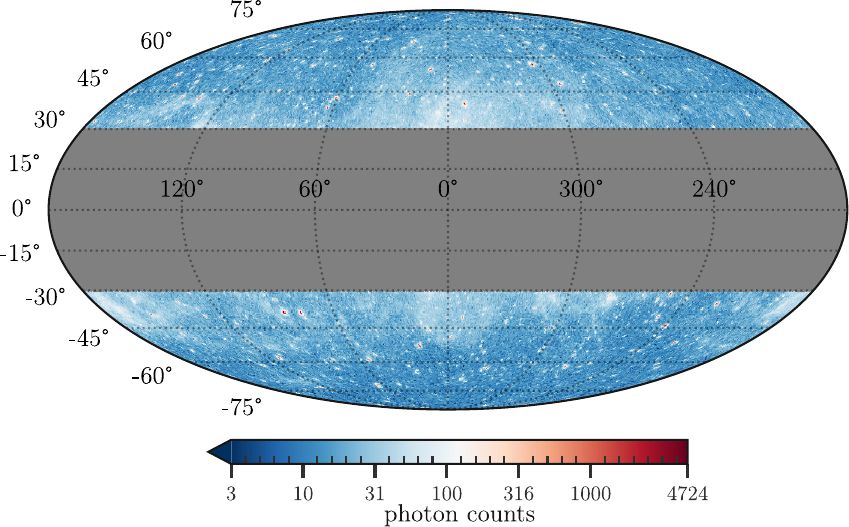}
    \caption{High-latitude ($|b| > 30^{\circ}$) 14-year \textit{Fermi}-LAT photon-counts map from 1 to 10 GeV using the \texttt{HEALPix} resolution of $N_{\mathrm{side}} = 128$, which corresponds to an average angular separation between adjacent pixels of about $0.5^{\circ}$.}
    \label{fig:fermi-data}
\end{figure}

Throughout this study, we work with binned all-sky maps following the \texttt{HEALPix} tessellation \cite{2005ApJ...622..759G} in the \texttt{NESTED} scheme and use a resolution of $N_{\mathrm{side}} = 128$, which corresponds to an average angular pixel distance of $0.5^{\circ}$. Events with Galactic latitudes $|b| \leq 30^{\circ}$ are masked. All inference tasks on simulated and real LAT data are based on a single energy bin that spans the entire range. However, we employ binned data to set up the gamma-ray simulator that we outline later in Sec.~\ref{sec:instrument}.  We show the binned LAT gamma-ray counts in our region of interest (ROI) in Fig.~\ref{fig:fermi-data}.

\section{Simulation framework}
\label{sec:model}

This section outlines the setup of our forward simulator of the high-latitude gamma-ray sky as seen by the \textit{Fermi} LAT. It is our SBI framework's core since the quality of every inference task crucially depends on its ability to accurately model the observed gamma-ray emission. The simulator draws from multiple components: it leverages data-driven components obtained via the \textit{Fermi} Science Tools (Sec.~\ref{sec:instrument}), a model and simulation prescription for point-like sources (Sec.~\ref{sec:ps_fm}), and a background model part (Sec.~\ref{sec:de_fm}). A schematic overview of our forward model is given in Fig.~\ref{fig:bayesian_model} supplemented by a definition of symbols in Tab.~\ref{tab:model-definitions}.

\subsection{Instrument- \& data-driven components}
\label{sec:instrument} 

The format of the gamma-ray simulator's output is designed to reproduce the choices made for the selected LAT data (cf.~Sec.~\ref{sec:data}). Hence, the simulator generates all-sky \texttt{HEALPix} maps with a resolution of $N_{\mathrm{side}} = 128$ for a single energy bin from 1 to 10 GeV. 

Predicting the observed gamma-ray counts emitted by discrete sources and high-energy astrophysical processes requires a convolution of their spectral energy distribution with the Instrument Response Functions (IRFs) of the LAT and the proper observation conditions such as exposure. The IRFs and exposure depend on the selected event class, event type, and true photon energy.~\footnote{See, e.g., the LAT performance page: \url{https://www.slac.stanford.edu/exp/glast/groups/canda/lat_Performance.htm}} Hence, we need to account for the instrumental effects that match our LAT data selection criteria while treating their
energy dependence 
correctly. To this end, we utilize multiple \textit{Fermi} Science Tools routines that readily yield the required quantities. In general, we aim for an efficient and accurate simulator that operates faster than similar Monte Carlo simulation routines of the \textit{Fermi} Science Tools. Our design choices reflect this objective.

\vspace{10pt} 
\paragraph{Exposure.} The LAT exposure is, roughly speaking, the product of its effective area -- the instrument's geometric surface capable of registering incident primary gamma-ray events -- and the observation time regarding a specific location in the sky. It takes into account time periods when the instrument did not point towards that location and periods of bad data quality due to, e.g., large zenith angles. We use the routine \texttt{gtexpcube2} regarding \texttt{FRONT}-type events~\footnote{Gamma rays of this event type converted in the first 12 layers of the LAT's particle tracker where photon-to-$e^+e^-$-pair conversions occur. These layers are thinner than the following 4 thick layers in the back of the tracker. In the front part, multi-scattering events are less likely than in the thicker part, which improves the angular resolution.} to generate \texttt{HEALPix} all-sky exposure maps evaluated at 16 logarithmically spaced energies between 1 and 10 GeV. The number of sample energies is sufficient to accurately model the energy dependence of the exposure (which is mild in this energy range). We note that due to the selection of \texttt{FRONT}-type events only, we lose about 50\% of the available photon statistics for the chosen event class. Yet, photons of this event type exhibit a more precise angular reconstruction, reducing the effect of the Point-Spread Function (PSF) of the LAT.

We process the exposure maps to derive an energy-averaged exposure $\bar{\mathcal{E}}$ as a function of position in the sky $\vec{p}$ by performing a weighted integral over the entire energy range using a power law $E^{-2.4}$ as weighting factor:
\begin{equation}
   \bar{\mathcal{E}}(\vec{p}) = \frac{\intop_{1\;\mathrm{GeV}}^{10\;\mathrm{GeV}} E^{-2.4} \mathcal{E}(E, \vec{p})\,\mathrm{d}E}{\intop_{1\;\mathrm{GeV}}^{10\;\mathrm{GeV}} E^{-2.4}\,\mathrm{d}E}.  
\end{equation}
The imposed power-law index of $-2.4$ is the average index of the 4FGL-DR4 sources within the energy range and ROI chosen in this study. The energy-averaged exposure $\bar{\mathcal{E}}$ retains its full dependence on the position in the sky, which we exploit later to generate gamma-ray sky maps. This goes beyond the exposure treatment in the 1p-PDF and NPTF studies where the sky is broken into a number of exposure regions. 

\vspace{10pt} 
\paragraph{Point-spread function.} Modeling and implementing a prescription for the PSF of the LAT is the most delicate part of the simulator. This prescription predominantly determines the speed of individual simulations. We devise two different ways of convolving the spectral energy distribution of a gamma-ray emission component with the PSF; one is applied to point-like sources, and the other one is applied to large-scale emission components and extended sources.

To handle point-like sources, we derive the angular profile of the LAT's PSF averaged over the selected dataset via the routine \texttt{gtpsf} at the fixed source location of $(\ell, b) = (90^{\circ}, 40^{\circ})$. We evaluate the PSF at 16 logarithmically spaced energies between 1 and 10 GeV to exhaustively sample its energy dependence. Lastly, we derive an energy-averaged PSF probability density function in complete analogy to the average exposure map. We verify that the dependence of the resulting 
energy-averaged PSF on the specified source location in the sky is very mild. Hence, choosing a representative position does not negatively impact the quality of our simulator. The technical details about how we use this PSF object for point-like sources are given in Sec.~\ref{sec:ps_fm}.

Convolving large-scale gamma-ray components with the LAT's PSF is conducted at the \texttt{HEALPix} all-sky map level. To this end, we simulate the expected gamma-ray counts of a point-like source at $(\ell, b) = (90^{\circ}, 40^{\circ})$ in a square ROI of $60^{\circ}\times60^{\circ}$ centered on the source's position\footnote{The number of spatial pixels in the horizontal and vertical direction is an odd number so that the source sits in the central pixel of the map.} with a spatial resolution of $0.05^{\circ}$ exhibiting a power-law spectrum $\propto E^{-2.4}$ with \texttt{gtmodel} in 16 logarithmically spaced energy bins between 1 and 10 GeV. We note that \texttt{gtmodel} does not perform a Monte Carlo simulation of single photons but generates the mean expected photon counts per pixel in the limit of infinite statistics, the so-called \textit{Asimov dataset} \cite{Cowan:2010js}. Therefore, the generated Cartesian photon count map is a smooth representation of how the PSF smears out a single pixel across the sky as a function of energy. Then, we compute the radial PSF profile in terms of photon counts along the vertical and horizontal axes through the source's location and take the arithmetic mean of both profiles. This time, we derive the energy-averaged PSF from the arithmetic mean by integrating it from 1 to 10 GeV without an energy weight factor because the PSF is computed from a counts map where this effect is already included. We then translate the obtained PSF profile into a \textit{beam window function} at a resolution of $N_{\mathrm{side}} = 512$ via the \texttt{python} package \texttt{healpy}, which can smooth input all-sky maps based on a given beam window function. We normalize the generated function by its maximal attained value and clip negative values. We verified that smoothing an all-sky map with a single non-zero pixel at the test source location by employing the custom beam window function reproduces the original simulated PSF profile of the point-like source. 

\begin{figure}
    \centering
    \includegraphics[width=\columnwidth]{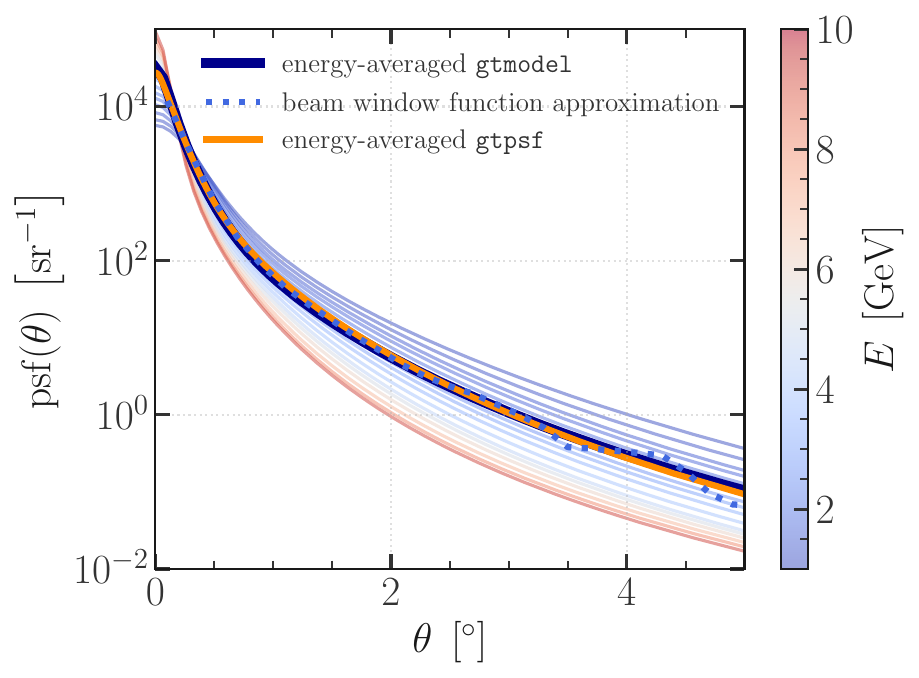}
    \caption{Comparison of the PSF profiles considered in this work. The solid blue line is the energy-averaged angular PSF profile obtained from \texttt{gtmodel}, simulating a Cartesian count map of a single point-like source with spectrum $\sim E^{-2.4}$ at $(\ell, b) = (90^{\circ}, 40^{\circ})$, while the transparent lines characterize the same PSF profile evaluated at different energies as stated in the colorbar. The dotted light blue profile is the result of employing the beam window function derived from the latter PSF profile to a \texttt{HEALPix} map with a single non-zero pixel (value 1) at the position of this source. It has been calculated by summing the emission normalized by the angular area computed from concentric annuli of width $0.2^{\circ}$ centered on the source. For comparison, the orange PSF profile was directly obtained via \texttt{gtpsf} for a source with the same properties. The latter PSF profile is utilized to model the PSF convolution for point-like sources in our simulator. All radial PSF profiles are normalized so that $2\pi\int\mathrm{psf}(\theta)\theta\,\mathrm{d}\theta = 1$ over the considered angular separation range from 0 to $30^{\circ}$ (the full range in which we extract the PSF from the \textit{Fermi} Science Tools).}
    \label{fig:psf_comparison}
\end{figure}

To demonstrate that the devised \texttt{healpy} approach satisfyingly mimics the LAT PSF for the selected dataset, we compare in Fig.~\ref{fig:psf_comparison} the obtained radial PSF profile with the original simulation results from the \textit{Fermi} Science Tools. The slightly transparent PSF profiles are the arithmetic mean of the horizontal and vertical profiles read off from the \texttt{gtmodel} output. Their color corresponds to the energy shown in the colorbar on the right. The dark blue line is the energy-averaged PSF profile resulting from the \texttt{gtmodel} approach. For comparison, the orange line provides the corresponding energy-averaged PSF obtained via \texttt{gtpsf}, which is used for point-like sources. Both energy-averaged PSFs agree with each other to a reasonable degree. The dotted light blue PSF profile emerges when we translate the dark blue curve into a \texttt{healpy} beam window function and apply it to a source located at $(\ell, b) = (90^{\circ}, 40^{\circ})$ in an all-sky \texttt{HEALPix} map of $N_{\mathrm{side}} = 512$.~\footnote{We essentially set all pixel values to zero except for the pixel containing $(\ell, b) = (90^{\circ}, 40^{\circ})$ to which the value 1 is assigned.} Then, we sum the total emission within concentric annuli of width $0.2^{\circ}$ centered on the active pixel and divide the result by the respective annulus' angular area. Indeed, the emerging synthetic PSF convolution generates a profile compatible with the dark blue original. We note that the profile starts at $\theta = 0.1^{\circ}$, the mean angular separation from the central source in the first annulus. However, we point out that there is a slight ringing feature towards large angular separations caused by the Fourier transforms that \texttt{healpy} exploits under the hood to perform the convolution. These artifacts should not strongly impact the realism of the simulator since about 98\% of the total gamma-ray emission is confined within the inner $2^{\circ}$ of the PSF profile. In this regime, the ringing is absent.

\vspace{10pt}
\paragraph{Energy dispersion.} We do not model the effect of energy dispersion, the statistical correlation between true photon energy and experimentally reconstructed energy. It is an important ingredient for studies that heavily rely on precise and accurate spectral analyses. Here, we study the gamma-ray sky in a single energy bin, thus removing all available spectral information. With only a single energy bin, the effective consequence of energy dispersion would be the rescaling of the flux of a given gamma-ray emission component; an effect we can capture in our forward model by construction.~\footnote{The authors of \cite{Linden:2016rcf} checked the impact of an explicit model for the LAT's energy dispersion with the NPTF approach on (Pass 8) LAT data of the Galactic center region without finding major differences.} The only exception would be a model component whose spatial morphology changes rapidly with energy. At high latitudes, such gamma-ray emitters are not expected.

\subsection{Forward model: Point-like sources}
\label{sec:ps_fm}

Forward modeling point-like sources is the most important part of our gamma-ray emission simulator since we aim to detect as many sources as possible individually while inferring the source-count distribution, i.e. the source number count in terms of flux, for the dim and unresolved parts of the population. 
Therefore, we need to specify three aspects:
\begin{itemize}[leftmargin=-0.18in]
    \item How many sources are there?
    \item What is their individual flux in the $(1,10)$ GeV range?
    \item Where are the sources?
\end{itemize}
As we are dealing with a single energy bin, we do not need to make assumptions about the intrinsic spectral energy distribution of each source, although it is straightforward to generalize our forward simulator to account for it. 

\vspace{10pt}
\paragraph{Flux sampling.} To answer the first two questions, we start from the source-count distribution parametrized as a multiply broken power law as a function of the source flux $S$ from 1 to 10 GeV with $N_b$ breaks at positions $\left\{S_{b,j}\right\}_{j=1}^{N_b}$,
\begin{widetext}
\begin{equation}
\label{eq:dNdS-parametric}
    \frac{\mathrm{d}N}{\mathrm{d}S}= A_S \times \left\{ \begin{array}{cc} \left(\frac{S}{S_0}\right)^{-n_1}, & S > S_{b,1} \\
    \left(\frac{S}{S_0}\right)^{-n_2}\left(\frac{S_{b,1}}{S_0}\right)^{-n_1+n_2}, & S_{b,1} \geq S > S_{b,2}\\
    \ldots & \\
     \left(\frac{S}{S_0}\right)^{-n_{N_b+1}}\ldots\left(\frac{S_{b,2}}{S_0}\right)^{-n_{2}+n_3}\left(\frac{S_{b,1}}{S_0}\right)^{-n_1+n_2}, & S_{b,N_b} \geq S 
    \end{array}\right.
\end{equation}
\end{widetext}
where the $\left\{n_i\right\}_{i=1}^{N_b+1}$ refer to the slopes in each power-law segment, while $S_0$ is a pivot flux which we fix to $S_0 = 3\times10^{-8}\;\mathrm{cm}^{-2}\,\mathrm{s}^{-1}$. The authors of \cite{Zechlin:2015wdz} verified the suitability and robustness of this choice for $S_0$. The parameter $A_S$ acts as an overall normalization, which is left as a free parameter in addition to the break positions and slopes (cf.~the schematic visualization in the upper left part of Fig.~\ref{fig:bayesian_model}). The set of all of these gamma-ray source population parameters is denoted by $\vec{\theta}$.

From this parametric form of the $\mathrm{d}N/\mathrm{d}S$, we can randomly draw the fluxes of the individual high-latitude sources corresponding to an input $\vec{\theta}$. To guarantee a finite total number of sources in the sky, we impose that $n_1 >2$ and $n_{N_b+1} < 2$. For practical purposes, we limit the range of fluxes from which we draw sources to $S\in\left[10^{-13}, 10^{-5}\right]\;\mathrm{cm}^{-2}\,\mathrm{s}^{-1}$. The \textit{benchmark parametrization} for all further studies will be the triple-broken power law (TBPL), which was also considered in previous analyses \cite{Zechlin:2015wdz, Amerio:2023uet}. To ascertain comparability between these publications and our work, we define very similar prior ranges for $\vec{\theta}$, detailed in the first part of Tab.~\ref{tab:params}. 

We employ log-uniform priors for $A_S$ and the three break positions. The ranges for the latter are defined as to avoid an overlap. The slope parameters all feature a uniform prior. In general, the reasoning behind the selected prior ranges is to ensure moderate variability of the parameters around their values suggested by the 4FGL source catalog, i.e.~with respect to the \textit{Fermi}-LAT sky. 

\vspace{10pt}
\paragraph{Spatial distribution.} Regarding the individual positions of the sources, we choose a \textit{uniform distribution across the sky} in line with the fact that most of the high-latitude sources are of extragalactic origin. We point out that a small number of Galactic sources such as millisecond pulsars and globular clusters may be present at higher latitudes. As concerns detected objects of these source classes, their spatial distribution is compatible with a uniform distribution \cite{Fermi-LAT:2023zzt} so that our setup covers them without additional refinements. We note that non-uniform distributions that align with the large-scale structure of the Universe have also been explored in the past \cite{Bartlett:2022ztj}.

\vspace{10pt}
\paragraph{Convolution with the LAT PSF.} The PSF convolution is realized as a direct Monte Carlo sampling of the PSF profile for each source. We adopt this approach from earlier works on gamma-ray data \cite{Mishra-Sharma:2016gis, List:2020mzd, List:2021aer}. Given a source with flux $S$ at location $\vec{x}$ in the sky, its emitted gamma rays including PSF effects are simulated as follows:
\begin{itemize}[itemindent=0.1in, leftmargin=0.2in]
    \item Compute the expected counts $\bar{C}$ in the range from 1 to 10 GeV by multiplying $S$ with the energy-averaged exposure evaluated at the source's location: $\bar{C} = S\times\bar{\mathcal{E}}(\vec{x})$.
    \item Draw a Poisson realization of the number of emitted gamma rays $C$ from $\bar{C}$.
    \item Determine the pixel location of the source in the \texttt{HEALPix} map with resolution $N_{\mathrm{side}} = 128$. Since the source could be anywhere inside this pixel of about $0.5^{\circ}$, we randomly draw its exact position within the parent pixel by upscaling it to $N_{\mathrm{side}} = 16384$ and selecting 
    any of the encompassed daughter pixels with equal probability as location $\hat{\vec{x}}$.
    \item Sample $C$ times from the energy-averaged PSF prepared in Sec.~\ref{sec:instrument} a radial angular displacement $\delta\psi$ from the true location $\hat{\vec{x}}$ of the source.
    \item Place each of the $C$ gamma-ray events at the sky's south pole. For each of them, draw an angle $\delta\varphi\in\left[0, 2\pi\right]$ from a 
    uniform distribution. Compute the new coordinates respecting the displacement by $\delta\varphi$ and $\delta\psi$ relative
    to the south pole for each event and rotate them
    onto the actual source's position $\hat{\vec{x}}$. In fact, uniformly sampling $\delta\varphi$ is admissible because we assume that the PSF profile is radially symmetric (neglecting, e.g., fish-eye effects).
    \item Repeat this procedure for each point-like source to generate the \textit{point-source map} $\boldsymbol{c}$. 
\end{itemize}
The advantage of this procedure is its parallelizability, allowing us to treat many sources at the same time and, thus, reducing the overall computation time to obtain 
$\boldsymbol{c}$. By parallelization, we here refer to a vectorization over all photon counts from all sources.

Moreover, this Monte Carlo-based approach produces a stochastic realization of discrete integer counts in our simulated maps, consistent with the \textit{Fermi} photon counts map, rather than a smooth deterministic distribution of fractional counts that would result from convolving the counts with the PSF profile. On average, given the $\mathrm{d}N/\mathrm{d}S$ parameters' priors, calculating $\boldsymbol{c}$ takes about 40 milliseconds on a single A100 GPU with 80 GB RAM. However, realizations of the source-count distribution with many bright point-like sources take longer to compute because of the growing number of single events to be rotated. The computation time (and memory in our vectorized implementation) scales linearly with the total number of photon counts produced by an emission component. This is the reason why we decided to treat the PSF convolution of large-scale gamma-ray components differently (cf.~Sec.~\ref{sec:de_fm}).

\vspace{10pt}
\paragraph{Rundown of the point-source map computation.} To conclude, we summarize the steps of our simulator to obtain a Poisson realization of the point-source contribution to the gamma-ray sky originating from the TBPL source-count distribution in Eq.~\eqref{eq:dNdS-parametric}. 
\begin{enumerate}
    \item Draw a realization of $\vec{\theta}$ 
    from the TBPL priors defined in Tab.~\ref{tab:params}.
    \item Compute the corresponding total number of sources in the entire sky in the flux range $S\in\left[10^{-13}, 10^{-5}\right]\;\mathrm{cm}^{-2}\,\mathrm{s}^{-1}$ by integrating the TBPL. To this end, we bin the full flux range and perform the integration per flux bin. We use a single bin $\left[10^{-7}, 10^{-5}\right]\;\mathrm{cm}^{-2}\,\mathrm{s}^{-1}$, 20 logarithmically spaced flux bins $\vec{\mathcal{S}}$ between $10^{-7}$ and $5\times10^{-12}\;\mathrm{cm}^{-2}\,\mathrm{s}^{-1}$ and five additional bins in the dim flux range: $\left[1\times10^{-12}, 5\times10^{-12}\right]$, $ \left[8\times10^{-13}, 1\times10^{-12}\right]$, $\left[5\times10^{-13}, 8\times10^{-13}\right]$, $\left[3\times10^{-13}, 5\times10^{-13}\right]$, $ \left[1\times10^{-13}, 3\times10^{-13}\right]\;\mathrm{cm}^{-2}\,\mathrm{s}^{-1}$.
    \item Regarding the 20 flux bins $\vec{\mathcal{S}}$, we first compute the flux-averaged $\mathrm{d}N/\mathrm{d}S$ in each bin and subsequently multiply the result by $S_c^2$, where $S_c$ is the central flux value of the respective bin. We store these values of $S^2\mathrm{d}N/\mathrm{d}S$ in units of $10^{-11}\;\mathrm{cm}^{-2}\,\mathrm{s}^{-1}\,\mathrm{deg}^{-2}$. This yields a bin-by-bin description $\vec{\theta}_S$ of the $\mathrm{d}N/\mathrm{d}S$, which we will use later in the parameter inference part (see Sec~\ref{sec:stat_nonparametric}). The authors of \cite{Amerio:2023uet} inferred the $\mathrm{d}N/\mathrm{d}S$ based on this representation. 
    \item Based on the number of sources per flux bin, we draw the number of realized sources from a Poisson distribution and sample their respective fluxes from the TBPL (by respecting the flux bin boundaries).
    \item We draw each point-like source's position $\vec{x}$ in terms of pixel indices directly from the all-sky \texttt{HEALPix} map via a multinomial distribution, where the placement of each source represents a single trial, and the mutually exclusive events correspond to the source being placed in any one of the $N_{\mathrm{pix}} = 12N_{\mathrm{side}}^2$ pixels. The probability for any pixel in the sky is  
    $1/N_{\mathrm{pix}}$ to guarantee a uniform placement. Note that while for an isotropic population, one could also draw the sources' latitudes and longitudes directly from a continuous distribution, this discrete approach is easily extended to non-isotropic point-source populations by replacing the uniform event probabilities with given spatial template values that sum up to unity, as done for the disk and bulge populations in Ref.~\cite{List:2021aer}, avoiding the need for slow rejection-based sampling in those cases.
    \item We remove all sources from the list with $|b|\leq25^{\circ}$. We note that this is a larger ROI than what we will later analyze. Simulating sources beyond the high-latitude ROI accounts for gamma-ray emission leaking into the smaller ROI due to the LAT's PSF, which we would otherwise miss. Conversely, photons from sources within the ROI can also scatter outside.
    \item Convolve each source with the LAT PSF as outlined in the previous paragraph to generate the final point-source map $\boldsymbol{c}$. For each simulation, we also store a point-source map without the PSF effect, where each source emits $C$ counts in its own pixel. This map will be relevant for training the point-source detection network (cf.~Sec.~\ref{sec:stat_detection}), as it allows for easy identification of pixels that contain sources above a given brightness threshold. We stress that this map without PSF convolution is a Poisson realization of the average emission expected for each point-like source. 
\end{enumerate}
This procedure is schematically shown in the upper part of Fig.~\ref{fig:bayesian_model}. 

\subsection{Forward model: Background emission}
\label{sec:de_fm}

Besides discrete point-like sources, the gamma-ray sky features a large variety of astrophysical gamma-ray emitters. An accurate gamma-ray simulator must incorporate these components. At high Galactic latitudes, the number of astrophysical processes producing gamma rays is not as rich as in other regions of the sky like the Galactic plane or center, which reduces the required model complexity. Broadly speaking, our ROI receives gamma-ray emissions from these components:
\begin{itemize}[itemindent=0.1in, leftmargin=0.2in]
    \item \textbf{Galactic diffuse emission and large-scale structures}: This component arises from cosmic-ray interactions with interstellar gas, producing gamma rays via hadronic processes (mainly $\pi^0$ decay) and bremsstrahlung. Its morphology follows the gas distribution, with decreasing intensity at high latitudes. Inverse Compton (IC) scattering of high-energy electrons on the interstellar radiation field (ISRF) and cosmic microwave background (CMB) also contributes, with CMB scattering dominating at high latitudes. Additional large-scale structures include the Fermi Bubbles \cite{Su:2010qj}, Loop I \cite{Wolleben:2007pq}, and gamma-ray emission from the Sun and Moon \cite{2011ApJ...734..116A, 2012ApJ...758..140A}. All these contributions are implemented via the official diffuse background model, \texttt{gll\_iem\_v07}\footnote{ \interlinepenalty=100000 See \url{https://fermi.gsfc.nasa.gov/ssc/data/analysis/software/aux/4fgl/Galactic_Diffuse_Emission_Model_for_the_4FGL_Catalog_Analysis.pdf}.}, of the \textit{Fermi}-LAT collaboration derived in connection with the 4FGL source catalog series. It is constructed with a data-driven approach to calculate the gas emissivity (see \cite{Fermi-LAT:2016zaq} for details) and combines interstellar emission and IC contributions. Loop I and the FBs are part of this model and located in a so-called `patch' region that contains smoothed residual emission from a fit to LAT gamma-ray data. This model is comprised of the energy-dependent two-dimensional spatial morphology (see the lower all-sky map on the left side in Fig.~\ref{fig:bayesian_model}) of the diffuse emission and its spectrum from MeV to TeV energies.
    
    \item \textbf{Diffuse isotropic gamma-ray background}: As stated in Sec.~\ref{sec:introduction}, the composition of the IGRB is an open question of gamma-ray astronomy. It encompasses unresolved discrete extragalactic sources but also truly diffuse processes such as ultra-high-energy cosmic-ray cascades \cite{Fermi-LAT:2016hfi} and the residual emission arising from charged cosmic rays (especially electrons and positrons) that were misclassified as gamma-ray events. Since we explicitly simulate point-like (resolved and unresolved) sources as detailed in Sec.~\ref{sec:ps_fm} and characterize them via their \dnds~(with a separate normalization parameter $A_S$), point-like non-Poissonian contributions to the IGRB are already accounted for. Yet, we need to model the remaining components of the IGRB that are, in nature, isotropic across the sky and constitute a diffuse isotropic background. To this end, we use a featureless spatial morphology, i.e., uniform, Poissonian background template, whose only free parameter is its normalization $A_{\mathrm{iso}}$.        
    
    The associated gamma-ray spectrum is adopted from the differential energy spectrum derived by the \textit{Fermi}-LAT collaboration in connection with the 4FGL diffuse background model. It depends on the selected event class and type, hence, we employ the quantitative assessment provided by the \textit{Fermi} Science Tools called \texttt{iso\_P8R3\_SOURCEVETO\_V3\_FRONT\_v1.txt}. We note that this characterization by the \textit{Fermi}-LAT collaboration is the differential energy spectrum of the entire IGRB (including the component we model via the \dnds).
    
    \item \textbf{Extended sources}: These include cosmic-ray acceleration sites such as supernova remnants and pulsar wind nebulae, mostly within the Galactic disk. At high latitudes, only a few extended sources are present, notably the Large and Small Magellanic Clouds (LMC, SMC) \cite{2010A&A...523A..46A, Fermi-LAT:2015bpm}.  To model the LMC and SMC, we pick the spatial templates provided by the \textit{Fermi}-LAT collaboration as supplementary material to 4FGL-DR4. The SMC is represented by a single template \texttt{SMC-Galaxy}, while the LMC is the composite emission from four distinct spatial templates: \texttt{LMC-FarWest}, \texttt{LMC-Galaxy}, \texttt{LMC-30DorWest} and \texttt{LMC-North}. For each of these components, we adopt the best-fitting differential energy spectrum listed in 4FGL-DR4. Their positions and flux intensity in the range from 1 to 10 GeV are shown in the two central all-sky maps in the leftmost column of Fig.~\ref{fig:bayesian_model}. Fornax A, a radio galaxy at 18.6 Mpc with a gamma-ray extension of $0.33^{\circ}$ \cite{Fermi-LAT:2016hfi}, is treated as point-like for our purposes, that is, it will be part of the inferred \dnds without explicit model. In fact, the chosen default pixel size of our \texttt{HEALPix} maps is larger than the extension of this object. We note that the addition of the LMC and SMC to our simulated background components is by no means necessary or impacts the \dnds~inference eventually. We added them to devise an exhaustive simulator.
\end{itemize}

\vspace{10pt} 
\paragraph{Background model template generation.} We integrate the above background components in our simulator based on the listed model (spatial profile and spectrum). 

We process each respective model input and compute their flux by integrating the adopted differential energy spectrum in the range from 1 to 10 GeV. The energy dependence of the Galactic diffuse emission model's spatial morphology is taken into account. This step yields the initial background model templates $\boldsymbol{m}_{\mathrm{diff}}$ for the Galactic diffuse emission, $\boldsymbol{m}_{\mathrm{iso}}$ for the diffuse isotropic background, $\boldsymbol{m}_{\mathrm{LMC}}$ and $\boldsymbol{m}_{\mathrm{SMC}}$ for the LMC and SMC, respectively. 

\vspace{10pt} 
\paragraph{Simulation of the background map $\boldsymbol{b}$.} The procedure to derive the final background map for a gamma-ray sky realization is schematically depicted in the central part of Fig.~\ref{fig:bayesian_model}. It involves the following steps:
\begin{itemize}[itemindent=0.1in, leftmargin=0.2in]
    \item Per component, we derive an all-sky map $\bar{\boldsymbol{b}}_t$ (the index $t$ indicates that it is a component map) containing the mean number of expected counts by multiplying the background model template $\boldsymbol{m}_t$ with the energy-averaged exposure all-sky map $\bar{\mathcal{E}}$. 
    \item To convolve this prediction with the LAT's PSF, we utilize the energy-averaged beam window function described in Sec.~\ref{sec:instrument}. We apply this custom beam window function with the routine \texttt{healpy.sphtfunc.smoothing}. Since this function was derived for $N_{\mathrm{side}}=512$, we first upscale the resolution of $\bar{\boldsymbol{b}}_t$ to this number of pixels and, after the PSF convolution, scale it down again to the original resolution. This generates a PSF smoothed all-sky map of the expected counts $\bar{\boldsymbol{b}}_t^{\mathrm{PSF}}$. We save $\bar{\boldsymbol{b}}^{\mathrm{PSF}}_t$ for all background components in the first run of the simulator. All consecutive runs load the pre-convolved maps from file to reduce the overall computation time of a realization. 
    \item Next, we draw a single normalization parameter $A$ that multiplies the mean expected counts from a log-uniform prior between 0.1 and 2 (cf.~also Tab.~\ref{tab:params}). Hence, our background model allows for global variations in the brightness of each individual component, producing a realization-dependent mean expectation $\bar{\boldsymbol{b}}^{\ast}_t$.
    \item In the last step, we add together all components $\bar{\boldsymbol{b}}^{\ast}_t$ multiplied by their normalization constants, yielding the mean background map $\bar{\boldsymbol{b}}$. The final background map $\boldsymbol{b}$ is a Poisson realization of $\bar{\boldsymbol{b}}$. 
\end{itemize}
The output of our simulator, $\boldsymbol{d}$, the observed photon count map, is the sum of the point-source map $\boldsymbol{c}$ and the background map $\boldsymbol{b}$ masked at Galactic latitudes $|b|\leq30^{\circ}$.

\subsection{Forward model extension: Background variations via Gaussian random fields.} 
\label{sec:bkg-model-GRF}

We pointed out above that the spatial morphology of the Galactic diffuse emission is directly related to the gas density distribution in the MW and, therefore, is subject to uncertainties in the gas reconstruction.~\footnote{The same is true for the IC component and the challenging derivation of the spatial structure of Galactic ISRFs \cite{Porter:2017vaa}.}  These uncertainties of the Galactic gas distribution can induce large- and, in particular, small-scale mis-modeling of the Galactic diffuse emission. In addition, the observationally inferred gas density might lack a certain fraction of \textit{dark gas} that cannot be traced with the methods typically applied in the field.~\footnote{These methods involve measuring the absorption or emission of light especially in the radio band, such as the 21-cm line of neutral hydrogen (H\textsc{i}) \cite{2023ARA&A..61...19M} or the emission from carbon monoxide (CO) molecules, which trace molecular hydrogen (H$_2$) \cite{Mertsch:2020qld, 2020A&A...643A.141M}.} Therefore, alternative techniques, such as dust emission \cite{Schlegel:1997yv} are employed to estimate the column density of this dark component \cite{2005Sci...307.1292G}. Small unaccounted gas clumps may resemble point-like sources and contaminate the sample of detected sources (which could be the case for some of the flagged sources in 4FGL).

We aim to phenomenologically treat the gas structure's uncertainty to increase the realism of our simulator. We resort to Gaussian Random Fields (GRFs) to effectively vary the Galactic diffuse emission on small and large scales over the entire sky. 
GRFs are mathematical functions that assign a normally distributed random value to each point in space, with spatial correlations governed by a covariance function $C(\vec{x}, \vec{y})$ or, equivalently (assuming spatial homogeneity and isotropy of the field), by the power spectrum $P(k)$,  where $k$ refers to the wave number.

In this work, we follow the algorithmic recipe reported in \cite{Dimitriou:2022cvc} and generate GRFs entirely on GPUs via the package \texttt{pytorch} by first creating white noise in Fourier space, then scaling this noise by the square root of a desired power spectrum (which sets the spatial variability) 
\begin{equation}
\label{eq:power-spectrum}
    P(k)= (k/2.5)^{-\gamma},
\end{equation}
where $\gamma$ is the slope of the power spectrum. The chosen pivot wave number of 2.5 translates to a pivot angular scale of around $0.05^{\circ}$. Finally, we transform the result back into position space using an inverse Fourier transform and take the real part of the result. 

\begin{figure*}[!t]
    \includegraphics[width=\textwidth]{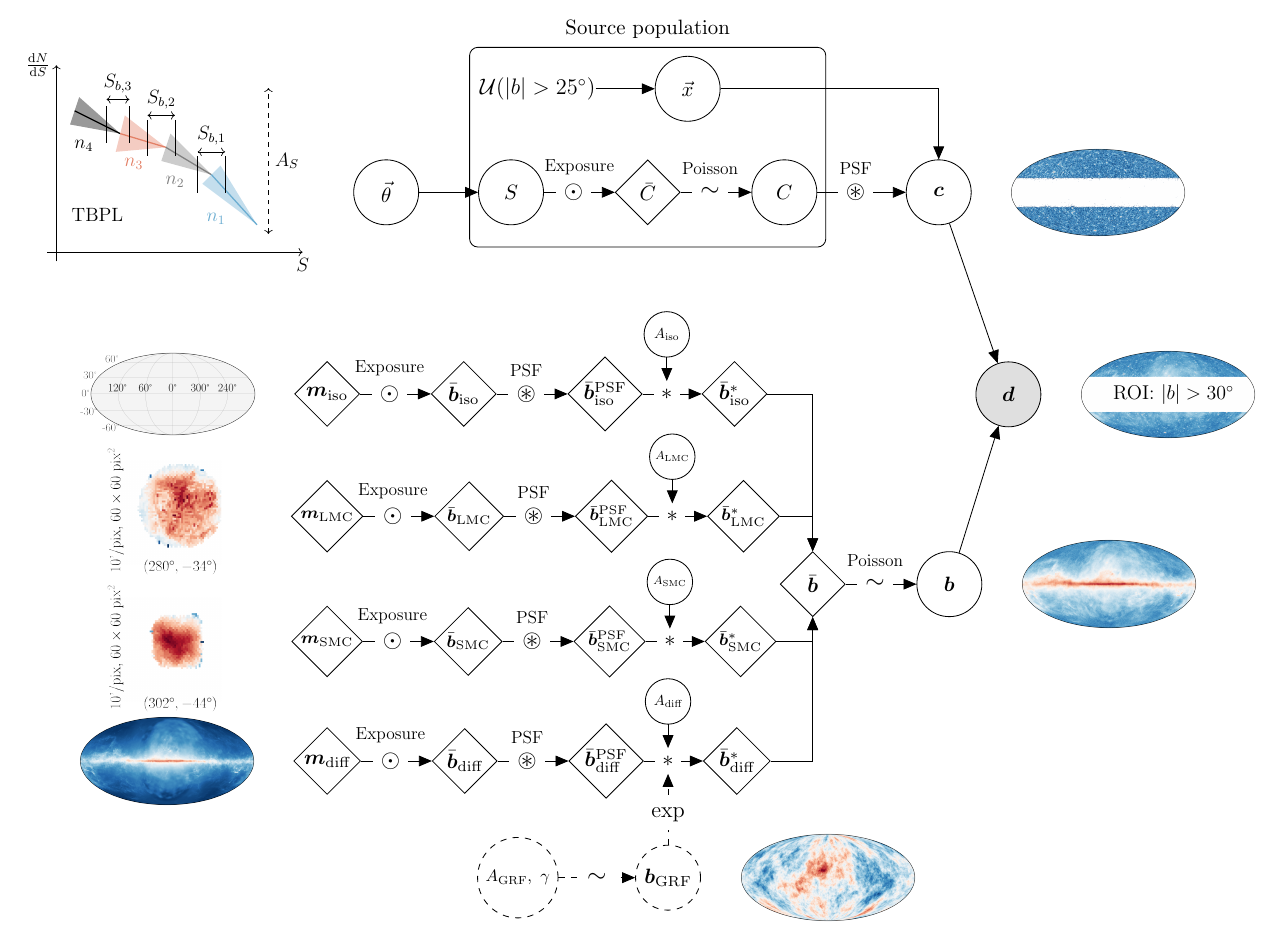}
    \caption{Schematic overview of the Bayesian hierarchical model implemented in our gamma-ray simulator. The definitions for all symbols are provided in Tab.~\ref{tab:model-definitions}. Boldface symbols refer to quantities at the sky-map level. The technical details of the steps visualized in the upper part of the sketch resulting in the point-source map $\boldsymbol{c}$ are outlined in Sec.~\ref{sec:ps_fm} while the simulation of the background map $\boldsymbol{b}$ in the lower part of the figure is described in Sec.~\ref{sec:de_fm}. Information about exposure maps and LAT PSF handling are given in Sec.~\ref{sec:instrument}. Diamond-shaped nodes represent deterministic quantities, circular nodes represent stochastic quantities, and nodes with dashed lines (the Gaussian Random Field component) denote components excluded (in Sec.~\ref{sec:results_woGRF}) or included (in Sec.~\ref{sec:results_dndS_wGRF}) explicitly in our analysis. 
    } 
    \label{fig:bayesian_model}
\end{figure*}

In practice, during a run of the simulator, the GRF calculation is performed on a rectangular wave number grid of $1440\times720$ points representing the two-dimensional Cartesian projection of an all-sky map at a resolution of $0.25^{\circ}$ in Fourier space. The (discrete) mapping from Fourier to real space is performed via the fast Fourier transform. The slope of the power spectrum is an additional parameter of the model, which we sample from a uniform prior between 1.5 and 3. We checked that this range of slopes allows for perceptible variations on small and large scales while ensuring that the power decreases sufficiently fast on small scales, preserving the physically expected correlations in the resulting random structures. Since the inverse fast Fourier transform generates a Cartesian projection of the sky, we project it onto the sphere with \texttt{HEALPix} pixelization ($N_{\mathrm{side}} = 128$) via the \texttt{python} package \texttt{gammapy} \cite{gammapy:2023}. To vary the amplitude of the small- and large-scale corrections modeled via the GRF, we draw a normalization $A_{\mathrm{GRF}}$ for this map from a uniform prior between 0 and 5. This guarantees that a (virtually) non-modulated diffuse foreground component is within the training sample, while it also allows for up to 10\% variations on the smallest scales. We then multiply 
this realization of the GRF $\boldsymbol{b}_{\mathrm{GRF}}$ 
with $A_{\mathrm{GRF}}$ and exponentiate it (cf.~the bottom all-sky map in Fig.~\ref{fig:bayesian_model} for an example). Since the exponentiation can lead to extreme amplitudes of the induced distortions, especially in the case of power spectrum slopes at the right end of the prior, we clip the achieved enhancement factor at 50 from above and $10^{-8}$ from below. The latter reduces artifacts caused by the projection onto the sphere.  The exponentiated GRF map is finally multiplied with the PSF-convolved template of the Galactic diffuse emission $\bar{\boldsymbol{b}}_{\mathrm{diff}}^{\mathrm{PSF}}$ and the normalization constant $A_{\mathrm{diff}}$. 

We note that we perform the GRF augmentation of the diffuse template after its PSF convolution for practical reasons, i.e.~to retain a fast gamma-ray simulator. Of course, gas density uncertainties should be injected in the model template first, while the PSF step follows. However, we would lose the speed gained by loading the pre-convolved background templates. Since we are only interested in an effective treatment of these observational uncertainties, this procedure still serves the purpose. We mention that this aligns with \cite{Mishra-Sharma:2020kjb, Mishra-Sharma:2021oxe}, where they use a Gaussian process to give additional freedom to background templates in order to enable a more robust characterization of the other modeled contributions. \textit{A posteriori}, we can also justify this approach in light of the results on real LAT data that detect a certain degree of diffuse mis-modeling on large angular scales above the PSF size at the considered energies (cf.~\ref{sec:results_dndS_wGRF}).

\begin{table}[]
\centering
\renewcommand{\arraystretch}{1.2}
\resizebox{\columnwidth}{!}{%
\begin{tabular}{@{}ccc@{}}
\toprule
    Component & Symbol & Definition \\ \midrule 
    & $\boldsymbol{d}$ & Observed photon count map  \\ \midrule
\parbox[t]{1mm}{\multirow{6}{*}{\rotatebox[origin=c]{90}{Point sources}}}
    & $\vec{\theta}$     & Source-population parameters \\
    & $\vec{x}$          & Source location       \\
    & $S$                & Source flux           \\
    & $\bar{C}$                & Expected number of counts      \\
    & $C$                & Actual number of counts      \\
    & $\boldsymbol{c}$   & Point-source map      \\ \midrule
\parbox[t]{1mm}{\multirow{7}{*}{\rotatebox[origin=c]{90}{Diffuse}}}
    & $\boldsymbol{m}_{\mathrm{diff}}$  & Diffuse background template in terms of flux  \\
    & $\bar{\boldsymbol{b}}_{\mathrm{diff}}$ & $-$\textsf{''}$-$ in terms of counts \\
    & $\bar{\boldsymbol{b}}_{\mathrm{diff}}^{\mathrm{PSF}}$ &  $-$\textsf{''}$-$ smoothed by PSF \\
    & $\bar{\boldsymbol{b}}_{\mathrm{diff}}^{*}$ &  $-$\textsf{''}$-$ scaled and smoothed by PSF; may include GRF \\
    & $A_{\mathrm{diff}}$               & Template normalization of diffuse background  \\
    & \color{darkgray} $A_{\mathrm{GRF}}$               & \color{darkgray} Gaussian random field normalization  \\
    & \color{darkgray} $\gamma$                      & \color{darkgray} Gaussian random field power-law index  \\
    & \color{darkgray} $\boldsymbol{b}_{\mathrm{GRF}}$   & \color{darkgray} Gaussian random field draw  \\ 
    \cmidrule(l){2-3}
    \parbox[t]{1mm}{\multirow{5}{*}{\rotatebox[origin=c]{90}{LMC}}}
    & $\boldsymbol{m}_{\mathrm{LMC}}$   & LMC template in terms of flux \\
    & $\bar{\boldsymbol{b}}_{\mathrm{LMC}}$ &  $-$\textsf{''}$-$ in terms of counts \\
    & $\bar{\boldsymbol{b}}_{\mathrm{LMC}}^{\mathrm{PSF}}$ &  $-$\textsf{''}$-$ smoothed by PSF \\
        & $\bar{\boldsymbol{b}}_{\mathrm{LMC}}^*$ &  $-$\textsf{''}$-$ scaled and smoothed by PSF\\
    & $A_{\mathrm{LMC}}$                & Template normalization of LMC  \\
    \cmidrule(l){2-3}
    \parbox[t]{1mm}{\multirow{5}{*}{\rotatebox[origin=c]{90}{SMC}}}
    & $\boldsymbol{m}_{\mathrm{SMC}}$   & SMC template in terms of flux \\
    & $\bar{\boldsymbol{b}}_{\mathrm{SMC}}$ &  $-$\textsf{''}$-$ in terms of counts \\
    & $\bar{\boldsymbol{b}}_{\mathrm{SMC}}^{\mathrm{PSF}}$ &  $-$\textsf{''}$-$ smoothed by PSF \\
        & $\bar{\boldsymbol{b}}_{\mathrm{SMC}}^*$ &  $-$\textsf{''}$-$ scaled and smoothed by PSF\\
    & $A_{\mathrm{SMC}}$                & Template normalization of SMC  \\
    \cmidrule(l){2-3}
    \parbox[t]{1mm}{\multirow{5}{*}{\rotatebox[origin=c]{90}{Isotropic}}}
    & $\boldsymbol{m}_{\mathrm{iso}}$   & Isotropic background template in terms of flux \\
    & $\bar{\boldsymbol{b}}_{\mathrm{iso}}$ &  $-$\textsf{''}$-$ in terms of counts \\
    & $\bar{\boldsymbol{b}}_{\mathrm{iso}}^{\mathrm{PSF}}$ &  $-$\textsf{''}$-$ smoothed by PSF \\
        & $\bar{\boldsymbol{b}}_{\mathrm{iso}}^*$ &  $-$\textsf{''}$-$ scaled and smoothed by PSF \\
    & $A_{\mathrm{iso}}$                & Template normalization of isotropic background  \\
    \cmidrule(l){2-3}
    & $\bar{\boldsymbol{b}}$                  & Expected background map  \\
    & $\boldsymbol{b}$                  & Actual background map  \\ \bottomrule
\end{tabular}%
}
\caption{Definition of symbols and quantities shown in Fig.~\ref{fig:bayesian_model} to illustrate the flow of our high-latitude gamma-ray emission simulator. Boldface symbols refer to quantities at the sky-map level. Quantities in gray are explicitly excluded (Sec.~\ref{sec:results_woGRF}) or included (Sec.~\ref{sec:results_dndS_wGRF}) depending on the analysis.}
\label{tab:model-definitions}
\end{table}

\begin{table}[h]
    \centering
    \renewcommand{\arraystretch}{1.6}
    \begin{tabular}{c c c}
        \toprule
        Component & \qquad Parameter & Prior  \\
        \midrule
        \parbox[t]{1mm}{\multirow{8}{*}{\rotatebox[origin=c]{90}{Source-count distribution}}}
        & $A_S$ & $\operatorname{Log}\mathcal{U}(1, 30)$ \\
        & $n_1$ & $\mathcal{U}(2, 3.5)$ \\
        & $n_2$ & $\mathcal{U}(1.7, 2.2)$ \\
        & $n_3$ & $\mathcal{U}(1.4, 2.3)$ \\
        & $n_4$ & $\mathcal{U}(-2, 2)$ \\
        & $S_{b,1} / (\mathrm{cm}^2\,\mathrm{s})$ & $\operatorname{Log}\mathcal{U}(3\times10^{-9}, 5\times10^{-8})$ \\
        & $S_{b,2} / (\mathrm{cm}^2\,\mathrm{s})$ & $\operatorname{Log}\mathcal{U}(2\times10^{-11}, 3\times10^{-9})$ \\
        & $S_{b,3} / (\mathrm{cm}^2\,\mathrm{s})$ & $\operatorname{Log}\mathcal{U}(3\times10^{-13}, 2\times10^{-11})$ \\
        \midrule
        \parbox[t]{1mm}{\multirow{6}{*}{\rotatebox[origin=c]{90}{Background}}}
        & $A_\mathrm{diff}$ & $\operatorname{Log}\mathcal{U}(0.1, 2.0)$  \\
        & $A_\mathrm{iso}$ &  $\operatorname{Log}\mathcal{U}(0.1, 2.0)$ \\
        & $A_\mathrm{LMC}$ &  $\operatorname{Log}\mathcal{U}(0.1, 2.0)$ \\
        & $A_\mathrm{SMC}$ &  $\operatorname{Log}\mathcal{U}(0.1, 2.0)$ \\
        & \color{darkgray} $A_\mathrm{GRF}$ &  \color{darkgray} $\mathcal{U}(0,5)$ \\
        & \color{darkgray} $\gamma$ &  \color{darkgray} $\mathcal{U}(1.5, 3.0)$ \\
        \bottomrule
    \end{tabular}\textsf{'}
    \caption{Benchmark choices for prior ranges and profiles regarding the TBPL parameters defining the source-count distribution in Eq.~\eqref{eq:dNdS-parametric} and background model parameters (cf.~Sec.~\ref{sec:de_fm}).
    Here, $\operatorname{Log}\mathcal{U}$ denotes the log-uniform distribution with base 10.
    }
    \label{tab:params}
\end{table}

\section{Method}
\label{sec:method}

In this work, we aim to robustly determine the source-count distribution, \dnds, of the Fermi-LAT sky at high latitudes by using three complementary methods within a unified neural simulation-based Bayesian inference framework. After a brief introduction to the adopted neural SBI framework (Sec.~\ref{sec:sbi}), we describe in Sec.~\ref{sec:stat} how each approach tackles the underlying statistical problem: inferring the source-count distribution \emph{(i)} from a detection map of the brightest detected sources, \emph{(ii)} from the inferred parameters of a broken power-law representation of the distribution, and \emph{(iii)} by directly estimating \dnds in each flux bin without imposing any functional form. 
Finally, Sec.~\ref{sec:nn_architecture} describes the neural network architectures that implement these methods.

\subsection{Simulation-based inference}
\label{sec:sbi}

Simulation-based inference (SBI) encompasses a suite of methods for performing inference of parameters, $\bm z$, given observed data, $\bm x$, in which the model is not represented by an explicit probability distribution $p(\bm x | \bm z)$ (the likelihood) but rather by a data-generating forward process, $\bm z, \bm x \sim p(\bm z, \bm x)$ (the simulator) \cite{Cranmer:2019eaq}.

This framework is particularly well suited for large forward models, as the one described in Sec.~\ref{sec:model} and shown in Fig.~\ref{fig:bayesian_model}, where a likelihood function cannot be used explicitly for practical purposes without further simplifications. Furthermore, SBI naturally marginalizes over nuisance parameters through sampling, eliminating the need for explicit parameter inference of the joint distributions or for analytically performing 
parts of the marginal integrals. By operating entirely with simulated samples, SBI enables the inclusion of complex physical processes and Bayesian reasoning elements that defy simple probabilistic descriptions. This flexibility allows for data-driven priors and nuisance spaces of arbitrary dimensionality and distribution, such as the GRF distortions we use in the background emission forward model (see Sec.~\ref{sec:de_fm}).

An important advantage of SBI over traditional techniques in the context of photon-count maps concerns the treatment of the PSF: since counts are leaked across pixel boundaries, the probability to observe a given number of counts in a pixel is not independent of the counts in the neighboring pixels. Likelihood-based approaches such as 1p-PDF \cite{Zechlin:2015wdz} and \texttt{NPTF} \cite{Mishra-Sharma:2016gis} account for the smearing of counts in an integrated way following the approach proposed in \cite[Sec.~2.2]{Malyshev2011StatisticsLimit} (see also \cite{Collin2022} for an improved treatment). Still, an \textit{exact} analytical description of the PSF would need to take into account all possible ways in which counts from a single pixel can be scattered to any other one, which is a combinatorial problem that is computationally intractable. Hence, these approaches typically model the likelihood as a product over the pixels once the pixel-wise PSF corrections have been included. In SBI, in contrast, likelihood information is extracted by a neural network, which can freely learn the inter-pixel correlations imprinted by the PSF.

Another difference between traditional likelihood-based methods and SBI in the context of gamma-ray analyses is the bias resulting from mis-modeling which, to some degree, is expected for any analysis of the sky. For approaches based on a pixel-wise likelihood, a mismatch between the employed templates and the real data requires the fit to accommodate a larger variance in the photon counts, which is more easily achieved by a point-like than a Poissonian component, potentially giving rise to spurious bias towards point-like emission \cite{Leane2020, Leane2020a}. By contrast, SBI methods for image-like data typically rely on (convolutional) neural networks, which make predictions by examining sky patches of different sizes, rather than considering each pixel in isolation. This suggests that these approaches might be more robust against large-scale mis-modeling (see Fig.~2 in Ref.~\cite{List:2021aer} for a schematic representation of this phenomenon, and Fig.~16 in that work for a demonstration of this behavior occurring in a toy example).

\subsubsection{Neural Ratio Estimation}
\label{sec:nre}

Among the various flavors within neural SBI \cite[see e.g.][for an overview]{Lueckmann:2021aa}, we use NRE \cite{cranmer2015approximating, Hermans:2019ioj}. We employ NRE here because it recasts density estimation as a binary classification problem using the so-called “likelihood-ratio trick” \cite{cranmer2015approximating}, which facilitates a relatively simple neural network design compared to other SBI approaches. In particular, targeting probability \textit{ratios} circumvents the need for tailored `probability-preserving' architectures such as normalizing flows \cite{papamakarios2021normalizing}, which are the 
leading choice for other SBI flavors.

Given an implicitly defined model $\bm x, \bm z \sim p(\bm x, \bm z) = p(\bm x\mid\bm z)p(\bm z)$, the core idea behind ratio estimation is to use a binary classifier to discriminate between two classes of drawn data-parameters pairs $(\bm x, \bm z)$ labeled by the binary variable $L$:
\begin{subequations}
\begin{align}
    p(\bm x, \bm z  \mid L = 1) &= p(\bm x, \bm z ) \\
    p(\bm x, \bm z  \mid L = 0) &= p(\bm x) p(\bm z ) \, .
\end{align}
\end{subequations}
These two distributions correspond, respectively, to sampling data and parameters jointly ($L = 1$) from the simulator, $\bm x, \bm z \sim p(\bm x, \bm z)$, and to sampling data and parameters independently of each other ($L = 0$), $\bm x, \bm z \sim p(\bm x) p(\bm z)$, by drawing an unrelated set of parameters from the prior versus data from the simulator (which can be easily constructed by scrambling joint pairs). 

If the classifier is trained on equal proportions of samples from both distributions, the  Bayes-optimal decision function \cite{Devroye:1996aa} for this classifier is:
\begin{equation} \label{eq:sbi-classifier}
    p(L = 1 \mid \bm x, \bm z) = \frac{p(\bm x, \bm z)}{p(\bm x, \bm z) + p(\bm x) p(\bm z)} \equiv \sigma[ \log r(\bm z; \bm x)] \, ,
\end{equation}
where $\sigma(y) \equiv 1 / (1 + e^{-y})$ is the sigmoid function. This formulation, often referred to as the ``likelihood-ratio trick" \cite[e.g.,][]{cranmer2015approximating}, reveals the relationship between the classifier and the ratio of posterior-to-prior, likelihood-to-evidence, or joint-to-marginal distributions ratio:
\begin{equation} \label{eq:sbi-ratio}
    r(\bm z;\bm x) \equiv \frac{p(\bm z \mid \bm x)}{p(\bm z)} = \frac{p(\bm x \mid \bm z)}{p(\bm x)} =  \frac{p(\bm x, \bm z)}{p(\bm x) p(\bm z)} \, .
\end{equation}

This result provides a direct way to estimate the posterior-to-prior ratio by training a classifier to distinguish between joint and marginal samples, both of which can be readily generated using a forward simulator, and subsequently use it for inference. In practice, if the prior is tractable, $r(\bm z;\bm x)$ allows direct computation of the posterior density via  $p(\bm z \mid \bm x)=r(\bm z;\bm x)p(\bm z)$.  Alternatively, if the prior lacks a closed-form expression, the ratio function can be used to re-weight prior samples, enabling posterior inference without explicit likelihood evaluation. 
As ratio estimation does not inherently provide posterior sampling functionalities -- unlike the normalizing flows used for neural posterior estimation \cite{papamakarios2021normalizing} -- we employ a nested sampling algorithm together with the estimated neural likelihood-to-evidence ratio to generate joint posterior samples in high ($>2$) dimensions. 

Since in this work both the data and parameter spaces are high-dimensional, a neural network is well-suited for learning the classification function.
Following \cite{Hermans:2019ioj}, in NRE, the classifier in Eq.~\eqref{eq:sbi-classifier} is implemented as a neural network,  $f_{\bm{\Phi}}(\bm x, \bm z)$, that takes parameters-data pairs as input and outputs an estimate of the ratio $r(\bm z;\bm x)$.
The network parameters ${\bm{\Phi}}$ are optimized via stochastic gradient descent \cite{Murphy:book} (see Tab.~\ref{tab:hyperparams} for the exact scheme) to minimize the binary-cross entropy loss \cite{mao2023cross}:
\begin{equation}\label{eq:sbi-bce}
\begin{split}
    \mathcal{L}[f_{\bm{\Phi}}(\bm x, \bm z)] = &-\int \dd \bm x  \dd\bm z \left\{ p(\bm x, \bm z) \log f_{\bm{\Phi}}(\bm x, \bm z) ] \right. \\
    & \left. + p(\bm x) p(\bm z) \log\left[ 1 - f_{\bm{\Phi}}(\bm x, \bm z) \right] \right\}\; .
\end{split}
\end{equation}
Therefore, by training a neural network $f_{\bm{\Phi}}(\bm x, \bm z)$ to estimate ${r}(\bm z; \bm x)$ through this supervised classification task, we obtain an estimate of the posterior through ${p}(\bm z \mid \bm x) = {r}(\bm z; \bm x) p(\bm z)$. 

For our implementation, we utilize \texttt{swyft} \cite{Miller:2020hua, Miller:2021aa, Miller:2022shs}, which is built on \texttt{pytorch-lightning} \cite{PyTorch_Lightning_2019}, to efficiently train and optimize the neural network. For posterior sampling, we use the \texttt{pytorch}-based nested sampler \texttt{torchns} \cite{AnauMontel:2023stj}.~\footnote{Available at \url{https://github.com/undark-lab/torchns}.}

\subsection{Statistical problem}
\label{sec:stat}

We employ three complementary approaches -- point source detection, parametric inference of the parameters $\vec{\theta}$ of a TBPL distribution, and non-parametric inference of the binned flux $\vec{\theta}_S$ -- to robustly infer the source-count distribution \dnds. Each approach is formulated in terms of NRE, as described in Sec.~\ref{sec:sbi}.

\subsubsection{Point source detection} \label{sec:stat_detection}
As first proposed in \cite{AnauMontel:2022ppb}, we perform source detection with NRE by estimating the following posterior-to-prior ratio: 
\begin{equation} \label{eq:ps-rSth}
    { r(\vec{x}, C_{\mathrm{th}}; \bm d)} 
    \equiv \frac{
    p(\mathbb{I}_{\bm d}(C \geq C_{\mathrm{th}})=1, \vec{x} \mid \bm d)
    }{
    p(\mathbb{I}_{\bm d}(C \geq C_{\mathrm{th}})=1, \vec{x})
    }\;.\footnote{
    \ Here, $\mathbb{I}_{\bm d}(C \geq C_{\mathrm{th}})$ is an indicator function over the data vector $\bm d$, that equals 1 if a point source at position $\vec{x}$ has counts $C \geq C_{\mathrm{th}}$ in the data $\bm d$, and 0 otherwise.
    }
\end{equation}
Here, the denominator represents the prior probability of having a source at position $\vec{x}$ with counts $C\geq C_{\mathrm{th}}$ that exceeds some threshold $C_{\mathrm{th}}$, which we choose to be $10\ \mathrm{counts}$.  Since we model the sources as being isotropically distributed, the prior distribution is in fact identical for all pixels in our ROI, i.e.\ $p(\mathbb{I}_{\bm d}(C \geq C_{\mathrm{th}})=1, \vec{x}) = p(\mathbb{I}_{\bm d}(C \geq C_{\mathrm{th}})=1)$. The numerator is the corresponding posterior given the observed data. The reasoning behind the choice of $C_{\mathrm{th}} = 10$ is as follows: We display via the gray band in Fig.~\ref{fig:detection_histogram} the flux corresponding to an average emission of 10 photons given the non-uniform LAT exposure of our 14-year dataset. The source-count distribution of detected 4FGL point-like sources (blue histogram in the same figure) is incomplete below and (for some flux range) above this threshold. The chosen threshold $C_{\mathrm{th}}$, thus, allows us to train on sources that have the chance to be recovered with traditional methods but that are also in a regime where these methods' sensitivity is insufficient to detect the full population. In principle, our SBI detection approach could identify sources not yet listed in the \textit{Fermi}-LAT collaboration's catalog. We deliberately refrained from using an arbitrarily small $C_{\mathrm{th}}$ during the training of the detection network because, depending on the realized \dnds, almost all pixels of the gamma-ray maps may contain a source (with a potentially low flux). In such a case, the network cannot learn any discriminative information about what is a point-like source and what is background. Labeling all pixels as sources would always be a good guess.

We implement the source-detection ratio estimator in Eq.~\eqref{eq:ps-rSth} as a map-to-map neural network that classifies each image pixel as either containing a point source with flux above the threshold counts or not (more details regarding the network architecture are given in Sec.~\ref{sec:nn_architecture} and App.~\ref{app:NN-architectures}). The simulated maps $\bm d$ serve as inputs to the classification network. In contrast, the labels for this pixel-wise classification task are binary masks indicating with ones the pixels that host sufficiently bright sources (i.e.~whose counts exceed the threshold, $C \geq C_{\mathrm{th}}$) and zeros elsewhere. To generate these labels we use the previously stored point-source maps without the PSF effect, i.e., where each source emits $C$ counts solely within its pixel (see Sec.~\ref{sec:ps_fm}).

\subsubsection{Parametric source-count distribution inference} \label{sec:stat_parametric}

As detailed in Sec.~\ref{sec:ps_fm}, we parametrize the source count distribution as a TBPL, described by eight parameters $\vec{\theta}$ in total (see Tab.~\ref{tab:params}). 

Estimating high-dimensional density ratios from a finite training set or with networks of limited capacity can lead to significant inaccuracies. To address this challenge, Ref.~\cite{AnauMontel:2023stj} proposes \emph{autoregressive} NRE, inspired by autoregressive normalizing-flow architectures \cite{huang2018neuralautoregressiveflows}. In this approach, the joint inference problem over all parameters is decomposed into an ordered series of low-dimensional ratio estimators, each learning a single parameter conditioned on the data and on the values of the ``previously” inferred parameters using the chain rule of probability, i.e.
\begin{equation}  \label{eqn:r_param}
\begin{split}
    r(\vec \theta ; \bm d) &=  \frac{p(\vec \theta\mid \bm d)}{\prod_{i=1}^d p(\theta_i)} \\
    & = \frac{p(\theta_1\mid \bm d)}{p(\theta_1)} \prod_{i=2}^d \frac{p(\theta_i\mid \bm d, \vec \theta_{1:i-1})}{ p(\theta_i)} \\
    & = 
    r(\theta_1;\bm d)  \prod_{i=2}^d  r(\theta_i;\bm d, \vec \theta_{1:i-1}) \; ,
\end{split}
\end{equation}
where we have introduced the compact notation $\vec \theta_{1:i-1} \equiv \{\theta_1, ..., \theta_{i-1}\}$. Thus, an autoregressive model can be defined by specifying the parametrization of each of the $d=8$ conditional distributions (more details regarding the network architecture are given in Sec.~\ref{sec:nn_architecture}). It is important to note that in Eq.~\eqref{eqn:r_param} we define the ratio between the posterior and the product of the marginal priors $\prod_{i=1}^d p(\theta_i)$, rather than the joint prior $p(\vec{\theta})$. Since all parameters have uniform priors (cf.~Tab.~\ref{tab:params}), this distinction makes no practical difference here.

The ratio defined in Eq.~\eqref{eqn:r_param} can be easily adapted to accommodate the estimation of the logarithm of the background parameters, $A_\text{diff}$, $A_\text{iso}$, $A_\text{LMC}$, and $A_\text{SMC}$, as defined in Tab.~\ref{tab:params}. Ultimately, we estimate the joint ratio over the TBPL parameters and the background parameters, $r(\vec \theta, \log_{10} A_\text{diff}, \log_{10} A_\text{iso}, \log_{10} A_\text{LMC}, \log_{10} A_\text{SMC} ; \bm d) $ in the scenario without GRFs.~\footnote{The scenario including GRF distortions on the Galactic diffuse emission model trivially adds the two parameters $A_{\mathrm{GRF}}$ and $\gamma$ to the list of inferred background parameters.}

Finally, we use a nested sampling algorithm with the estimated neural ratio to draw joint posterior samples, $p(\vec \theta, \log_{10} A_\text{diff}, \log_{10} A_\text{iso}, \log_{10} A_\text{LMC}, \log_{10} A_\text{SMC} | \bm d) $.

\subsubsection{Non-parametric source-count distribution inference} \label{sec:stat_nonparametric}
 
Rather than inferring parameters $\vec{\theta}$ that control the TBPL shape, we can estimate the source-count distribution $\mathrm{d}N/\mathrm{d}S$ directly in flux bins $\vec{\theta}_S$ (see Sec.~\ref{sec:ps_fm} for the definition of the bins). Using NRE, we infer the one-dimensional marginal distribution for the source counts in each bin, i.e. $p(\theta_{S,i} | \bm d)\ \forall \theta_{S,i} \in \vec{\theta}_S$. As in the parametric case, we also estimate the logarithm of the background parameters $A_\text{diff}$, $A_\text{iso}$, $A_\text{LMC}$, $A_\text{SMC}$ (and $A_{\mathrm{GRF}}$ and $\gamma$ when using GRF distortions), as defined in Tab.~\ref{tab:params}, enabling us to reconstruct the one-dimensional marginal distribution for both the source-count distribution's intervals and the background scaling. 

\subsection{Neural network architecture}
\label{sec:nn_architecture}
For the source-detection network, we employ a U-Net architecture \cite{ronneberger2015u}, composed of an encoder and a decoder part, which are connected by a skip connection at each level. During the encoder pass, the photon-count maps undergo progressive downsampling in spatial resolution, while the number of abstract feature channels increases. The bottleneck is reached at a spatial resolution corresponding to $N_{\mathrm{side}} = 2$, by which point the number of channels has reached $512$.
The decoder then reverses this process, restoring the spatial resolution while reducing the number of channels. This continues until the output reaches the full resolution of $N_{\mathrm{side}}= 128$ with a single channel, representing the log-prior-to-posterior ratio $r(\vec{x}, C_{\mathrm{th}}; \bm d)$, as described above.

The parameter inference network uses an encoder, followed by a simple feed-forward network that interprets the feature maps extracted by the encoder and further processes them to yield summary statistics. These are then mapped to log-prior-to-posterior ratios, either for the parameters $\vec{\theta}$ defining the TBPL in the parametric approach, or directly for the bin-wise $\vec{\theta}_{S}$ values in the non-parametric approach. 

Since we consider a large region of interest in the high-latitude sky, accounting for the curved sky geometry is indispensable. This prevents the use of standard convolutional neural network architectures, which typically assume Euclidean geometry. In the recent years, several spherical neural network architectures have been proposed (e.g.\ \cite{esteves2018learning, cohen2018spherical}). Herein, we employ the \texttt{DeepSphere} architecture \cite{Perraudin_2019,Defferrard2020}, which models the \texttt{HEALPix} sphere as a graph. In order to define the convolution operation on the graph, one leverages the fact that graphs admit a Fourier basis, spanned by the eigenvectors of the graph Laplacian operator. Thus, the graph convolution in \texttt{DeepSphere} is implemented as a product with a convolutional kernel in (graph) Fourier space. For the pooling and unpooling (or upsampling) operations, the hierarchical structure of the \texttt{HEALPix} pixelization is harnessed. Specifically, each pixel at resolution level $N_{\mathrm{side}}$ can be decomposed into four pixels at the next finer resolution level $2 N_{\mathrm{side}}$. Similarly to the Euclidean case, where the pooling operation typically aggregates values over $2 \times 2$ blocks via a permutation-invariant function such as the mean or maximum, the pooling in \texttt{DeepSphere} operates on sets of four subpixels. For the unpooling operation, we use nearest neighbor interpolation; i.e., we replicate all values four times in order to obtain a map at the next higher resolution. In the \texttt{NESTED} ordering of \texttt{HEALPix}, the four subpixels have indices $4 r, \ldots, 4 r + 3$ for some $r \in \mathbb{N}_0$, which makes the implementation of these operations straightforward.

In practice, given that half of the sky area ($|b| \leq 30^\circ$) is masked in our analysis, we provide compressed maps to our neural networks, which contain only the values in our ROI. At each level of the resolution hierarchy, we then build a \texttt{HEALPix} graph from the pixels that fall within our ROI. The graph convolutions employed by \texttt{DeepSphere} do not rely on the specific \texttt{HEALPix} tesselation and hence do not need to be adjusted. However, we customize the (un-)pooling operations to only take ROI pixels into account.

For the source-detection network, the decoder output is directly trained to produce a map of log-ratios at each pixel according to Eq.~\eqref{eq:ps-rSth}. To achieve this, pixels in the maps without the PSF effect exceeding the threshold are masked before calculating the loss. Following the binary classification logic behind NRE, the training procedure penalizes the network if it assigns a low probability to genuine sources (those with $C\geq C_{\mathrm{th}}$) or a high probability to pixels containing only background (or sub-threshold sources). In the inference tasks on the TBPL $\vec{\theta}$ or the bin-wise $\vec{\theta}_S$, each log-prior-to-posterior ratio (so also each of the conditional ratios in Eq.~\eqref{eqn:r_param}) is modeled with a residual neural network~\cite{He:2015wrn} with 2 blocks with 128 hidden features each (as implemented in \texttt{swyft}). 

All further details on the respective neural network architectures and the chosen training hyper-parameter settings for each inference task are reported in App.~\ref{app:NN-architectures}. We note that for the training of the detection network, we perform \textit{background resampling}, which refers to a data augmentation technique that scrambles the background maps $\boldsymbol{b}$ of a batch at each epoch. In this way, we enhance the background variations that the detection network sees so that it may better generalize the learned features. For the detection analysis, this approach is feasible since the learning objective is the distribution of sources above $C_{\mathrm{th}} = 10$ independent of the background, whereas it is not applicable when we aim for parameter reconstruction for obvious reasons. In all scenarios, the training -- according to these hyper-parameter definitions -- is conducted on a single NVIDIA A100 80 GB GPU.

\section{Results}
\label{sec:results}

In this section, we present and interpret the results we obtained after applying the SBI framework to the selected \textit{Fermi}-LAT dataset from a total of 14 years. We start in Sec.~\ref{sec:results_det} with the uncovered population of significant point-like sources derived within our source-detection approach. We also compare the findings using a simulator with and without GRFs to vary the diffuse MW foreground. In Sec.~\ref{sec:results_woGRF}, we report on the quality of the reconstructed source-count distribution and background parameters in the absence of GRF modulations of the diffuse background, whereas Sec.~\ref{sec:results_wGRF} covers the results obtained in the scenario when GRFs are included in the model. 

Additional technical details and the performance of the respective networks on simulated data are given in App.~\ref{app:simulations}.

\subsection{Source detection}
\label{sec:results_det}

\begin{figure*}[t]
    \includegraphics[width=\linewidth]{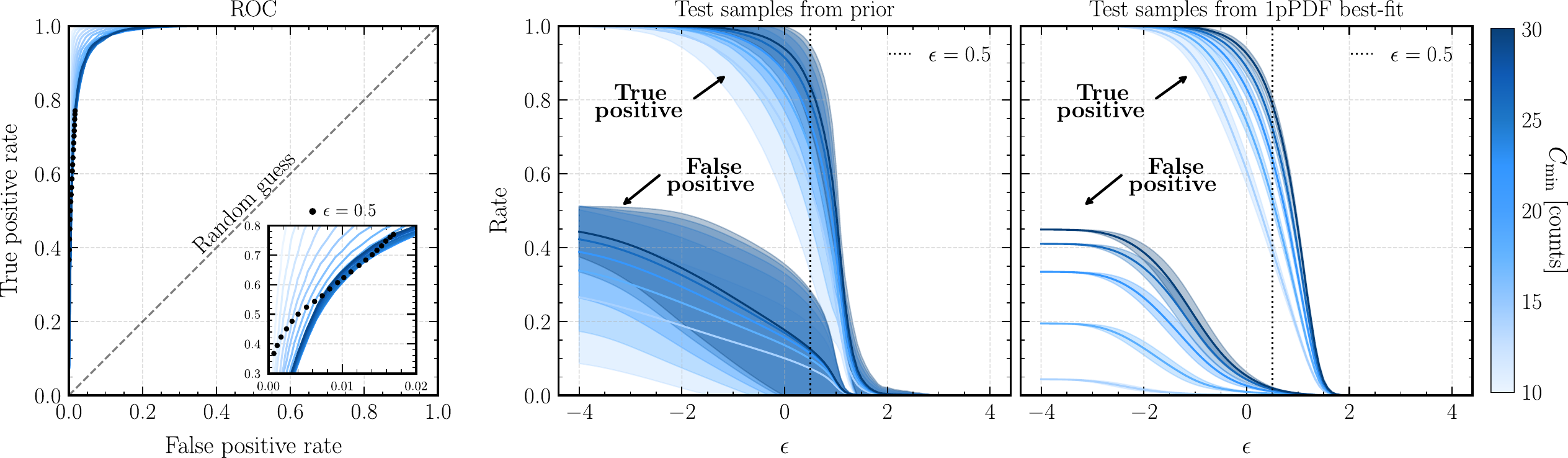}
    \caption{Detection map ROC and threshold selection (see text for description). \textbf{Left:} ROC curve for a representative test simulation, displaying the trade-off between the true-positive rate (completeness) and the false-positive rate (contamination). The inset zooms in on the region for the specific true and false positive rates achieved at the chosen threshold $\epsilon=0.5$. \textbf{Center:} Variation of both rates with the threshold parameter across 30 test simulations based on the full prior. \textbf{Right:} The same plot for 30 simulations was generated using the best-fit source-count distribution parameters from a 1p-PDF analysis; the background parameters are all set to one. The different color‐coded lines correspond to excluding sources with expected counts below a minimum threshold of $C_{\mathrm{min}}$. In other words, these lines reflect the true and false positive rates evaluated on maps that contain point-like sources emitting at least $C_{\mathrm{min}}$ photons.  Based on these results, $\epsilon=0.5$ has been selected as an optimal compromise, ensuring a small false-positive rate while recovering most of the genuine sources. 
    }
    \label{fig:roc}
\end{figure*}

\begin{figure}[h]
    \includegraphics[width=\columnwidth]{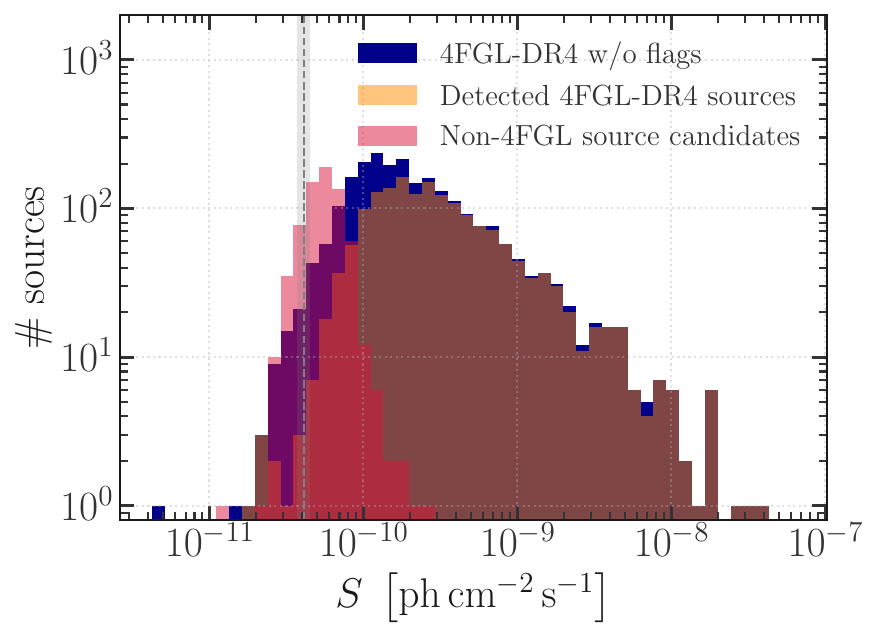}
    \caption{Comparison of the source-count distribution of detected point-like sources taken from 4FGL-DR4 (without flagged sources; blue histogram) and the population of point-like sources detected by our SBI detection framework applied to the 14-year LAT data sample. We divide the significant detections of our network into sources that could be matched with the positions and fluxes of 4FGL-DR4 sources (orange histogram) and non-4FGL candidate sources that do not have a counterpart in the \textit{Fermi}-LAT collaboration's catalog (red histogram) based on a source association step, as explained in the main text. The vertical dashed gray line marks the flux $S$ corresponding to a mean expectation of 10 photons per source. The shaded gray band illustrates the variance of this threshold due to the non-uniform LAT exposure.~\footnote{In practice, the gray band should be even wider since we train the detection network on labels that are Poisson realizations of the expected counts. Hence, it adds a factor of $\sqrt{10}\sim3$.}
    }
    \label{fig:detection_histogram}
\end{figure}

\begin{figure*}[t]
    \includegraphics[width=0.9\linewidth]{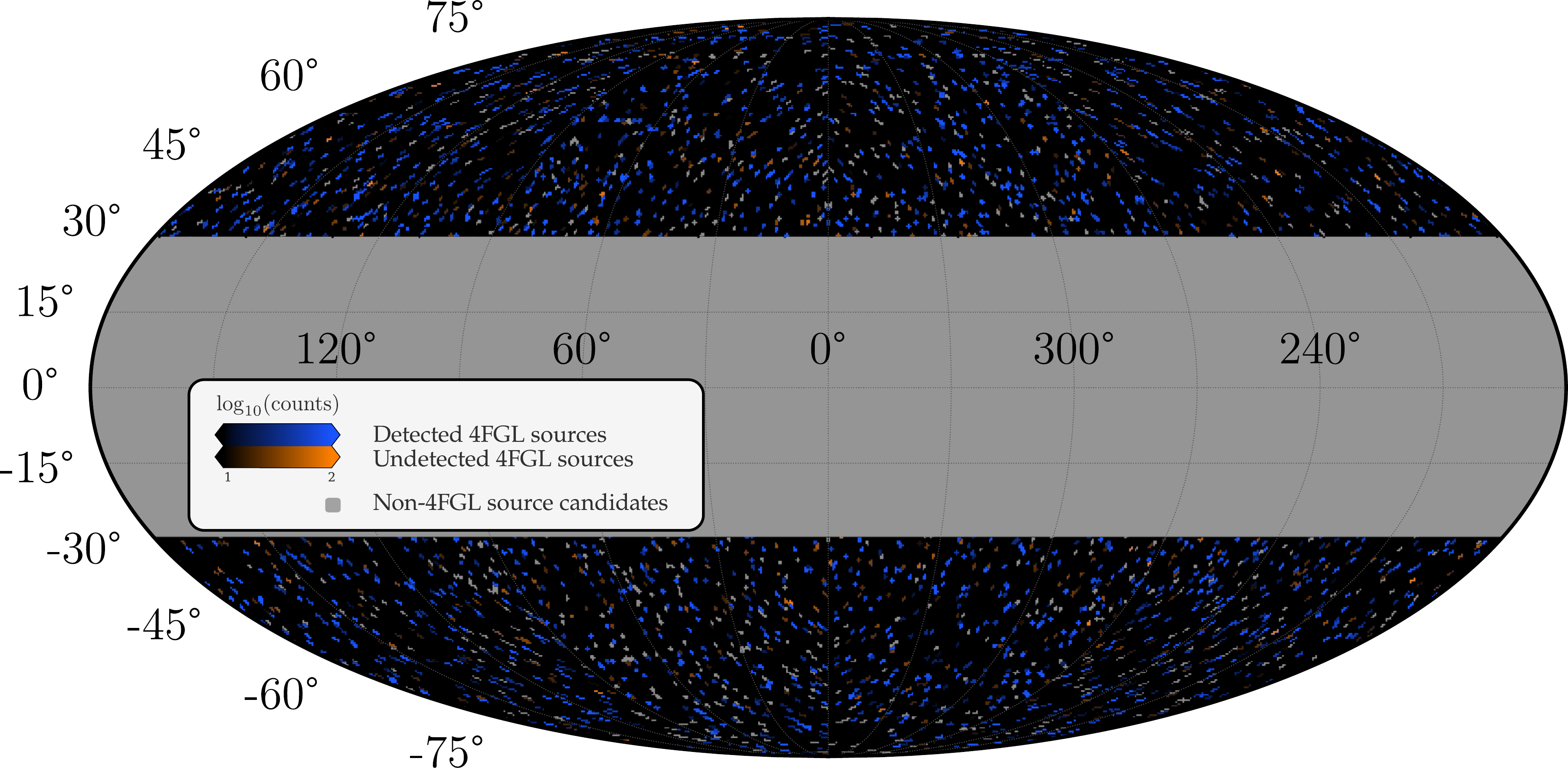}
    \caption{\textit{Fermi} source detection map (with a sensitivity threshold of $\epsilon = 0.5$) for the high-latitude sky generated by our SBI framework \emph{not including} variations of the diffuse MW foreground generated by the GRF component. 4FGL sources are colored blue if they have been recovered by our method, otherwise orange. The lightness of the pixels indicates the brightness of the source. In addition, detected sources that are not part of the 4FGL-DR4 catalog are shown in gray (with constant lightness). Note that for better visibility, this map is downscaled to a resolution of $n_{\mathrm{side}} = 64$, whereas the neural network operates at $n_{\mathrm{side}} = 128$ (and we count predicted sources in direct neighbor pixels of 4FGL source pixels at the latter resolution as a recovered source).}
    \label{fig:detection_map}
\end{figure*}

Before applying the trained point-source detection network (Sec.~\ref{sec:stat_detection}) to the target \textit{Fermi}-LAT 14-year dataset, we must define what constitutes a `detected' source based on the posterior-to-prior ratio in Eq.~\eqref{eq:ps-rSth}. We say a source is `detected' if a pixel in the detection map exhibits a detection significance exceeding a user-defined threshold $\epsilon$, i.e., if $\log_{10} r(\vec{x}, C_\text{th}; \bm d) > \epsilon$. 
Because we assume a uniform spatial prior (see Sec.~\ref{sec:ps_fm}), the ratio alone determines the probability that a given pixel contains a source that emitted counts $C\geq C_\text{th}$. Under a non-uniform prior, one would need to threshold the posterior itself instead. 

We select a threshold $\epsilon$ by balancing the fraction of genuine sources recovered (often called completeness or true-positive rate) against the spurious ones (contamination or false-positive rate) in test data. This trade-off is illustrated by the ROC (receiver operating characteristic) curve (left panel of Fig.~\ref{fig:roc}), which plots the true-positive rate against the false-positive rate for one randomly selected test sample. The different color‐coded lines correspond to excluding sources with observed counts $C_{\mathrm{min}}$ (Poisson realization) below a minimum threshold (with the colorbar starting at $C_{\mathrm{min}} = 10$ counts). For the false-positive rates, the lower count threshold $C_{\mathrm{min}}$ is applied based on the intensity of the background in the pixel. The central panel of Fig.~\ref{fig:roc} shows how each rate varies with $\epsilon$ for the test dataset, which contains 30 simulations generated from the full priors, illustrating how increasing the threshold reduces spurious detections at the expense of completeness. The right panel of Fig.~\ref{fig:roc} is identical to the central one, but for a test dataset which contains 30 simulations generated using the source-count distribution parameters' best-fit values from a 1p-PDF analysis \cite{Zechlin:2015wdz}, while the background parameters are set to one.~\footnote{To be precise, we adopt the mean values of the TBPL \dnds~reported in Tab.~2 of this work obtained via the profile-likelihood approach.} We note that, for fixed $\epsilon$, the false-positive rate increases with higher $C_\text{min}$ thresholds, reflecting the fact that the probability of our network mistakenly identifying a background pixel as hosting a source grows with higher background counts.
Based on these results, we adopt $\epsilon = 0.5$ as an optimal compromise, maintaining a false-positive rate below $5\%$ while recovering most of the true sources in our test simulations.

\vspace{10pt} 
\noindent\textbf{Results without GRFs.} Now -- concerning a gamma-ray emission simulator without GRFs -- we apply this empirically motivated definition of detection threshold $\epsilon$ to the detection significance map obtained from the 14-year LAT dataset. Fig.~\ref{fig:detection_histogram} illustrates the population of detected sources in terms of a source-count histogram, i.e.~the number of sources per flux bin regarding the energy range from 1 to 10 GeV. The blue histogram represents all 4FGL-DR4 sources (without flags) listed in our ROI (for more details about the catalog processing see App.~\ref{sec:data-4fgl}). This set of sources is the one with which we compare our SBI results as it serves as the ground truth for defining a successful reconstruction of a known gamma-ray source and the detection of a potential point-like-source candidate regarding the 14-year LAT dataset. However, to state it already here, the situation is different from the findings shown in Fig.~\ref{fig:roc} because the 4FGL-DR4 catalog is incomplete below a certain flux and does not contain the full source population in contrast to simulated catalogs. Therefore, we do not refer to the detected sources as true or false positives. In particular, instead of `false positive', we use the label `candidate' (and variants thereof) because there is a chance that there is a genuine point-like source at this position that is not part of 4FGL-DR4. 

To declare an association between a 4FGL-DR4 source and a significant pixel in our SBI detection significance map ($\log_{10} r(\vec{x}, C_\text{th}; \bm d) > 0.5$), we proceed as follows: Starting from the obtained detection significance map, we loop through the \texttt{HEALPix} pixel indices fulfilling the log-ratio criterion above and browse the 4FGL-DR4 catalog for a source in the same pixel. If there is one, we accept the pixel as a recovered source, assigning a flux to it exactly matching the flux predicted by the 4FGL-DR4 parameters.~\footnote{The current implementation of the detection network is such that it characterizes only the position of sources above a count threshold $C_{\mathrm{th}}$, but it does not predict their flux. This should be a straightforward extension of the current framework (e.g.~by adding an additional output channel to the UNet to return the predicted flux value), which we leave for future work. In this work, we associate a flux with recovered 4FGL sources based on the 4FGL-DR4 catalog.} If more than one 4FGL-DR4 object is within the respective pixel, we pick the source with the largest total flux as the association target. If 4FGL-DR4 does not list a source in that pixel, we consider the eight nearest neighbors of the pixel and check for each of them if a 4FGL-DR4 source is located in it. If true, the original pixel is associated with the neighboring 4FGL-DR4 object. Otherwise, it is declared a candidate source. Any point-like source listed in 4FGL-DR4 can only be assigned once to a firing pixel to avoid double counting. The step to also consider nearest-neighbor pixels is motivated by the combined effect of coarse binning ($N_{\mathrm{side}} = 128$ corresponds to an average pixel separation of $0.5^{\circ}$), PSF smearing and Poisson scatter that might lead to some uncertainty in the exact localization of the source. The obtained population of SBI-uncovered point-like sources is displayed as an orange histogram in Fig.~\ref{fig:detection_histogram}. The remaining candidate sources are depicted as a red histogram in the same figure. To derive their total flux in the 1 to 10 GeV band, we take the measured photon counts $C$ in the respective pixel, subtract from it the best-fit contribution of the astrophysical background emission (reported in more detail in Sec.~\ref{sec:results_bkg_woGRF}) and divide the residual counts by the exposure. We stress that the results of the detection network on the LAT dataset are 4FGL-dependent since we assume 4FGL-DR4 to be the ground truth. 

We obtained encouraging results: Our SBI approach can recover around 97\% of all 4FGL-DR4 sources with $S>3\times10^{-10}$ cm$^{-2}$s$^{-1}$. Based on the mean exposure in our ROI, this corresponds to sources that emit $\sim75$ gamma rays on average in the chosen energy band. For lower fluxes, the 4FGL-source recovery rate drops, and we start to accumulate non-4FGL candidates. Therefore, our SBI method (given the employed training size and network capacity) does not reach the efficiency of the detection approach employed to construct 4FGL-DR4. Yet, this comparison is not entirely fair and on the same footing because the source-detection algorithm used by the \textit{Fermi}-LAT collaboration involves more steps (exploiting analyses with more refined prior information in patches of the sky) than our SBI source detection approach. Quantitatively, we can recover around 72\% of all 4FGL-DR4 point-like sources without analysis flags. We note, however, that our efficiency depends on the chosen threshold $\epsilon$. We checked that by reducing $\epsilon$, we improve the 4FGL recovery rate and also retrieve those bright 4FGL-DR4 sources we miss with the benchmark threshold. In that sense, Fig.~\ref{fig:detection_histogram} 
represents the high-quality sample of detected point-like sources resulting in 685 non-4FGL candidates in total, compared to 1720 recovered 4FGL point-like emitters. This statement is further corroborated by the fact that we do not find any non-4FGL candidate detection to which we would assign a flux above $3\times10^{-10}$ cm$^{-2}$s$^{-1}$. 

What about the chances that the obtained candidate sources are indeed genuine point-like gamma-ray emitters? We know that the set of non-4FGL candidates is a conservative assessment since the 4FGL catalog itself is not complete around a flux of $10^{-10}$ cm$^{-2}$s$^{-1}$.~\footnote{We checked that none of these candidates coincides with the location of a flagged 4FGL-DR4 source.} 
Leveraging the intuition we gained from simulated realizations of the 1p-PDF best-fit TBPL in the right panel of Fig.~\ref{fig:roc}, the false positive rate (because it has been derived from simulated data) for this threshold choice should be rather small $<5\%$. This may indicate that those candidates emitting more than ten photons on average (gray line and band in Fig.~\ref{fig:detection_histogram}) uncovered in the real LAT data could be actual sources. Exploring this possibility requires dedicated gamma-ray analyses of a small ROI around the respective firing pixels, which is beyond the scope of this work. 

A spatial 
visualization of our detection findings is given in Fig.~\ref{fig:detection_map}, where we plot the locations of recovered 4FGL (blue) and non-4FGL candidate (gray) as well as not recovered 4FGL-DR4 sources (orange) in an all-sky map. The brighter the color of the respective pixel, the more photons it emits on average. This style of presentation demonstrates that there is no preferred direction in the sky for recovered or missed 4FGL-DR4 sources. 

\vspace{10pt}
\noindent\textbf{Results with GRFs.} The addition of GRFs to the gamma-ray simulator enhances the variability of the Galactic diffuse emission that the detection network eventually learns to marginalize over. These variations on small and large angular scales have the potential to render the source detection more robust against background fluctuations and, thus, to reduce the amount of non-4FGL candidates at a given threshold cut on $\log_{10} r(\vec{x}, C_\text{th}; \bm d)$. We observe exactly this effect when applying the detection network to the 14-year LAT dataset. While the full discussion of the obtained results, including figures, is given in App.~\ref{app:lat-detection-wGRF}, we restrict ourselves here to reporting 
the main findings in comparison with the results obtained using 
the detection network trained on samples without GRFs.
\begin{itemize}[itemindent=0.1in, leftmargin=0.2in]
    \item The ROC curve confirms the exquisite sensitivity to point-like sources with observed total counts $C > 10$ that we also observed in the left panel of Fig.~\ref{fig:roc}. Moreover, when evaluated on simulated test data, the true- and false-positive rates have a larger scatter.
    \item The evaluation on simulated target datasets following the best-fit TBPL \dnds~reported in \cite{Zechlin:2015wdz} shows that the source detection is more conservative at a given threshold $\log_{10} r(\vec{x}, C_\text{th}; \bm d) > \epsilon$. In other words, at the same threshold cut, we recover fewer true positives while also reducing the number of false positives. In addition, the false positive rate increases less steeply with decreasing cut on $\log_{10} r(\vec{x}, C_\text{th}; \bm d)$.
    \item Applied to the 14-year LAT high-latitude sky, the detection network confirms the expectations gained from the 1p-PDF test maps. We can reduce the threshold to $\log_{10} r(\vec{x}, C_\text{th}; \bm d) > 0$ to recover more than 98\% of all 4FGL-DR4 sources without flags above a flux of $S = 3\times10^{-10}$ cm$^{-2}$s$^{-1}$. Over the full flux range, we find about 70\% of the 4FGL-DR4 sources. Fig.~\ref{fig:detection_histogram} shows that 4FGL-DR4 starts being incomplete around sources with fluxes of $S\sim1\times10^{-10}$ cm$^{-2}$s$^{-1}$. Hence, our detection algorithm exhibits a detection efficiency that is only about a factor of three worse than the \textit{Fermi}-LAT collaboration's catalog, which uses a more involved analysis pipeline.
    \item Moreover, with a lower threshold, the case with GRFs yields 387 non-4FGL candidates. This is considerably lower than the 685 candidates we identified without GRFs.   
\end{itemize}

\subsection{Source-count distribution parameter inference without Milky Way foreground variations}
\label{sec:results_woGRF}

We apply the SBI parameter inference techniques described in Sec.~\ref{sec:method} to the target \textit{Fermi}-LAT dataset covering 14 years of observations of the high-latitude sky. Here, we report on the baseline simulation setup results that do not include any variations of the diffuse MW foreground generated by the GRF component. We examine and quantify the gamma-ray emission mis-modeling that this baseline scenario exhibits in App.~\ref{app:svdd}. We present our findings for the source-count distribution and background model components separately. Since our main interest is the assessment of the LAT sky's \dnds, we report only briefly on the findings regarding the background parameters. A more detailed discussion can be found in App.~\ref{app:lat-data-bkg}. All results were obtained with a single training round, yielding an amortized\footnote{In an `amortized' method the trained model learns a function that maps directly from input data to the solution, rather than solving the problem for each input 
individually. In the SBI case, the learned model can perform inference across multiple datasets within the prior's support without needing retraining. This is especially useful when repeated inferences are needed, as the one-time training cost is amortized over many uses.
} NRE.

\subsubsection{Objective: Source-count distribution}
\label{sec:results_dndS_woGRF}

\begin{figure*}[t]
    \includegraphics[width=\textwidth]{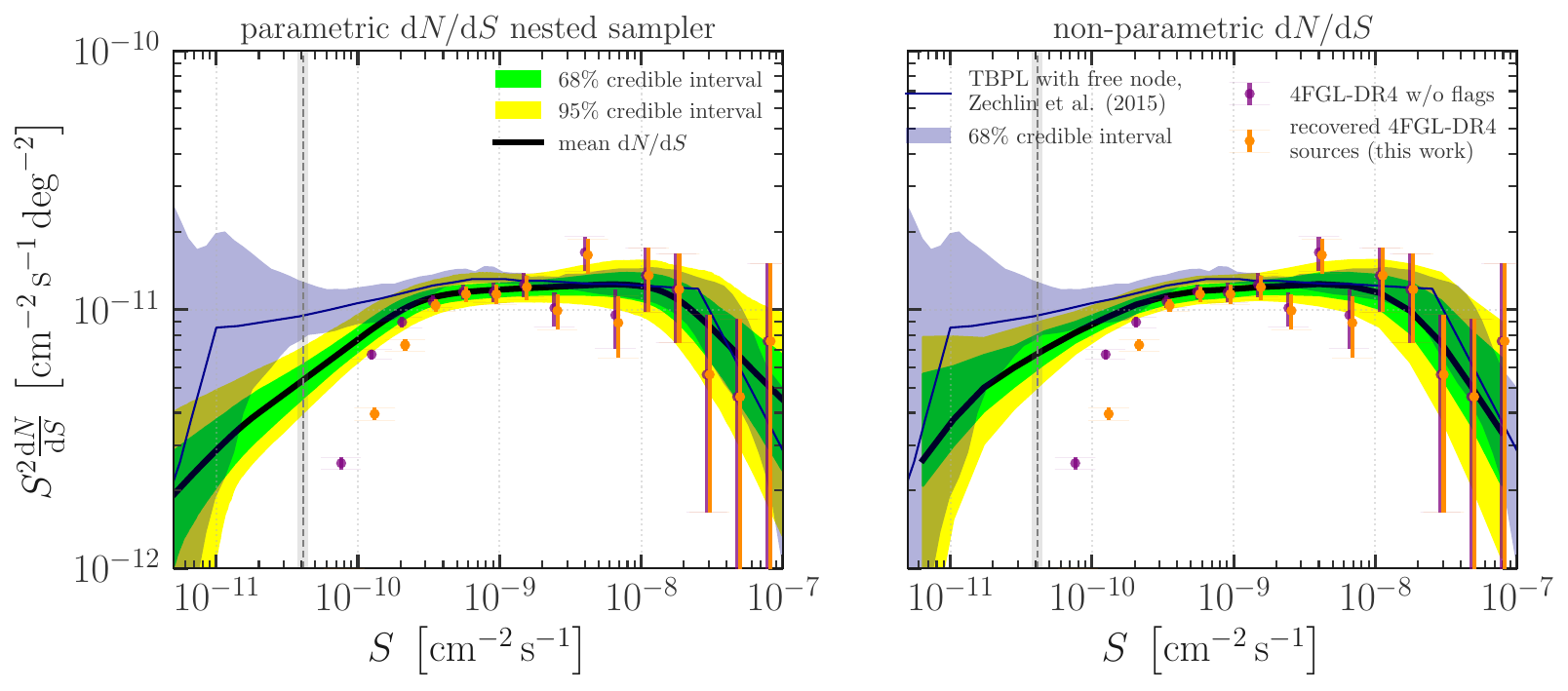}
    \caption{Comparison of the differential source-count distribution $\mathrm{d}N/\mathrm{d}S$ obtained from 14 years of \textit{Fermi}-LAT data at $|b|>30^{\circ}$ in a parametric way via an autoregressive model and nested sampling (\emph{left}) and non-parametrically by inferring its value in 20 flux bins (\emph{right}). The presented results do \emph{not include} variations of the diffuse MW foreground generated by the GRF component. In both cases, the simulated point-like sources follow the triple-broken source-count distribution. We denote the mean $\mathrm{d}N/\mathrm{d}S$ as a black solid line while the 68\% and 95\% credible intervals are displayed as green and yellow bands, respectively. We overlay our findings with the profile-likelihood results of \cite{Zechlin:2015wdz} (a TBPL ``hybrid'' approach for $N_{\mathrm{side}} = 128$), denoted by a solid blue line and corresponding blue band referencing the 68\% credible interval. In addition, we show two results of detected gamma-ray sources in the high-latitude sky: 4FGL-DR4 sources without analysis flags (purple; cf.~App.~\ref{sec:data-4fgl} for details) and 4FGL-DR4 sources recovered with our SBI detection network (orange; cf.~Fig.~\ref{fig:detection_histogram}). The latter points are slightly shifted to the right simply for visualization purposes. The analogous plots on simulated data with a \dnds~following the 1p-PDF best-fit profile are displayed and discussed in App.~\ref{app:simulation-1pPDF-parametric} and Fig.~\ref{fig:dNdS-1pPDF-woGRF}. We also add the vertical gray line from Fig.~\ref{fig:detection_histogram} representing the flux that corresponds to a source emitting 10 photons on average. Note that the legend is shared between the two plots.
    \label{fig:dNdS-woGRF}}
\end{figure*}

In accordance with the outlined parameter inference techniques, we infer the source-count distribution in the high-latitude sky in a parametric and non-parametric manner. The obtained mean $\mathrm{d}N/\mathrm{d}S$ profiles are shown in Fig.~\ref{fig:dNdS-woGRF} together with the respective 68\% and 95\% credible intervals derived within the framework of the respective method. To guide the eye, we also provide in purple the source-count distribution of detected gamma-ray sources from the 4FGL-DR4 catalog (without analysis flags) as detailed in App.~\ref{sec:data-4fgl} as well as in orange those 4FGL-DR4 sources recovered by our detection network (cf.~Sec.~\ref{sec:results_det}). 

We observe that both inference approaches yield $\mathrm{d}N/\mathrm{d}S$ profiles that are compatible with each other at the $1\sigma$-level. The most pronounced deviation between both methods is present in the dim-flux region, where the non-parametric source-count distribution predicts slightly more sources on average. The credible bands of the parametric method are slightly narrower than the corresponding bands of the non-parametric approach. Since the parametric results rely on the full joint posterior distribution that takes into account all correlations among the parameters, we expect this assessment to produce the most stringent credible intervals. At the same time, it indicates that the non-parametric parameter estimation is not overconfident. Such a scenario was conceivable because the simulated data was generated from a triple-broken $\mathrm{d}N/\mathrm{d}S$, which induces correlations among adjacent flux bins, which the neural networks might have picked up. 

Our findings are consistent with the detected part of the gamma-ray source population at high latitudes. We stress that this detected population reflects 4FGL-DR4 sources without analysis flags. Those flags are predominantly caused by potential background contamination. A better agreement with non-flagged objects was also reported in \cite{Calore:2021jvg} regarding gamma-ray emission from the central region of the MW employing traditional statistical methods to derive the source-count distribution. As an additional cross-check between the direct \dnds~inference and the detected part of the high-latitude gamma-ray sources, we added the recovered 4FGL-DR4 sources and non-4FGL candidates obtained via our SBI point-source detection network (cf.~Fig.~\ref{fig:detection_histogram}) together and corrected the found numbers by the expected completeness in that flux range according to the right panel of Fig.~\ref{fig:roc} (derived from simulated data of a similar \dnds). We observed that these completeness-corrected source catalog points are below the reconstructed \dnds~profiles but compatible at the $2\sigma$-level. Hence, even if all non-4FGL candidates were genuine gamma-ray sources, we would not overshoot the measured source-count distribution of the LAT sky. Despite this sanity check being rather rough, it demonstrates that the inferred LAT sky \dnds~is a plausible assessment of reality and that our detection network does not overpredict non-4FGL candidates, e.g., caused by background fluctuations.

Fig.~\ref{fig:dNdS-woGRF} also allows for a comparison between our SBI results and the likelihood-based 1p-PDF technique for the same sky fraction. We display as blue lines and bands (68\% credible interval) the analysis results of \cite{Zechlin:2015wdz} for 6 years of \textit{Fermi}-LAT observations by selecting the setup that most closely resembles our parametric and non-parametric approach. It uses our \texttt{HEALPix} resolution and a `hybrid' approach that slightly increases the flexibility of the TBPL assumption. The bright and intermediate flux ranges are fully consistent with each other, whereas the 1p-PDF technique finds more power in the dim flux regime where 4FGL-DR4 starts being incomplete. We demonstrate in App.~\ref{app:simulations} that our parametric inference method can reliably reconstruct a $\mathrm{d}N/\mathrm{d}S$ following the best-fit result of \cite{Zechlin:2015wdz} even in the dim regime. Therefore, this difference seems to be related to the sensitivity and performance of the applied approaches with respect to the utilized \textit{Fermi}-LAT dataset. The findings of the non-parametric approach exhibit a better agreement with the 1p-PDF \dnds, being consistent at the $1\sigma$-level. Therefore, it could be that the rigid parametrization of the source-count distribution as a TBPL is not ideal for capturing the $\mathrm{d}N/\mathrm{d}S$ profile realized in nature and more flexible parametrizations perform more reliably.

Further, but less critical, aspects of the obtained posterior distributions are discussed in App.~\ref{app:realData-material}. We demonstrate in App.~\ref{app:coverage} via coverage tests that the quoted posterior distributions and credible intervals are well calibrated. Moreover, and this is true for all parameter inference results, we base our reported results on the inference network state that yields the best calibration among all training epochs (cf.~also App.~\ref{app:NN-architectures} for more details).

\subsubsection{Objective: Background parameters}
\label{sec:results_bkg_woGRF}

The background parameters are treated equally in both $\mathrm{d}N/\mathrm{d}S$ parametrization scenarios. Yet, the inference target varies between the parametric (full point posterior) and non-parametric approach (one-dimensional marginals). We provide the inference results of both cases in Fig.~\ref{fig:bkg-woGRF-wGRF} of App.~\ref{app:lat-data-bkg} where we also discuss the details of the obtained findings. Here, we give a brief summary of the highlights.

\begin{itemize}[itemindent=0.1in, leftmargin=0.2in]
    \item The parametric approach constrains all four background normalization parameters, while the SMC remains unconstrained in the non-parametric case. 
    \item The inferred normalization parameters for LMC and diffuse MW foreground are consistent between approaches, while the isotropic background component is reconstructed with a higher flux in the parametric case. This is reasonable because we obtained a lower predicted source-count distribution at low fluxes in the parametric approach, which necessitates an increased isotropic background normalization to fit total gamma-ray emission.
    \item The inferred normalization of the diffuse isotropic background in the parametric approach is $A_{\mathrm{iso}}^{\ast} = 0.91^{+0.04}_{-0.036}$, leading to a total integrated flux of $S_{\mathrm{iso}} = 4.16^{+0.19}_{-0.16}\times10^{-7}\;\mathrm{cm}^{-2}\,\mathrm{s}^{-1}\,\mathrm{sr}^{-1}$.
    \item Unlike standard template-based fits, our analysis explicitly models unresolved sources via the source-count distribution, leading to a lower inferred $A_{\mathrm{iso}}^{\ast}$ compared to the best-fit IGRB model by \textit{Fermi}-LAT.~\footnote{We emphasize that referring to the IGRB in connection with the input spectrum from the \textit{Fermi}-LAT collaboration is appropriate and intentional.}    
    
    \item Comparison with \cite{Zechlin:2015wdz} shows our inferred diffuse isotropic flux is a factor of three higher, which can be explained by differences in event selection and non-photon contamination levels between Pass 7 \texttt{CLEAN} and Pass 8 \texttt{SOURCEVETO} event classes. The 1p-PDF results on the same ROI and energy range with Pass 8 \texttt{ULTRACLEANVETO} LAT data reported in \cite{Manconi:2019ynl} yield the same difference in the diffuse isotropic flux' normalization when compared to our work. This observation renders the choice of event class the most likely explanation for the enhanced diffuse isotropic contribution that we obtain.  
\end{itemize}

\subsection{Impact of Milky Way foreground distortions on the parameter inference}
\label{sec:results_wGRF}

Having examined the results obtained from our NRE approach trained on simulations of the high-latitude sky without GRFs, we now turn towards the scenario where the diffuse MW foreground is allowed to vary in a much more flexible manner than a mere rescaling via $A_{\mathrm{diff}}$. Distorting the original Galactic diffuse model employed in our gamma-ray simulator with GRFs is an effective method to broaden the sample space of this component and to train the NRE network on a wider class of background models to overcome potential (spatial) background mis-modeling. In App.~\ref{app:mismodeling-GDE}, we purposefully generate simulated training data that feature such mis-modeling of the Galactic diffuse emission and quantify how well the networks trained on simulations with GRFs can mitigate these effects. In this section, we limit ourselves to the presentation of the results on the real LAT gamma-ray sky. 

\subsubsection{Objective: Source-count distribution}
\label{sec:results_dndS_wGRF}

\begin{figure*}[t]
    \includegraphics[width=\textwidth]{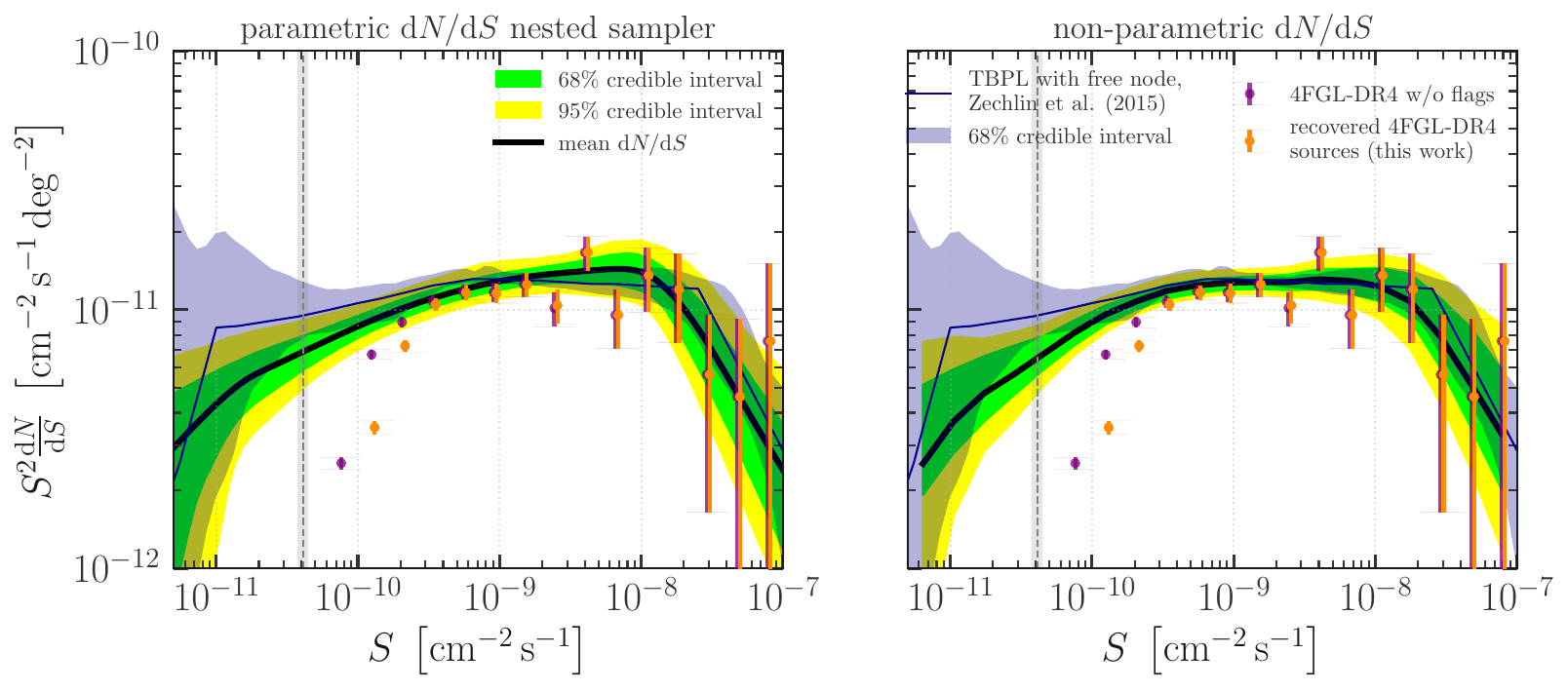}
    \caption{Same as Fig.~\ref{fig:dNdS-woGRF} derived via training on simulated high-latitude gamma-ray maps featuring GRF-modulated MW foreground emission. The recovered 4FGL-DR4 point-like sources were derived based on the detection network trained on simulations with GRFs and a detection threshold of $\log_{10} r(\vec{x}, C_\mathrm{th}; \bm d) > 0$ (cf.~Fig.~\ref{fig:lat-detection-wGRF} in App.~\ref{app:lat-detection-wGRF}). We note that the difference between the non-parametric \dnds~profiles with and without GRFs is very small, hinting at the robustness of this inference result. 
    \label{fig:dNdS-wGRF}}
\end{figure*}

In Fig.~\ref{fig:dNdS-wGRF}, we show the inferred \dnds~profiles obtained with our parametric (left panel) and non-parametric approach (right panel).~\footnote{A quantitative assessment of the calibration and coverage of the posteriors derived with both approaches is given in App.~\ref{app:coverage}.} In direct comparison with the analogous results based on networks trained on simulated data without GRFs in Fig.~\ref{fig:dNdS-woGRF}, these source-count distributions do not exhibit any noteworthy deviations, and agree on the $1\sigma$ level. This is also true between the parametric and non-parametric findings. When confronted with the 1p-PDF profile from \cite{Zechlin:2015wdz}, we observe an intriguing difference with respect to the scenario without GRFs: The parametrically and non-parametrically obtained \dnds~profiles follow the 1p-PDF equivalent more closely in the bright and intermediate source regimes. On one side, this is an indicator that our method works as intended; being in line with the efficacy of traditional likelihood-based methods (which do not implement a comparable mis-modeling mitigation scheme). On the other side, it lends credence to the characterization of the high-latitude source-count distribution presented here and in the original 1p-PDF work. 

In summary, training on simulations with GRFs yielded a better agreement between the inferred \dnds~profiles of parametric and non-parametric approaches. Both source-count distributions remain below the 1p-PDF result though fully consistent at the $2\sigma$ level. As demonstrated in App.~\ref{app:simulation-1pPDF-parametric} on simulated data, there is no reason to believe that our \dnds~inference algorithm is introducing an early break in the source-count distribution due to a lack of sensitivity to the dim regime. It appears to be a feature of the real gamma-ray sky. Corroborative evidence for this statement can be found, e.g., in \cite{Manconi:2019ynl}. Given the predictions from a simple blazar luminosity function, i.e.~the dominant extragalactic gamma-ray source population, our \dnds~profile matches well the expected low number of dim blazar-like sources in the flux range  $S<10^{-10}\;\mathrm{cm}^{-2}\,\mathrm{s}^{-1}$ obtained from the 1p-PDF technique. 

\subsubsection{Objective: Background parameters}
\label{sec:results_bkg_wGRF}

Besides inferring the source-count distribution, we also obtain the one-dimensional marginal posteriors concerning the enlarged set of background components and parameters. We report these results in the bottom rows of Fig.~\ref{fig:bkg-woGRF-wGRF} of App.~\ref{app:lat-data-bkg} displaying the posterior distributions from the parametric approach (gray histograms and colored credible intervals) with the respective posteriors derived in the non-parametric approach (black lines). A plot of the two-dimensional marginal posteriors is provided in Fig.~\ref{fig:2Dmarginal-LAT-combined} of App.~\ref{app:realData-material}. While more details are provided in App.~\ref{app:lat-data-bkg}, we summarize here the highlights we can deduce from the background inference:
\begin{itemize}[itemindent=0.1in, leftmargin=0.2in]
    \item Parametric and non-parametric approaches now better agree on the posterior of the dominant background components, the Galactic diffuse emission and the diffuse isotropic background concerning the posterior's mean and width. The inferred mean values $A_{\mathrm{diff}}^{\ast}$ and  $A_{\mathrm{iso}}^{\ast}$ are consistent with their equivalents obtained without GRFs.
    \item The uncertainties regarding $A_{\mathrm{diff}}$ and $A_{\mathrm{iso}}$ are larger than before (emphasized by the identical parameter range used for the panels associated with parameters common among results with and without GRFs). This is expected since the diffuse MW foreground exhibits much more variability than before enlarging the model space.
    \item The posteriors for the normalizations of LMC and SMC yield compatible mean values and credible bands regarding the findings without GRFs within the parametric approach. 
    \item The non-parametric approach is now capable of predicting $A_{\mathrm{SMC}}$ whose posterior is in $2\sigma$ agreement with the parametric posterior.
    \item We were able to infer loosely constrained posterior distributions for the GRF parameters realized in the real data. The best-fit values of power spectrum slope and normalization parameter hint at mis-modeling of the Galactic diffuse emission component present at scales smaller than the FBs or Loop I. We note that on simulated data, the GRF parameters can be reconstructed rather precisely. To emphasize this fact, we point to Fig.~\ref{fig:inference-diff-confusion-bkg} of App.~\ref{app:mismodeling-GDE}. There, we show the posterior profile of the resulting GRF parameters for a synthetically generated mis-modeling of the Galactic diffuse emission by using an alternative background template missing the FBs and Loop I. Even though the created mis-modeling is not a GRF by construction, the posteriors are sharply defined.
\end{itemize}

\section{Discussion}
\label{sec:discussion}

We want to highlight and elaborate on several points regarding the inference results from the gamma-ray sky as seen by the \textit{Fermi} LAT.
\\

\noindent\textbf{Compatibility and cross-talk between point-source detection and inference networks.} We employed two distinct ways to derive the source-count distribution of the \textit{Fermi}-LAT high-latitude sky: \emph{(i)} via an SBI point-source detection approach yielding a catalog of resolved discrete gamma-ray sources tracing the \dnds~until its intrinsic detection threshold and consequent loss of detection efficiency and, \emph{(ii)}, an SBI inference approach reconstructing the \dnds~profile directly from the target dataset. We demonstrated in Figs.~\ref{fig:dNdS-woGRF} and~\ref{fig:dNdS-wGRF} that both methods yield consistent characterizations of the high-latitude source-count distribution. More quantitatively, our inferred median \dnds corresponds to 799 expected sources above a flux of $S > 3 \times 10^{-10} \, \mathrm{cm}^{-2} \, \mathrm{s}^{-1}$ in our ROI (for the parametric \dnds), which excellently matches the 800 4FGL sources in this flux range ($>$98\% of which, recall, have been detected by our detection network).
At this point, we emphasize that the source-detection approach offers the advantage of resolving concrete sources, or more precisely, of inferring the probability that a certain pixel of the target dataset contains a point-like source. While we were able to associate a fraction of these high-probability pixels with 4FGL-DR4 sources, others could not be matched. Yet, among these additional firing pixels (red histograms in Figs.~\ref{fig:detection_histogram} and~\ref{fig:lat-detection-wGRF}) could be potential point-like sources not yet listed in the \textit{Fermi}-LAT collaboration's catalog. It would require dedicated gamma-ray analyses of each possible source candidate to confirm them as genuine discrete point-like gamma-ray emitters, which is beyond the scope of this work.

We observe the mutual consistency of both methods, which underlines the robustness and aptness of our SBI framework to reconstruct the high-latitude source-count distribution. We stress, there is no cross-talk between the two methods. In particular, the inference networks do not receive information from the output of the point-source detection network, i.e., the resolved part of the respective gamma-ray sky realization.\\

\begin{figure}
    \centering
    \includegraphics[width=\columnwidth]{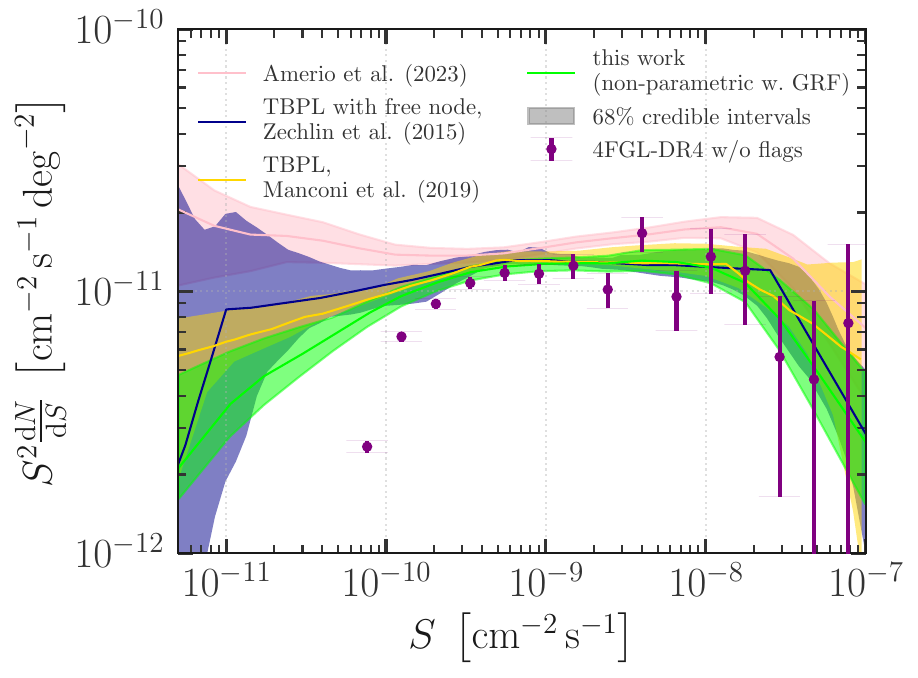}
    \caption{Comparison of the source-count distribution of the high-latitude ($|b|>30^{\circ}$) gamma-ray sky in the range from 1 to 10 GeV derived in this study (green; non-parametric approach with GRFs) with similar characterizations from the available literature. In blue, we display the adopted best-fit result of the 1p-PDF study \cite{Zechlin:2015wdz} while the yellow profile is an updated 1p-PDF assessment based on Pass 8 LAT data taken from \cite{Manconi:2019ynl}. The
    pink line and band refer to the convolutional-neural-network-based analysis presented in \cite{Amerio:2023uet}. We stress that all displayed bands denote the respective 68\% credible (or containment) intervals.}
    \label{fig:literature-comparison}
\end{figure}

\noindent\textbf{Comparison and relation to previous analyses in the literature.} Throughout Sec.~\ref{sec:results}, we confronted the source-count distribution derived from our SBI approach with similar findings obtained via the likelihood-based 1p-PDF technique \cite{Zechlin:2015wdz}. In Fig.~\ref{fig:literature-comparison}, we make this comparison more explicit adhering to the established color code by plotting our result in shades of green and the one of \cite{Zechlin:2015wdz} in shades of blue. 

Although the exposure gap between the 1p-PDF analysis and our work is larger than a factor of two,~\footnote{The 1p-PDF study also relies on different LAT data selection criteria and a different template describing the Galactic diffuse emission.} we always found a good agreement between the inferred \dnds~profiles. The degree of agreement is larger in the case of networks trained on simulations with GRFs, which also improves the compatibility between our parametric and non-parametric approaches. We note that the enhanced level of agreement between parametric and non-parametric approaches is a sign of a robust inference because we see this rise in improvement while increasing the degrees of freedom of the simulator/gamma-ray emission model. We have also demonstrated in App.~\ref{app:simulation-1pPDF-parametric} that our SBI approach can reliably recover a \dnds~profile following one of the best-fit results in \cite{Zechlin:2015wdz} regarding point-like source detection and \dnds~parameter inference.

An update of this first 1p-PDF assessment of the source-count distribution at high latitudes was reported in \cite{Manconi:2019ynl}, which is displayed as a yellow profile and band in Fig.~\ref{fig:literature-comparison}. This update is based on about 10 years of \texttt{ULTRACLEANVETO} LAT data following the Pass 8 event reconstruction criteria; thus, bringing it closer to our data setup. The level of agreement in the bright and intermediate flux regime remains high and further improves in the dim flux range with respect to the original 1p-PDF work. We now agree at the $1\sigma$ level with the 1p-PDF technique across the full range of considered fluxes. Therefore, the slight deviation from the original 1p-PDF results was most likely the effect of the different quality standards in processing the raw LAT events. 

We already referred to a third, more recent study on the same ROI with which our work shares most of the data selection criteria \cite{Amerio:2023uet}. The study leverages convolutional neural networks to reconstruct the source-count distribution of the LAT sky in a non-parametric approach (in the same sense as we do). On a technical level, however, there are a few differences: Their
convolutions are by default not defined on \texttt{HEALPix} all-sky maps as we implemented them in our analysis, but rather on rectilinear two-dimensional patches of \texttt{HEALPix} maps. Later, the authors employ convolutions with a spherical neural network (not based on \texttt{DeepSphere}) as a cross-check. Additionally, their training data contains the same Galactic diffuse emission model that we adopted but with a fixed normalization according to the best-fit value derived via a prior likelihood fit to the high-latitude LAT sky including the IGRB component and by masking all 4FGL sources. The best-fit \dnds~derived in \cite{Amerio:2023uet} is shown in Fig.~\ref{fig:literature-comparison} together with the best-fit 1p-PDF result and one of our characterizations of the source-count distribution (non-parametric with GRFs). 

First, we note that there is a considerable discrepancy contrasting \cite{Amerio:2023uet} from the other two findings. While we agree with \cite{Amerio:2023uet} in the bright and intermediate flux regime at the $2\sigma$ level, this is only marginally true in the dim flux range. The 1p-PDF study and ours identify a downward trend in the number of dim sources whereas the results of \cite{Amerio:2023uet} seem to prefer an upturn. The containment/credible bands open up in this flux regime no matter the respective analysis so that this behavior is not \textit{per se} inconceivable given the limited sensitivity of all three approaches (and, for that matter, of \textit{any} analysis method in view of point-source emission becoming degenerate with Poisson emission in the ultrafaint limit). However, it is not clear why the source-count distribution of \cite{Amerio:2023uet} does not fit the 4FGL-DR4 data points very well (even when including sources with analysis flags). The authors of \cite{Amerio:2023uet} have shown that their network produces correct results using simulated data. Yet, this does not exclude the possibility that there is a reality gap between their training data and the actual LAT sky. The simplification made to fix the normalization of the Galactic diffuse emission component could have led to such a gap. In our case, we have shown in App.~\ref{app:svdd} using an anomaly test that the real LAT sky is part of the model space spanned by our gamma-ray emission simulator. 
We also point out that the gamma-ray flux attributed to the diffuse isotropic background reported in \cite{Amerio:2023uet} for the \dnds~shown in Fig.~\ref{fig:literature-comparison} (with an error of about 10\% of the reported value) is almost identical to the value the authors report to have obtained in the initial likelihood fit that fixed the diffuse MW foreground's template normalization and refers to the IGRB emission. This is a curious observation requiring an interpretation: The initial likelihood fit was conducted in the selected high-latitude ROI with masked 4FGL(-DR3) sources and a model comprised of the Galactic diffuse emission and IGRB. Therefore, all gamma-ray emissions from sub-threshold point-like sources must have been absorbed by those two background components since this part of the source-count distribution is not part of the fit model. When inferring the \dnds, in contrast, this contribution to the gamma-ray sky is part of the model. As the Galactic diffuse emission model is fixed and the dim flux part of the \dnds~absorbs a certain amount of photon counts, the gamma-ray flux attributed to the diffuse isotropic background should decrease consistently. As stated in Sec.~\ref{sec:de_fm}, a sizeable fraction of the IGRB is the emission of unresolved point-like sources. This tension was mentioned by the authors of \cite{Amerio:2023uet}. In fact, our SBI analysis yields a total diffuse isotropic flux lower than the value of \cite{Amerio:2023uet} (cf.~Sec.~\ref{sec:results_bkg_woGRF}). \\


\noindent\textbf{Degeneracy between the dim regime of the \dnds~and a diffuse isotropic background.} Our gamma-ray simulator generates gamma-ray emission that contributes to the IGRB in two ways: Via the faint part of sources drawn from the \dnds~and a truly Poissonian diffuse isotropic background -- scaled by the normalization $A_{\mathrm{iso}}$. Let us briefly comment on the statistical degeneracy between these two quantities. The faint part of the \dnds~becomes formally indistinguishable from the diffuse isotropic background as the number of observed counts from each source tends to zero. This degeneracy is a concern in any count-map analysis (and perhaps even more so in our non-parametric case, where the \dnds~values in the lowest few flux bins float freely), see e.g.~Ref.~\cite{Collin2022}. First, we note that the total flux summed over the lowest four bins where $S \leq 3.6 \times 10^{-11} \, \mathrm{cm}^{-2} \, \mathrm{s}^{-1}$ as estimated by the non-parametric \dnds is $2.8 \times 10^{-8} \, \mathrm{cm}^{-2} \, \mathrm{s}^{-1} \, \mathrm{sr}^{-1}$ and hence more than a magnitude below the value of $S_\mathrm{iso}$, which is not surprising given the considerable cosmic-ray contamination in the \texttt{SOURCEVETO} class. Thus, even if there were significant cross-talk between the point-like and diffuse isotropic background components, our inferred value of $S_\mathrm{iso}$ would change only modestly. Conversely, one might worry that the Poissonian component affects the low flux end of our inferred \dnds. However, the center of the lowest flux bin in our non-parametric analysis corresponds to $\lambda \approx 1.5$ counts per source, which is still in the regime where the increased variance of point-like emission as compared to Poisson emission (specifically by a factor of $1 + \lambda = 2.5$) is sufficient for discerning point-like from Poissonian emission (see e.g.\ Fig.~9 in \cite{List:2021aer}). Further evidence for the ability of our network to distinguish between these components is provided by the corner plot in Fig.~\ref{fig:2Dmarginal-LAT-combined}, which shows only minor correlations between $A_{\mathrm{iso}}$ and the \dnds parameters. Suppose one is interested in extending source-count distributions to even lower fluxes. In that case, an alternative is to model both point-like and Poisson emission with a single unified point-like template, with \dnds~priors that reach far into the degenerate regime, as done in Ref.~\cite{List:2020mzd}. This avoids potential bias either towards point-like or Poisson emission in the degenerate (and hence prior-driven) regime arising from the fact that the priors for the \dnds~parameters translate very nonlinearly to priors for the associated total flux, see e.g.~\cite[Fig.\,2,][]{Collin2022}. In that case, one can determine a lower limit for the discriminatory power of the method in terms of flux, and then needs to remain agnostic regarding the point-like or Poissonian nature of the emission below that limit.\\   

\noindent\textbf{Background mis-modeling and the role of Gaussian random fields.} We designed the gamma-ray simulator employed in this work with two objectives in mind: Being fast and realistic. To accommodate both guiding principles, we sometimes had to make compromises that could have impacted the realism of the generated gamma-ray sky samples; examples being the effective energy-averaged PSF implementation or the modulation of the Galactic diffuse emission component after convolution with the PSF. Therefore, we evaluated the model space spanned by our simulator in various ways to ascertain a low degree of mis-modeling. 

The most informative test of this sort is described in App.~\ref{app:svdd}, which invokes the `One-Class Deep Support Vector Data Description' method \cite{pmlr-v80-ruff18a, Caron:2021wmq}, a neural-network-based anomaly test. In simple terms, the underlying neural network of the anomaly test learns to map the simulated gamma-ray maps onto a user-defined manifold (here: a vector of a certain length). The trained network can then be applied to any dataset of the same format (and not generated by our simulator) to test how well it can be mapped to this predefined manifold. The distance of the resulting vector to the training manifold is a score for how anomalous the data under scrutiny is. We employed this prescription to simulated high-latitude gamma-ray maps with and without GRFs. In both cases, the distribution of anomaly scores generated by a validation dataset created with the same simulator always encompasses the anomaly score derived for the real LAT data. This indicates that our simulator produces gamma-ray maps that resemble reality. It also corroborates the robustness of our findings regarding source detection and the shape of the obtained source-count distribution.

Furthermore, we probed in App.~\ref{app:mismodeling-GDE} how explicit mis-modeling of the Galactic diffuse emission component can impact the obtained profile of the \dnds. To this end, we devised an alternative gamma-ray sky featuring a different diffuse MW foreground model (adopted from \cite{Fermi-LAT:2014ryh}). It differs less in the spatial profile of the MW's gas distribution but it lacks a FBs and Loop I contribution. Both, the FBs and Loop I are part of our ROI and their mis-modeling can impact the inference results. However -- judging from the results of this test -- it does not seem to be a significant source of confusion as the ground truth of the injected \dnds~profiles was correctly reconstructed. Nonetheless, we found evidence for background mis-modeling in this alternative dataset when we applied the parameter inference network trained on simulations with GRFs. There was a clear signal for the necessity of a GRF correction hinting at an enhancement of power on scales similar to those of the FBs and Loop I (cf.~Fig.~\ref{fig:inference-diff-confusion-bkg}). We interpret this as evidence of the effectiveness of our GRF implementation to detect and correct background mis-specification. In this sense, we consider the \dnds~inference results based on simulations with GRFs as the most robust findings (within the capabilities of our method) we presented in this study.

The interpretation of the best-fit GRF parameters regarding the real LAT sky is not straightforward. Based on the results with the alternative Galactic diffuse emission model and the explicit non-inclusion of FBs and Loop I, we were able to give a qualitative interpretation of the best-fit real-sky GRF parameters in App.~\ref{app:lat-data-bkg}. Since the LAT sky prefers a harder GRF power spectrum than this constructed example, it calls for modulations on scales slightly smaller than those of FBs and Loop I. However, the obtained posteriors are less constrained than in the alternative dataset with explicit model mis-specification -- potentially indicating a small degree of mis-modeling (as we would expect from our anomaly test). It would be interesting to assess at what exact scale the current diffuse MW foreground models should be modulated to improve the fit to the real sky because such information may reveal the existence of dark gas clumps not accounted for in contemporary models. GRFs are probably not the best tool to answer this question since they only allow for an effective distortion of the baseline model. We defer to future work the investigation of the best alternative.\\

\noindent\textbf{Future developments.} We have demonstrated that the current framework operates reliably at high latitudes. This opens the possibility to answer more physics-related questions about the high-latitude gamma-ray sky and its connection to extragalactic sources. An interesting scientific question is the contribution to the IGRB of different source classes like blazars, misaligned active galactic nuclei, star-forming galaxies, or more exotic processes.

To conduct such a focused IGRB study, it could already be beneficial to extend our SBI framework along the lines we discussed above: 
\begin{itemize}[itemindent=0.1in, leftmargin=0.2in]
    \item Extending the source detection method to operate conditioned on a given flux range, thereby avoiding the introduction of a fixed $C_{\mathrm{th}}$.
    \item Allowing for multiple energy bins and intrinsic source spectra with their associated parameters and priors. This includes the implementation of an energy-dependent PSF and energy dispersion effects.
    \item Adding a source classification module that assigns to each detected source a probability of belonging to a certain source population present in the high-latitude sky. This information can also be consistently injected into the parameter inference network similar to the detection results.
\end{itemize}  
On a longer timescale, these modifications to our current SBI framework will come in handy to address scientific problems of gamma-ray astronomy that are relevant at lower latitudes and along the Galactic plane. 

Lastly, our SBI framework does not need to stay limited to \textit{Fermi}-LAT observations. In the coming years, the focus of gamma-ray astronomy will shift to TeV energies and from space to ground via the Cherenkov Telescope Array Observatory (CTAO) \cite{CTAConsortium:2017dvg}. CTAO will produce high-precision data rendering it necessary to analyze them with physical models of equal quality and detail. An SBI approach like ours adapted to CTAO observations can make a difference in fully harnessing the potential of this next-generation instrument in light of its key science programs. 

\section{Conclusion}
\label{sec:conclusion}

Our study demonstrates the potential of simulation-based inference for gamma-ray astronomy via the example of the GeV high-latitude gamma-ray sky as observed by the \textit{Fermi} LAT. We developed the two main ingredients that compose and determine the power of any SBI method: a realistic simulator of the physics pertinent to the scientific objective and a neural network-supported statistical inference framework adapted to the problem at hand. We designed the gamma-ray simulator with the idea in mind to enable fast generation of simulated data using effective methods to incorporate the instrument response functions of the LAT. The realized physics incorporates the emissions from discrete gamma-ray sources as well as extended and large-scale gamma-ray backgrounds. Among those, the Galactic diffuse emission is treated with special care allowing for variations on a continuous spectrum of angular scales via Gaussian random fields. Our inference framework leverages neural ratio estimation as the underlying SBI flavor. The employed neural networks, given the geometry of the problem, are based on convolutions on the sphere to implement the task of point-like source detection and parameter inference.

Having devised both, the gamma-ray simulator and the statistics toolset, we derived the source-count distribution, \dnds, of discrete point-like gamma-ray sources at Galactic latitudes of $|b|>30^{\circ}$ from 14 years of \textit{Fermi}-LAT data in the range from 1 and 10 GeV. The reconstruction of the source-count distribution in the real data was performed in two decoupled ways: \emph{(i)}, explicitly creating the \dnds~from the population of intermediate and bright point-like sources detected by our SBI detection framework and, \emph{(ii)}, inferring the \dnds~from the \texttt{HEALPix} all-sky map of the binned LAT photon counts in Fig.~\ref{fig:fermi-data}. To realize the latter, we trained an SBI parameter inference network on simulated data and labels specifying the ground truth parameters of the respective \dnds~and backgrounds realized in the samples. 

The key achievements and results of our study are the following:
\begin{itemize}[leftmargin=0pt]
    \item \textbf{Development of gamma-ray simulator and SBI inference framework:} We devised a novel fast gamma-ray simulation software for the high-latitude gamma-ray sky that implements realistic LAT PSF effects on binned gamma-ray sky maps and that adds variability to the Galactic diffuse emission via GRFs. In parallel, we created an SBI inference framework capable of source detection and parameter inference. In contrast to likelihood-based approaches, we can easily account for simulation details that are hard or impossible to incorporate into a likelihood function, such as the treatment of background variations with GRFs. 
    \item \textbf{Source detection:} For the first time, we applied an ML-based gamma-ray detection algorithm to actual \textit{Fermi}-LAT data from a large ROI (instead of simulated data like, e.g., in \cite{Panes:2021zig, vandenOetelaar:2021cot, 2024A&A...692A..38L}). Above a flux of $S=3\times10^{-10}\;\mathrm{cm}^{-2}\,\mathrm{s}^{-1}$ our SBI detection approach can recover up to 98\% of all 4FGL-DR4 point-like gamma-ray sources (without analysis flags). Below this flux value, 4FGL itself becomes gradually incomplete. In this flux range, we detected potential non-4FGL candidates whose nature needs to be examined in dedicated analyses. The detection network trained on simulations with GRFs improves the 4FGL-source recovery rate while also reducing the number of non-4FGL candidates by almost a factor of 2 compared to a network trained on samples without GRFs.
    \item \textbf{Reconstructed high-latitude \dnds:}  We approached the reconstruction of the high-latitude source-count distribution with a parametric and a non-parametric inference network. With and without GRFs, the obtained \dnds~profiles from both strategies agree with each other at the $1\sigma$ level. They also agree well with the source-count distribution following from the 4FGL-DR4 catalog as well as the literature result of \cite{Zechlin:2015wdz, Manconi:2019ynl} based on a traditional likelihood-based method. In light of the many sanity and robustness checks we performed throughout this work, we declare the non-parametric \dnds~profile obtained including GRFs as the best estimate for the high-latitude source-count distribution.  
    \item \textbf{Performance on simulated data:} We evaluated the performance of our \dnds~inference networks on simulated data demonstrating their faithfulness of the \dnds~reconstruction even in the dim-flux regime where point-like sources emit less than 10 photons on average (see Appendix \ref{app:simulation-1pPDF-parametric}). We also showed with coverage tests that the derived marginal posterior distributions are well-calibrated in the sense of Bayesian statistics (see Appendix \ref{app:coverage}).
    \item \textbf{Realism of the gamma-ray simulator:} By allowing for distortions of the Galactic diffuse emission via GRFs, we introduced a larger background variance in the simulated gamma-ray maps. We found that the \dnds~inference results are not strongly impacted by this addition; they remained compatible with each other. Including GRFs even improved the agreement with literature results. Moreover, we checked via a neural-network-based anomaly test that the LAT sky is part of the model space spanned by our gamma-ray simulator with and without GRFs (see Appendix \ref{app:svdd}).
    Both observations point to the robustness and reliability of our findings. 
\end{itemize}
The success of our methodology reveals the great potential of SBI in gamma-ray astronomy. We could reproduce the literature results (which exploit traditional likelihood-based techniques) obtained from the same fraction of the sky. We demonstrated the reliability, calibration, and robustness of our approach. Therefore, we are equipped with a framework to address open questions concerning high-energy gamma-ray astrophysics like the composition and anisotropy of the isotropic gamma-ray background, the source-count and spatial distribution of Galactic gamma-ray bright pulsars, as well as the composition of the gamma-ray emissions in the environment of the Milky Way's Galactic center. The flexibility of SBI allows us to adapt the framework we developed in this study to these more complex challenges while having the certainty that we start from a foundation proven to work. 

\section*{Data Availability}
High-level data products of the SBI analysis, e.g., to reproduce the figures presented in this paper will be provided by the authors upon reasonable request.

\begin{acknowledgments}
We thank Aurelio Amerio for providing us with the numerical data to reproduce the best-fit \dnds~of \cite{Amerio:2023uet}. We thank Silvia Manconi for her continued suggestions and discussions during the work phase of the project and her constructive comments that improved the manuscript. In addition, we thank Dmitry Malyshev, Nick Rodd, Fabian Schmidt and Eiichiro Komatsu for their helpful comments on the draft of the manuscript. FL also thanks Nick~Rodd for many insightful conversations. 
This work has been done thanks to the facilities offered by the Univ.~Savoie Mont Blanc - CNRS/IN2P3 MUST computing center. 
CE acknowledges support by the ``Agence Nationale de la Recherche'', grant n.~ANR-19-CE31-0005-01 (PI: F.~Calore). This publication is supported by the European Union's Horizon Europe research and innovation programme under the Marie Skłodowska-Curie Postdoctoral Fellowship Programme, SMASH co-funded under the grant agreement No. 101081355. The operation (SMASH project) is co-funded by the Republic of Slovenia and the European Union from the European Regional Development Fund. The work of NAM and CW was supported by a project that has received funding from the European Research Council (ERC) under the European Union’s Horizon 2020 research and innovation program (Grant agreement No.~864035 – UnDark). The authors gratefully acknowledge the HPC RIVR consortium \href{https://www.hpc-rivr.si}{(www.hpc-rivr.si)} and EuroHPC JU \href{https://eurohpc-ju.europa.eu/}{(eurohpc-ju.europa.eu)} for funding this research by providing computing resources of the HPC system Vega at the Institute of Information Science \href{https://www.izum.si/en/home/}{(www.izum.si)}.
\end{acknowledgments}

\clearpage
\appendix

\noindent \textbf{APPENDICES}

\vspace{10pt}
\noindent We specify in App.~\ref{sec:data-4fgl} the details of the 4FGL-DR4 source catalog. In App.~\ref{app:NN-architectures} we detail our neural network architectures and training strategies. App.~\ref{app:realData-material} provides additional material supporting our inference on LAT data, while App.~\ref{app:simulations} demonstrates the performance of our framework using simulated data for which the ground truth is known. Furthermore, in App.~\ref{app:gof} we validate the calibration of our posteriors using Bayesian coverage plots and assess simulator realism via anomaly detection. Finally, in App.~\ref{app:mismodeling}, we examine the performance of our framework on synthetic data purposely designed with mis-modeled features in the Galactic diffuse emission and the source-count distribution.

\section{Gamma-ray source catalog}
\label{sec:data-4fgl}

The selected LAT data period is identical to the choice reported in the \textit{Fermi}-LAT collaboration's 4FGL-DR4 gamma-ray source catalog \cite{Ballet:2023qzs}. Although the event selection criteria differ, we can use this list of detected gamma-ray sources to gauge the performance of our SBI framework regarding source detection and the inference of the source-count distribution.

While the comparison of 4FGL to our catalog of detected sources is straightforward, we need to process the information in 4FGL to obtain the source-count distribution. To this end, we select all sources satisfying $|b|>30^{\circ}$ and calculate their flux in the energy band from 1 to 10 GeV. The derivation of the flux uses the best-fitting spectral profile per source stored in the \texttt{SpectrumType} key. Roughly speaking, the spectrum can either be a power law, a log-parabola or an exponentially cutoff power law. From the list of sources, we exclude all objects associated with the Large Magellanic Cloud (LMC) and Small Magellanic Cloud (SMC) since they will be part of our background model. 

As the 4FGL catalog series is a high-level data product, its content depends on the employed data selection and analysis details. All reported sources are statistically significant in terms of a particular test statistic defined by the \textit{Fermi}-LAT collaboration. Yet, some sources are assigned a \textit{flag} indicating potential contamination by background emission or being too close to a bright bordering source (for the exact flag criteria see \cite{Fermi-LAT:2019yla}). To account for the inherent uncertainty of such \textit{flagged} sources, we consider two versions of 4FGL-DR4: with and without flagged sources.

\section{Details on neural network architectures}
\label{app:NN-architectures}
In Tab.~\ref{tab:nn_architecture}, we detail our neural network architectures for source detection and $\mathrm{d}N/
\mathrm{d}S$ inference. The input to our networks is given by photon-count maps (either generated by our forward model as detailed in Sec.~\ref{sec:model} or, once the training has finished, from \textit{Fermi}), standardized to have zero mean and unit variance.
The encoder and decoder together make up the U-Net employed for source detection (which is framed as a pixel-wise classification task). To highlight that the encoder gradually reduces the spatial resolution and increases the number of channels while the decoder reverses the process, we number the encoder and decoder layers in increasing and decreasing order, respectively. Specifically, each pooling (unpooling) layer halves (doubles) the $N_{\mathrm{side}}$ resolution parameter. At the level of the bottleneck in Layer VII, the spatial information has been coarsened down to $N_{\mathrm{side}} = 2$. Note that the number of pixels does not exactly change by a factor of 4 (as would be the case for the entire sky when changing $N_{\mathrm{side}}$ by a factor of 2), as we only propagate values for the `valid' pixels that overlap with our ROI at each resolution. For instance, at our full resolution of $N_{\mathrm{side}} = 128$, this is the case for exactly half of the pixels in the sky, i.e.\ $N_{\mathrm{pix}} / 2 = 6 \times 128^2 = $ 98,304 pixels (where $N_{\mathrm{pix}} = 12 \times N_{\mathrm{side}}^2$ in the \texttt{HEALPix} tesselation \cite{2005ApJ...622..759G}), while 40 out of the 48 pixels at $N_{\mathrm{side}} = 2$ touch our ROI of $|b| \geq 30^\circ$. In particular, the neural network graph representing the \texttt{HEALPix} sphere is built using valid pixels only, and we implemented custom pooling functions that only take these pixels into account. For the graph convolutions, the filters are defined as Chebyshev polynomials, and we take their maximum degree (which is akin to the kernel size in the Euclidean setting) to be $K = 5$.

For the $\mathrm{d}N/\mathrm{d}S$ network, we use an independent encoder with the same structure (apart from leaving out batch normalization, which slightly reduces the total parameter number of the encoder to 3,496,512) as the network head. Since this network is tasked with the inference of global (rather than pixel-wise) quantities, no decoder is needed. The flattened output of the encoder is combined with the mean and standard deviation of the input map and processed through a linear layer to further reduce the dimensionality of the feature vector, which is then fed into the $\mathrm{d}N/\mathrm{d}S$ ratio estimator. In \texttt{swyft} notation, we apply the \texttt{LogRatioEstimator\_1dim} to estimate one-dimensional posterior-to-prior ratios and the \texttt{LogRatioEstimator\_Autoregressive} for higher-dimensional cases, as detailed at the bottom of the table.

\begin{table*}[]
\caption{Layer architecture of our neural networks. The building blocks consist of graph convolutions (GC), maximum pooling (MaxPool) and unpooling (Unpool), batch normalization (BN), and the rectified linear unit (ReLU) activation function. The skip connections are realized by concatenating (Concat) the associated encoder output at each level (I $-$ VI) along the channel dimension. In the $\mathrm{d}N/\mathrm{d}S$ ratio estimator network, which is coupled to an encoder, `P' and `NP' stand for the `parametric' and `non-parametric' case, respectively, and numbers in brackets refer to the case with a GRF. For the parametric variant, the labels \textit{1D} and \textit{12/14D} indicate whether the marginal 1D or the full joint distributions are the inference target.
}
\centering
\resizebox{\textwidth}{!}{%
\begin{tabular}{lclccr}
\cline{2-6}
                                                    & Layer  & Operations                                 & \begin{tabular}[c]{@{}c@{}}Output shape \\ (pixels $\times$ channels)\end{tabular} & Output $N_{\text{side}}$ & \begin{tabular}[c]{@{}r@{}}Trainable parameters\\ (Conv. kernel + bias + BN)\end{tabular} \\ \cline{2-6} 
\multirow{10}{*}{\rotatebox[origin=c]{90}{Encoder}} & Input  &                                            & 98,304 $\times$ 1   & $128$                    & --                                                                                        \\
                                                    & \cellcolor{Layer1}I      & \textcolor{ReLUcolor}{ReLU} $\circ$ \textcolor{BNcolor}{BN} $\circ$ \textcolor{GCcolor}{GC}                 & 98,304 $\times$ 16  & $128$                    & 80 + 16 + 32 $=$ 128                                                           \\
                                                    & \cellcolor{Layer1}       & \textcolor{ReLUcolor}{ReLU} $\circ$ \textcolor{BNcolor}{BN} $\circ$ \textcolor{GCcolor}{GC}                 & 98,304 $\times$ 32  & $128$                    & 2,560 + 32 + 64 $=$ 2,656                                                      \\
                                                    & \cellcolor{Layer2}II     & \textcolor{ReLUcolor}{ReLU} $\circ$ \textcolor{BNcolor}{BN} $\circ$ \textcolor{GCcolor}{GC} $\circ$ \textcolor{MaxPoolcolor}{MaxPool} & 24,832 $\times$ 64  & $64$                     & 10,240 + 64 + 128 $=$ 10,432                                                    \\
                                                    & \cellcolor{Layer3}III    & \textcolor{ReLUcolor}{ReLU} $\circ$ \textcolor{BNcolor}{BN} $\circ$ \textcolor{GCcolor}{GC} $\circ$ \textcolor{MaxPoolcolor}{MaxPool} & 6,272 $\times$ 128  & $32$                     & 40,960 + 128 + 256 $=$ 41,344                                                  \\
                                                    & \cellcolor{Layer4}IV     & \textcolor{ReLUcolor}{ReLU} $\circ$ \textcolor{BNcolor}{BN} $\circ$ \textcolor{GCcolor}{GC} $\circ$ \textcolor{MaxPoolcolor}{MaxPool} & 1,600 $\times$ 256  & $16$                     & 163,840 + 256 + 512 $=$ 164,608                                                \\
                                                    & \cellcolor{Layer5}V      & \textcolor{ReLUcolor}{ReLU} $\circ$ \textcolor{BNcolor}{BN} $\circ$ \textcolor{GCcolor}{GC} $\circ$ \textcolor{MaxPoolcolor}{MaxPool} & 416 $\times$ 512    & $8$                      & 655,360 + 512 + 1,024 $=$ 656,896                                                \\
                                                    & \cellcolor{Layer6}VI     & \textcolor{ReLUcolor}{ReLU} $\circ$ \textcolor{BNcolor}{BN} $\circ$ \textcolor{GCcolor}{GC} $\circ$ \textcolor{MaxPoolcolor}{MaxPool} & 112 $\times$ 512    & $4$                      & 1,310,720 + 512 + 1,024 $=$ 1,312,256                                            \\
                                                    & \cellcolor{Layer7}VII    & \textcolor{GCcolor}{GC} $\circ$ \textcolor{MaxPoolcolor}{MaxPool}                         & 40 $\times$ 512     & $2$                      & 1,310,720 + 512 + 0 $=$ 1,311,232                                                             \\ \cline{6-6} 
                                                    &        &                                            &                     &                          & \textit{Total:}                                   3,499,552                             \\                                   
\multicolumn{6}{l}{}                                                                                                                                                                                                                                                                                                          \\

\multirow{20}{*}{\rotatebox[origin=c]{90}{Decoder}} & \cellcolor{Layer6}VI     & \textcolor{ReLUcolor}{ReLU} $\circ$ \textcolor{BNcolor}{BN} $\circ$ \textcolor{GCcolor}{GC} $\circ$ \textcolor{Unpoolcolor}{Unpool}  & 112 $\times$ 512    & $4$                      & 1,310,720 + 512 + 1,024 $=$ 1,312,256                                            \\
                                                    & \cellcolor{Layer6}       & \cellcolor{Layer6}\textcolor{black}{Concat-VI}                                  & 112 $\times$ 1,024  & $4$                      & --                                                                                        \\
                                                    & \cellcolor{Layer6}       & \textcolor{ReLUcolor}{ReLU} $\circ$ \textcolor{BNcolor}{BN} $\circ$ \textcolor{GCcolor}{GC}                 & 112 $\times$ 512    & $4$                      & 2,621,440 + 512 + 1,024 $=$ 2,622,976                                            \\
                                                    & \cellcolor{Layer5}V      & \textcolor{ReLUcolor}{ReLU} $\circ$ \textcolor{BNcolor}{BN} $\circ$ \textcolor{GCcolor}{GC} $\circ$ \textcolor{Unpoolcolor}{Unpool}  & 416 $\times$ 512    & $8$                      & 1,310,720 + 512 + 1,024 $=$ 1,312,256                                            \\
                                                    & \cellcolor{Layer5}       & \cellcolor{Layer5}\textcolor{black}{Concat-V}                                   & 416 $\times$ 1,024  & $8$                      & --                                                                                        \\
                                                    & \cellcolor{Layer5}       & \textcolor{ReLUcolor}{ReLU} $\circ$ \textcolor{BNcolor}{BN} $\circ$ \textcolor{GCcolor}{GC}                 & 416 $\times$ 512    & $8$                      & 2,621,440 + 512 + 1,024 $=$ 2,622,976                                            \\
                                                    & \cellcolor{Layer4}IV     & \textcolor{ReLUcolor}{ReLU} $\circ$ \textcolor{BNcolor}{BN} $\circ$ \textcolor{GCcolor}{GC} $\circ$ \textcolor{Unpoolcolor}{Unpool}  & 1,600 $\times$ 256  & $16$                     & 655,360 + 256 + 512 $=$ 656,128                                                \\
                                                    & \cellcolor{Layer4}       & \cellcolor{Layer4}\textcolor{black}{Concat-IV}                                  & 1,600 $\times$ 512  & $16$                     & --                                                                                        \\
                                                    & \cellcolor{Layer4}       & \textcolor{ReLUcolor}{ReLU} $\circ$ \textcolor{BNcolor}{BN} $\circ$ \textcolor{GCcolor}{GC}                 & 1,600 $\times$ 256  & $16$                     & 655,360 + 256 + 512 $=$ 656,128                                                \\
                                                    & \cellcolor{Layer3}III    & \textcolor{ReLUcolor}{ReLU} $\circ$ \textcolor{BNcolor}{BN} $\circ$ \textcolor{GCcolor}{GC} $\circ$ \textcolor{Unpoolcolor}{Unpool}  & 6,272 $\times$ 128  & $32$                     & 163,840 + 128 + 256 $=$ 164,224                                                \\
                                                    & \cellcolor{Layer3}       & \cellcolor{Layer3}\textcolor{black}{Concat-III}                                 & 6,272 $\times$ 256  & $32$                     & --                                                                                        \\
                                                    & \cellcolor{Layer3}       & \textcolor{ReLUcolor}{ReLU} $\circ$ \textcolor{BNcolor}{BN} $\circ$ \textcolor{GCcolor}{GC}                 & 6,272 $\times$ 128  & $32$                     & 163,840 + 128 + 256 $=$ 164,224                                                \\
                                                    & \cellcolor{Layer2}II     & \textcolor{ReLUcolor}{ReLU} $\circ$ \textcolor{BNcolor}{BN} $\circ$ \textcolor{GCcolor}{GC} $\circ$ \textcolor{Unpoolcolor}{Unpool}  & 24,832 $\times$ 64  & $64$                     & 40,960 + 64 + 128 $=$ 41,152                                                    \\
                                                    & \cellcolor{Layer2}       & \cellcolor{Layer2}\textcolor{black}{Concat-II}                                  & 24,832 $\times$ 128 & $64$                     & --                                                                                        \\
                                                    & \cellcolor{Layer2}       & \textcolor{ReLUcolor}{ReLU} $\circ$ \textcolor{BNcolor}{BN} $\circ$ \textcolor{GCcolor}{GC}                 & 24,832 $\times$ 64  & $64$                     & 40,960 + 64 + 128 $=$ 41,152                                                    \\
                                                    & \cellcolor{Layer1}I      & \textcolor{ReLUcolor}{ReLU} $\circ$ \textcolor{BNcolor}{BN} $\circ$ \textcolor{GCcolor}{GC} $\circ$ \textcolor{Unpoolcolor}{Unpool}  & 98,304 $\times$ 32  & $128$                    & 10,240 + 32 + 64 $=$ 10,336                                                    \\
                                                    & \cellcolor{Layer1}        & \cellcolor{Layer1}\textcolor{black}{Concat-I}                                   & 98,304 $\times$ 64  & $128$                    & --                                                                                        \\
                                                    & \cellcolor{Layer1}       & \textcolor{ReLUcolor}{ReLU} $\circ$ \textcolor{BNcolor}{BN} $\circ$ \textcolor{GCcolor}{GC}                 & 98,304 $\times$ 32  & $128$                    & 10,240 + 32 + 64 $=$ 10,336                                                    \\
                                                    & \cellcolor{Layer0}Output & \textcolor{BNcolor}{BN} $\circ$ \textcolor{GCcolor}{GC}                              & 98,304 $\times$ 1   & $128$                    & 160 + 1 + 2 $=$ 163                                                            \\ \cline{6-6} 
                                                    &        &                                            &                     &                          & \textit{Total:}                                    9,614,307     \\
\multicolumn{6}{l}{}                                                                                                                                                                                                                                                                                                          \\
\multirow{6}{*}{\rotatebox[origin=c]{90}{d$N/$d$S$ estimator}} & \multicolumn{1}{l}{}                                                         & Flatten                                    & 20,480               & \multicolumn{1}{l}{}     & --                                                                                        \\
 & \multicolumn{1}{l}{}                                                         & Append mean \& STD of the map                  & 20,482               & \multicolumn{1}{l}{}     & --                                                                                        \\
 & \multicolumn{1}{c}{\multirow{3}{*}{\rotatebox[origin=c]{90}{P}}}     & \quad Linear                                     & 1,706                & \multicolumn{1}{l}{}     & 34,942,292 + 1,706 + 0 $=$ 34,943,998                                                     \\
 & \multicolumn{1}{l}{}                                                         & \quad LogRatioEstimator                          & 12 (14)                  & \multicolumn{1}{l}{}     & \begin{tabular}[c]{@{}r@{}}{\it 1D:} 1,518,348 (1,771,406) \\ {\it 12D:} 6,896,664 ({\it 14D:} 8,053,276) \end{tabular}    \\
 & \multicolumn{1}{c}{\multirow{2}{*}{\rotatebox[origin=c]{90}{NP}}} & \quad Linear                                     & 426                 & \multicolumn{1}{l}{}     & 8,725,332 + 426 $=$ 8,725,758                                                             \\
 & \multicolumn{1}{l}{}                                                         & \quad LogRatioEstimator                          & 24 (26)                  & \multicolumn{1}{l}{}     & \begin{tabular}[c]{@{}r@{}}{\it 1D:} 1,070,616 (1,159,834) \end{tabular} 
\\ \bottomrule
\end{tabular}%
}
\label{tab:nn_architecture}
\end{table*}

In addition to the definition of each network for the conducted inference tasks, we list in Tab.~\ref{tab:hyperparams} the training hyper-parameters that we employ in each case. As described in the main text's Sec.~\ref{sec:nn_architecture}, the background resampling is a data augmentation technique that scrambles the background maps $\boldsymbol{b}$ among the samples of a batch to increase the variability of the background on-the-fly.

\begin{table*}[t]
    \centering
    \renewcommand{\arraystretch}{1.6}
    \begin{tabular}{l@{\hskip 0.2in}c@{\hskip 1.5in}c}
        \toprule
        Hyper-parameter & Source Detection & \dnds \\
        \midrule
        Number of samples & \multicolumn{2}{c}{200,000}\\
        (Training/Validation) ratio & \multicolumn{2}{c}{(80/20)\%}\\
        Batch size & 64 & 128 \\
        Optimizer & \multicolumn{2}{c}{AdamW \cite{2017arXiv171105101L}}\\
        Initial learning rate & $7.5\times10^{-4}$ & $1\times10^{-4}$\\
        Learning rate scheduler  & \multicolumn{2}{c}{\texttt{ReduceLROnPlateau}}\\
        Learning ratio schedule decay factor & \multicolumn{2}{c}{0.1}\\
        Learning ratio schedule patience & \multicolumn{2}{c}{8}\\
        Learning ratio schedule criterion & \multicolumn{2}{c}{minimal validation loss}\\
        Maximal number of training epochs & 75 & $100$ \\
        Early stopping patience & \multicolumn{2}{c}{$15$} \\
        Early stopping criterion & \multicolumn{2}{c}{minimal validation loss} \\
        Background resampling & \texttt{true} &  \texttt{false}\\
        \bottomrule
    \end{tabular}
    \caption{Summary of training hyper-parameters for each inference task to be conducted within the scope of this study.}
    \label{tab:hyperparams}
\end{table*}

\section{Additional material supplementing the LAT data inference}
\label{app:realData-material}

In this section, we list and collect additional inference results we obtained from the application of the trained networks on the 14-year LAT dataset. They supplement the discussion of our findings and help to provide a better understanding of their implications.\\

\subsection{Detection results with Gaussian random fields}
\label{app:lat-detection-wGRF}

As anticipated in the main text in Sec.~\ref{sec:results_det}, we observe an improved sensitivity of the detection algorithm compared to the results without GRFs. We attribute this improvement to the enlarged variability of the Galactic diffuse emission component and the ability of the detection network to better recognize background fluctuations. Effectively, this lowers the false positive rate for a given value of $\log_{10} r(\vec{x}, C_\text{th}; \bm d)$.

\begin{figure*}[t]
    \includegraphics[width=\textwidth]{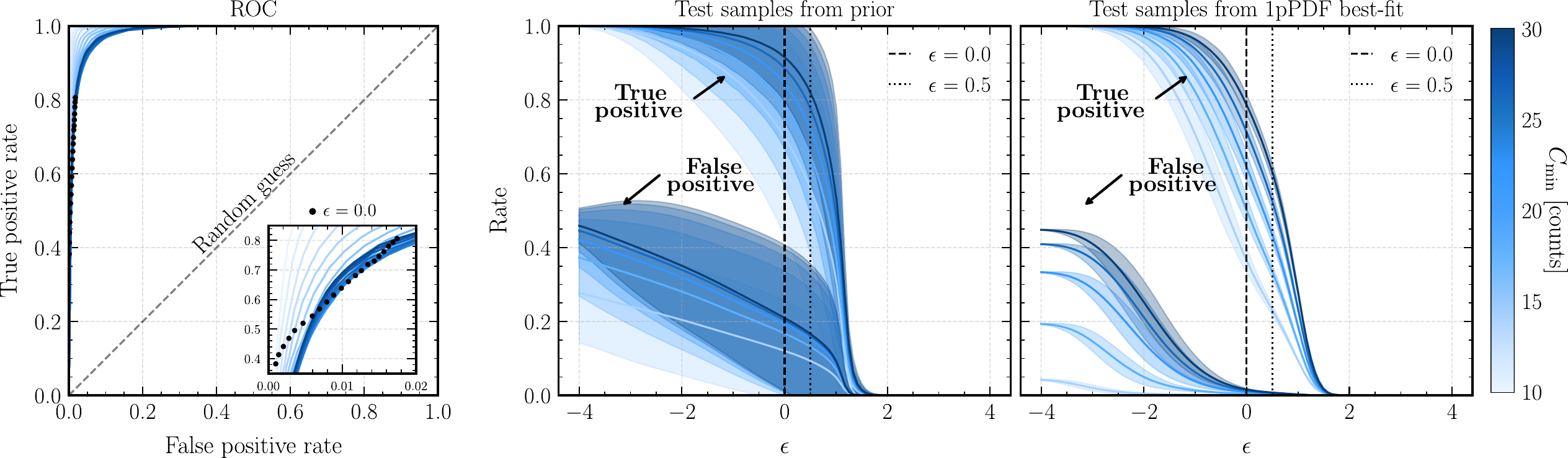}
    \caption{Same as Fig.~\ref{fig:roc} regarding the detection network trained on samples including background distortions via GRFs.}
    \label{fig:roc-wGRF}
\end{figure*}

To gauge the threshold we impose on $\log_{10} r(\vec{x}, C_\text{th}; \bm d)$ when we analyze the LAT dataset, we study the performance of the detection network on simulated data. Fig.~\ref{fig:roc-wGRF} displays these checks in the same manner as Fig.~\ref{fig:roc} in the main text. The extracted ROC curve resembles its analog without GRFs, which indicates that the inclusion of GRFs does not inhibit the performance of the detection network. The central panel of this figure compares true and false positive rates on simulated sky maps sampled from the full prior as a function of the chosen threshold $\epsilon$ on $\log_{10} r(\vec{x}, C_\text{th}; \bm d)$. As a general observation, the inferred scatter of the multiple profiles is wider than it was in the case without GRFs, whereas the mean of these curves is affected only mildly. This outcome is not unexpected since the variability of the simulated gamma-ray sky maps is increased. In particular, the background emission may get enhanced significantly, which renders it harder to recover some of the injected point-like sources.

In contrast, we do not expect a strong enhancement of the Galactic diffuse emission in the actual LAT data; in fact, the default model we are using is a fit to gamma-ray data. Therefore, we show in the right panel of Fig.~\ref{fig:roc-wGRF}, the true and false positive rates regarding simulated test data that follows a TBPL \dnds~with the best-fit parameters of the original 1p-PDF work \cite{Zechlin:2015wdz}. The four background component normalizations are fixed to 1, and no GRF distortion is applied to the Galactic diffuse emission template. Hence, this test case corresponds to a situation where the simulated data does not contain GRFs but the detection network is trained in their presence. We stress, however, that the simulator with GRFs encompasses the possibility of $A_{\mathrm{GRF}} = 0$. We observe a sizeable shift of the true positive curve to the left in contrast to Fig.~\ref{fig:roc}. Consequently, at a fixed threshold $\epsilon$, fewer point-like sources are correctly recovered from the injected ground truth. Even for the sample of bright sources ($C = 30$) only 60\% of the sources can be retrieved at $\epsilon = 0.5$ while the analogous positive rate was around 80\% in the scenario without GRFs. Yet, the false positive rate is shifted even further to the left, implying that we can expect fewer false positives at $\epsilon = 0.5$ or lower values. Hence, it is well-motivated to shift the threshold for detection in the scenario with GRFs to $\epsilon = 0$. This threshold achieves a similar true positive rate of 80\% for sources with $C = 30$. The false positive rate is even lower than in the case without GRFs evaluated at $\epsilon = 0.5$. 

\begin{figure*}[t]
    \includegraphics[width=\columnwidth]{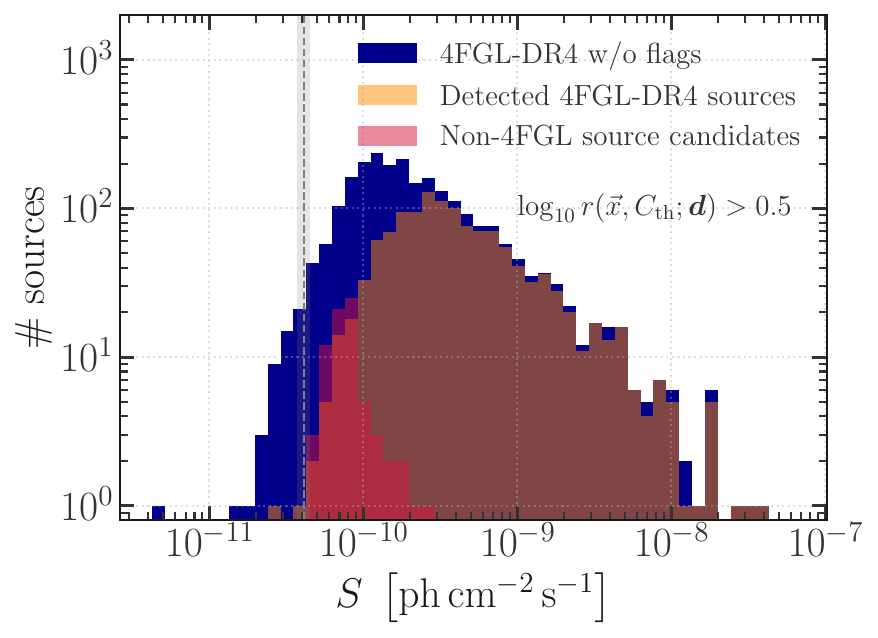}\hfill\includegraphics[width=\columnwidth]{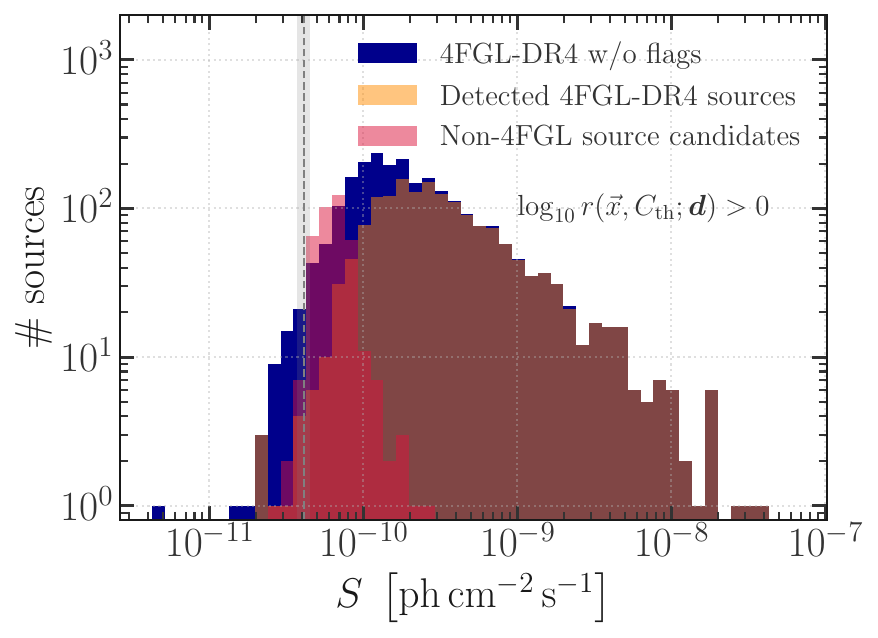}
    \caption{Same as Fig.~\ref{fig:detection_histogram} representing the source-count distribution obtained by applying the detection network trained on samples including background distortions via GRFs on the 14-year LAT dataset. We provide the source counts for \textbf{Left:} $\log_{10} r(\vec{x}, C_\text{th}; \bm d) > 0.5$ and \textbf{Right:} $\log_{10} r(\vec{x}, C_\text{th}; \bm d) > 0.0$.}
    \label{fig:lat-detection-wGRF}
\end{figure*}

Applied to the real data, we obtain a catalog of detected sources that we present in Fig.~\ref{fig:lat-detection-wGRF}. To illustrate how the idealistic test on simulated data translates to real-data application, we show two histograms of the source-count distribution. The fluxes of non-4FGL candidates are, again, computed by taking the counts observed in the respective pixel and subtracting the background contribution according to the best-fit values reported in Fig.~\ref{fig:bkg-woGRF-wGRF} (without modulating the Galactic diffuse emission component via a GRF realization). In the left panel, we employ the previous threshold of $\epsilon = 0.5$ while in the right panel, we adopt the value of $\epsilon = 0$ that is more suited for the scenario with GRFs. In general, the real data findings confirm the intuition we gained from simulated data. The threshold of $\epsilon = 0.5$ leads to fewer recovered 4FGL sources even in the bright part of the catalog in contrast to the scenario without GRFs. Additionally, the number of potential source candidates is markedly reduced, too. Interestingly, we recover only two sources that are slightly below the cut $C_{\mathrm{th}} = 10$ used to train the networks on point-source positions. The same is true in the right panel with a reduced detection threshold $\epsilon = 0$: While the number of recovered 4FGL and non-4FGL candidate sources are sizeable above this cut (gray vertical line), they quickly drop for fluxes below the threshold. This observation corroborates the efficacy of our method and demonstrates that it works as intended. In this setup, it is remarkable that we are able to recover about 98\% of all sources with $S>3\times10^{-10}$ cm$^{-2}$s$^{-1}$ and 70\% of all 4FGL-DR4 point-like sources without analysis flags. These numbers are not so different from the case without GRFs but they are achieved at a larger detection threshold (which yields a lower false-positive rate in simulated data). Moreover, with a lower threshold, the case with GRFs yields 387 non-4FGL candidates. This is considerably lower than the 685 candidates we identified without GRFs. This observation demonstrates that the increased variability of the Galactic diffuse emission improves the overall performance of our detection approach. It would be interesting to investigate whether the retained 387 non-4FGL candidates are, on average, more likely to be a genuine gamma-ray source compared to the 685 candidates obtained without GRFs. The answer would corroborate our expectation that training on modulated diffuse MW foreground realizations reduces the detection of spurious background fluctuations. 

\subsection{Background parameter estimation and one-dimensional marginal posteriors}
\label{app:lat-data-bkg}

\begin{figure*}[t]
    \includegraphics[width=\textwidth]{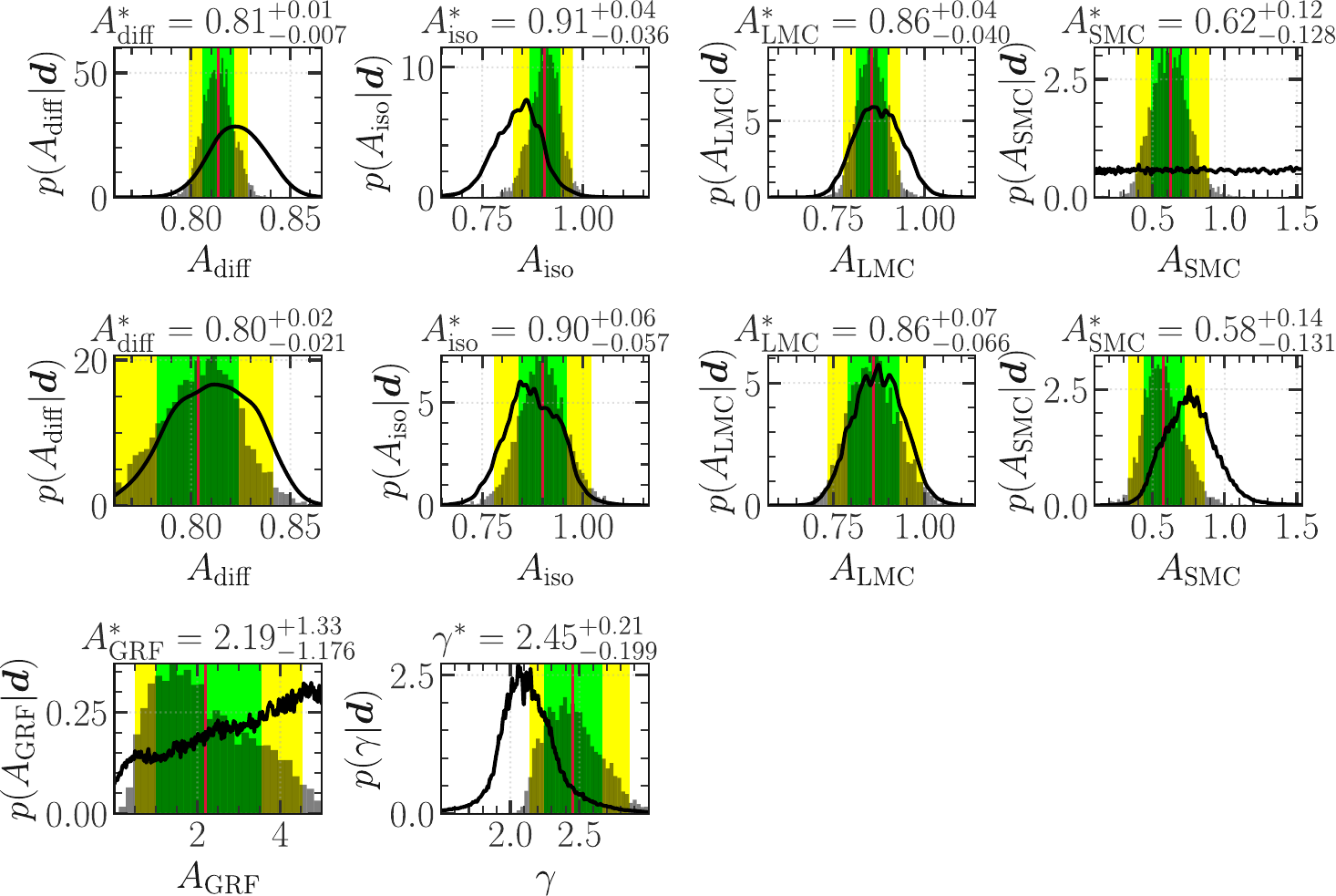}
    \caption{Comparison of the posterior probability density functions for the four background components of our gamma-ray emission model (see Tab.~\ref{tab:params}) obtained from 14 years of \textit{Fermi}-LAT data at $|b|>30^{\circ}$. We display as histograms the marginal one-dimensional posteriors generated by the nested sampler applied in the parametric $\mathrm{d}N/\mathrm{d}S$ approach. A vertical red line denotes the mean of these posteriors, while 68\% and 95\% credible intervals are overlaid in green and yellow, respectively. The panels' titles state the numerical values of the mean and 68\% credible region. The corresponding result of the non-parametric $\mathrm{d}N/\mathrm{d}S$ approach is shown as black solid profiles. \textbf{Top row:} inference networks trained on simulations without GRF; \textbf{Bottom rows:} Analogous results based on simulations with GRFs, including the GRF parameters $A_\text{GRF}$ and $\gamma$. Panels of parameters common between the simulations with and without GRFs share the same horizontal axis range. Note that the panels show a zoom into the parameter region with high posterior probability which, in some cases, is significantly narrower than the prior range. For instance, recall that we adopted a log-uniform prior for $A_{\mathrm{diff}}$ and $A_{\mathrm{iso}}$ between 0.1 and 2.0.}
    \label{fig:bkg-woGRF-wGRF}
\end{figure*}

This section supplements the Secs.~\ref{sec:results_bkg_woGRF} and ~\ref{sec:results_bkg_wGRF} of the main text by providing the one-dimensional marginal posterior distributions of the background parameters we inferred in the parametric and non-parametric approach using simulations with and without GRFs. The summary plot is shown in Fig.~\ref{fig:bkg-woGRF-wGRF}.

The one-dimensional marginal posterior distributions obtained within the parametric $\mathrm{d}N/\mathrm{d}S$ scenario are given as histograms of the samples generated by the nested sampler. Their mean values are denoted by a vertical red line, while the 68\% and 95\% credible intervals are denoted by yellow and green bands, respectively. The resulting best-fit values with 68\% uncertainty are stated in the respective panel's title. For comparison, we overlay the corresponding posterior derived with the non-parametric $\mathrm{d}N/\mathrm{d}S$ framework as solid black profiles. \\

\noindent\textbf{Details about the background parameter inference without GRFs.} The parametric approach allows us to determine the normalization parameters of all four background components relatively precisely, whereas the SMC remains unconstrained in the non-parametric case. As stated above, the parametric approach relies on the full joint posterior and, thus, parameter correlations, which enhances the sensitivity to all parameters of the model. This loss of information is likely to be the reason why we cannot infer the SMC parameter in the non-parametric scenario, as it is only trained on one-dimensional ratios. We checked that even the non-parametric approach can infer this normalization once we introduce \textit{sequential inference}. This method goes beyond what we describe in this study as it exploits multiple training rounds on the same target data. The general idea is, first, to use the NRE network to obtain the posterior distributions for all parameters regarding the target dataset. With this information, the original priors are truncated in a particular truncation scheme suitable for the problem where the posteriors have essentially zero support. The subsequent training round then utilizes training samples that were generated from the truncated priors, hence, being much closer to the target data than the first training set. This procedure can be repeated several times until a certain convergence condition is reached. Sequential inference and truncation are core features of \texttt{swyft}. Since the SMC parameter is, in principle, accessible, improving the network capacity and/or increasing the number of training samples is potentially enough to obtain an informed posterior on $A_{\mathrm{SMC}}$ already with a single training round.

The posteriors of the normalizations of LMC and diffuse MW foreground agree. The isotropic emission, in contrast, is predicted with a larger flux in the parametric scenario than in the non-parametric case; however, the posteriors are compatible. This behavior is consistent with the fact that the source-count distribution in the dim regime is also different in both approaches. The isotropic component is, to some degree, degenerate with this part of the $\mathrm{d}N/\mathrm{d}S$. Since the parametric approach predicts fewer sources in this flux range, the isotropic emission needs to be larger to fit the total gamma-ray emission in the data. 

At first glance, it may seem counter-intuitive that the normalization of the Galactic diffuse emission is reconstructed to $A^{\ast}_{\mathrm{diff}} = 0.81^{+0.01}_{-0.01}$, that is, lower than 1. As we pointed out in Sec.~\ref{sec:de_fm}, this model was derived from a fit to the \textit{Fermi}-LAT sky and should hence fit the expectations after rescaling it with the proper exposure for the selected observation period. However, the fit to gamma-ray data was iteratively performed in different partitions of the full sky, utilizing energy-binned templates and explicitly including energy dispersion effects. Since we are considering only a single energy bin and, thus, not accounting for the LAT's energy dispersion, we expect a renormalization of the model to match the observational data (also cf.~Sec.~\ref{sec:instrument}). A similar reasoning applies to the recovered normalization constants of LMC and SMC because all source spectra in 4FGL were obtained with energy dispersion corrections. 

Valuable information is contained in the inferred normalization of the diffuse isotropic background, $A_{\mathrm{iso}}^{\ast} = 0.91^{+0.04}_{-0.036}$ (parametric approach). With this inferred value, we can compute the total integrated flux of this background normalized to the angular extent of the considered ROI to be $S_{\mathrm{iso}} = 4.16^{+0.19}_{-0.16}\times10^{-7}\;\mathrm{cm}^{-2}\,\mathrm{s}^{-1}\,\mathrm{sr}^{-1}$. The total diffuse isotropic background flux is relevant since, in standard template-based fits, this component will be the entire IGRB since it necessarily receives a contribution from unresolved point-like sources. However, the unresolved-source contribution is explicitly modeled via the source-count distribution and, hence, it should not be part of $S_{\mathrm{iso}}$ in our analysis. This expectation is met, on one hand, because the obtained normalization $A_{\mathrm{iso}}^{\ast}$ is below 1, which corresponds to the best-fit IGRB model derived by the \textit{Fermi}-LAT collaboration. On the other hand, we can compare this number to the value reported in the 1p-PDF study, which also explicitly models the unresolved source flux. For the case shown in Fig.~\ref{fig:dNdS-woGRF}, the authors of \cite{Zechlin:2015wdz} report an diffuse isotropic flux of $S_{\mathrm{iso}} = 1.4^{+0.3}_{-0.4}\times10^{-7}\;\mathrm{cm}^{-2}\,\mathrm{s}^{-1}\,\mathrm{sr}^{-1}$. The value is about a factor of three lower than our total diffuse isotropic contribution, but this observation needs to be contextualized properly. The 1p-PDF study was performed on reprocessed Pass 7 LAT data for photon events classified as \texttt{CLEAN}. The closest equivalent to this event class regarding the Pass 8 standard is the \texttt{ULTRACLEANVETO} class, whose contamination by residual non-photon events is a factor of 5 to 6 lower than for the \texttt{SOURCEVETO} event class that we selected for our study. Those residual non-photon events are the other considerable component of the IGRB. With this in mind, it is normal that our total diffuse isotropic flux is larger than the one reported in \cite{Zechlin:2015wdz}. In fact, the authors of \cite{Manconi:2019ynl} provide an update of the original 1p-PDF assessment based on \texttt{ULTRACLEANVETO} Pass 8 LAT data. They report a similar diffuse isotropic total flux as \cite{Zechlin:2015wdz}. Hence, it is very likely that the higher diffuse isotropic flux we inferred is the result of having chosen the \texttt{SOURCEVETO} event class.\\

\noindent\textbf{Details about the background parameter inference in the presence of GRFs.} In addition to the four component normalization parameters, we have an assessment of the GRFs' power spectrum slope $\gamma$ and amplitude $A_{\mathrm{GRF}}$ that characterize the background model mis-specification between our simulator and reality. Hence, we obtain information about what angular scales drive the discrepancies between the \textit{Fermi} diffuse background model and the real high-latitude sky. Since GRFs with a power-law power spectrum by construction inject power at all angular scales, the value of  $\gamma^{\ast}$ does not yield a single scale that predominantly causes the mis-modeling. According to the parametric approach, the real LAT data hints at the existence of mis-modeling on larger angular scales since the power spectrum slope's posterior peaks towards the soft end of the scanned prior range. For example, given the best-fit GRF parameters, angular scales of $l=10^{\circ}$ are enhanced by a factor of about 60 with respect to angular scales of $l=2^{\circ}$. In App.~\ref{app:mismodeling-GDE}, we inspect the case of purposefully mis-modeled Galactic diffuse emission and whether the GRFs pick up the scales and normalization at which the alternative diffuse model deviates from our benchmark choice. The findings of this sanity check help us here to formulate a qualitative interpretation of the best-fit GRF parameters: While the alternative diffuse MW foreground model features a similar spatial morphology traced by the gas density distribution, it lacks the FBs and Loop I. We find that the best-fit GRF parameters produce exponentiated GRF realizations that correct these discrepancies. In that case, the obtained negative power spectrum slope is larger than that derived from the real data, hinting at mis-modeling emerging at angular scales smaller than those of Loop I and the FBs. Interestingly, the respective posterior for $\gamma$ in the non-parametric approach predicts slightly harder power spectra, which are consistent with the autoregressive NRE findings at the $2\sigma$ level. This $2\sigma$ discrepancy is most likely caused by the differing sensitivity to this specific parameter the NRE network's training target exhibits, i.e., either the one-dimensional marginal distribution or the full joint distribution. We verified that the parametric approach trained on such one-dimensional marginals yields a posterior for $\gamma$ compatible with its non-parametric analog (also regarding $A_{\mathrm{GRF}}$). Hence, the additional sensitivity of the joint posterior (achieved through added information about parameter correlations) leads to the observed shift of the mean of the posterior. In principle, applying sequential inference might improve the outcome and eventually reconcile the inferred posterior properties.

We performed a cross-check in the parametric approach where we trained the autoregressive NRE on the same training dataset but by marginalizing over all background parameters except for $A_{\mathrm{diff}}$ and $A_{\mathrm{iso}}$. As shown in Fig.~\ref{fig:2Dmarginal-LAT-combined} of App.~\ref{app:lat-2D-marginals}, these background components do not exhibit a strong correlation with any of the parameters of interest. Therefore, we do not expect a largely different \dnds~profile or large-scale diffuse normalizations. However, we still incorporate the variability of the Galactic diffuse emission provided by the GRFs. Eventually, our expectations are met. The obtained reconstructed \dnds~profile is in full agreement with the left panel of Fig.~\ref{fig:dNdS-wGRF}. The same applies to the one-dimensional marginal posteriors for the diffuse isotropic background and the Galactic diffuse emission. 

\subsection{Parameter correlations in the parametric \dnds~inference approach}
\label{app:lat-2D-marginals}

\begin{figure*}[t]
    \includegraphics[width=\textwidth]{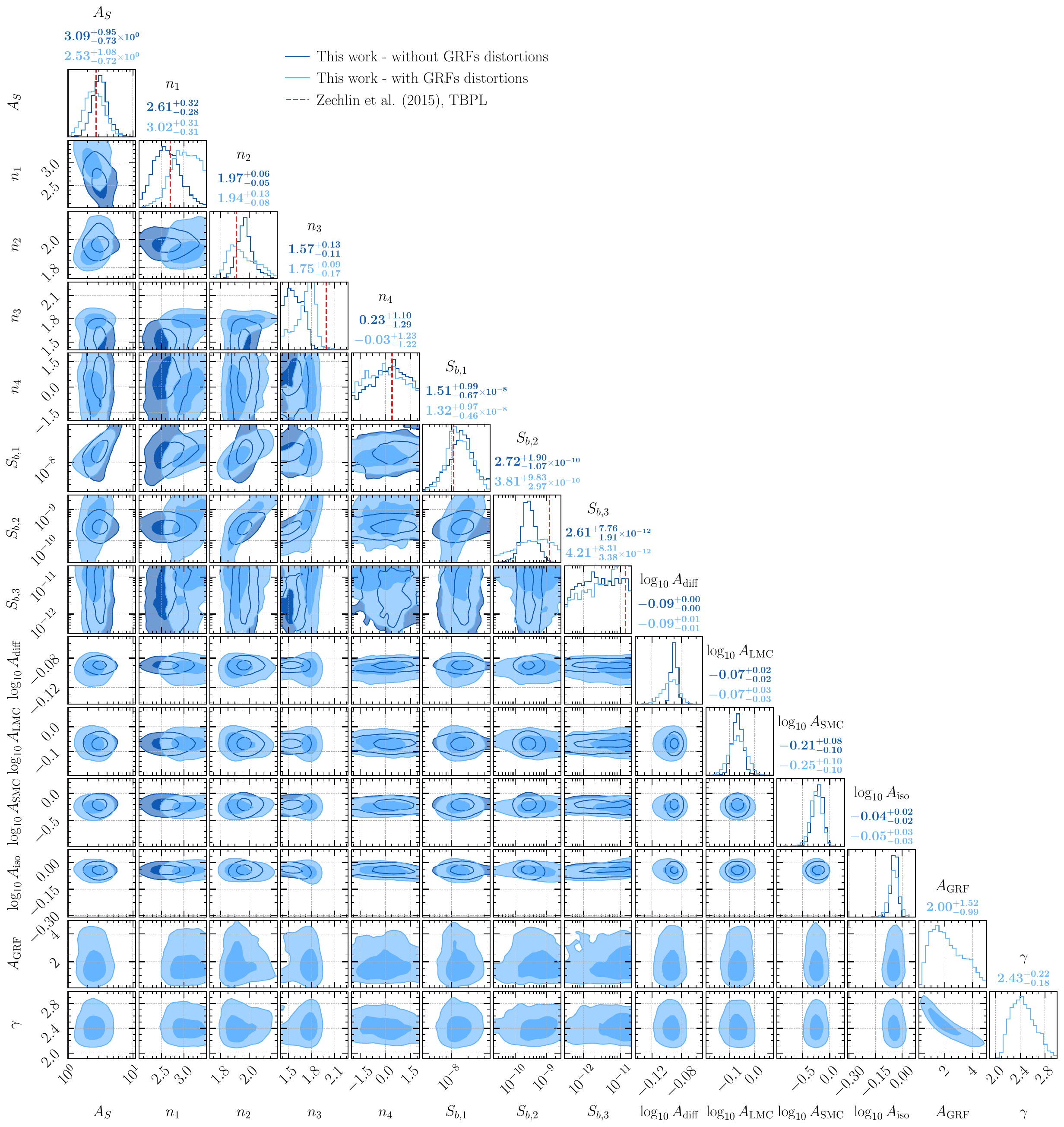}
    \caption{Corner plot of the two-dimensional marginal posterior distributions obtained from applying the trained network of Sec.~\ref{sec:results_woGRF} to the 14-year LAT dataset. The parameters were reconstructed within the parametric \dnds~inference approach using autoregressive NRE and nested sampling. We show in dark blue results without including variations from the GRFs, and in light blue including them. The off-diagonal panels show the contours corresponding to 1- and 2-$\sigma$ respectively for each respective parameter pair, whereas the diagonal panels indicate the one-dimensional marginal posteriors of each parameter. In the titles, we report the median of the distribution with the 68\% credible interval. The vertical dashed red lines mark the best-fit TBPL parameters determined via the 1p-PDF profile-likelihood approach of \cite{Zechlin:2015wdz}.}
    \label{fig:2Dmarginal-LAT-combined}
\end{figure*}

\noindent\textbf{Results without GRFs.} In the main text, we did not elaborate on all the information that we can extract from the autoregressive NRE approach and the parametric \dnds~inference. It provides us with the full joint posterior distributions and, thus, we can inspect the correlations among different parameters, e.g., based on the two-dimensional marginal posterior distributions that we display in Fig.~\ref{fig:2Dmarginal-LAT-combined} in dark blue shades. At first glance, we observe that all marginal posterior distributions are, to some degree constrained by the data, although their predictiveness varies; it is, in particular, reduced for parameters determining the dim fraction of the source population ($n_4$ and $S_{b,3}$). In contrast, the posteriors of all astrophysical background components are relatively precise. The diagonal of this plot shows the one-dimensional marginal posteriors of each parameter. For comparison, we add as a vertical dashed red line the respective best-fit value of the TBPL 1p-PDF result (profile likelihood; \cite{Zechlin:2015wdz}) in all panels referencing \dnds-parameters (there are no equivalents regarding the astrophysical background). Most of these best-fit parameters fall within the 68\% credible intervals of the obtained posteriors. The exceptions are $n_3$ and $S_{b,2}$, which are responsible for the reduced number of dim sources we uncovered compared to the 1p-PDF technique.    

There are a few prominent correlations and degeneracies: the amplitude $A_S$ of the \dnds~is correlated with the position of the first break $S_{b,1}$. This is intuitively clear, as two \dnds~realizations with the same amplitude but different first break positions will predict different amounts of point-like sources in the intermediate flux range where the inference algorithm is most sensitive (see, e.g., the precision of the posterior of $n_2$). The scenario with a break at higher fluxes would predict fewer intermediate sources so that, when applied to an actual dataset, the amplitude needs to compensate for this effect to allow for the same amount of sources. Although less pronounced, the normalization of the diffuse isotropic background is correlated with the third break position $S_{b,3}$ and technically anti-correlated with $n_4$. The third break position marks the transition to the faint flux regime where sources emit only a few gamma rays on average so that their cumulative emission generates an isotropic background. Then, the total brightness of the source population's faint fraction is determined by the last \dnds~slope $n_4$. The higher the slope, the more faint sources one expects, consequently lowering the normalization of the diffuse isotropic background required to fit the data. Yet, the effect is washed out by our limited sensitivity to this flux regime and the fact that there is only a mild correlation between this background component and the faint part of the \dnds.\\

\noindent\textbf{Results with GRFs.} Concerning the scenario with GRF distortions, we display with light blue shades in Fig.~\ref{fig:2Dmarginal-LAT-combined} the two-dimensional marginal posterior distributions obtained with the autoregressive NRE approach on the 14-year LAT dataset. Qualitatively, there are no striking differences between the parameter correlations present in this scenario compared to the case without GRFs. We note that the added parameters governing the GRFs are not degenerate with any of the other parameters except for a mutual correlation. The softer the GRF power spectrum (larger $\gamma$), the smaller the required amplitude of the random field $A_{\mathrm{GRF}}$. This result is expected because a soft power spectrum means a lot of power on large scales that produce large amplitudes due to the chosen normalization of the power spectrum in Eq.~\ref{eq:power-spectrum}. As there is a fixed degree of model mis-specification in the simulated data, $\gamma$ and $A_{\mathrm{GRF}}$ must be anti-correlated. 

\section{Inference results on simulated data}
\label{app:simulations}

In this appendix section, we list and comment on multiple inference tests on simulated data to demonstrate the reliable performance of our SBI framework. 

\subsection{Recovery of the 1p-PDF best-fit \dnds}
\label{app:simulation-1pPDF-parametric}

In the main text's Fig.~\ref{fig:dNdS-woGRF}, we confronted the parametric and non-parametric \dnds~inference results on real data with similar findings obtained with the 1p-PDF technique. A common observation between our two SBI inference methods was a reduced number of dim sources in comparison with the 1p-PDF best-fit TBPL (which itself comes in different but consistent flavors). One might be worried that this observation is the consequence of our SBI approach being biased or insensitive to a large population of relatively dim sources beyond the detection threshold of the LAT. 

\begin{figure}[h]
    \includegraphics[width=\columnwidth]{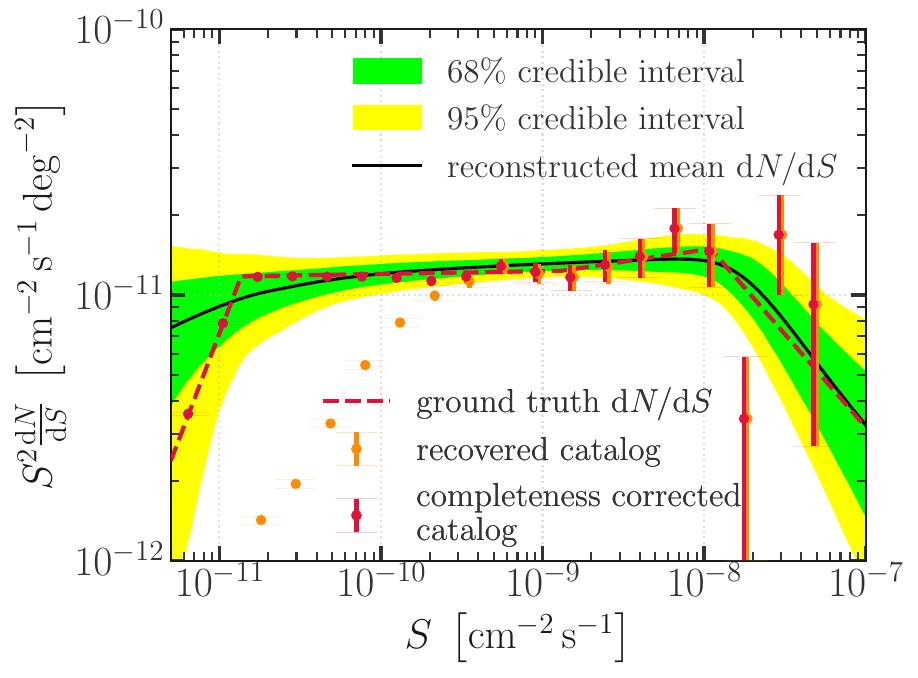}\\
    \includegraphics[width=\columnwidth]{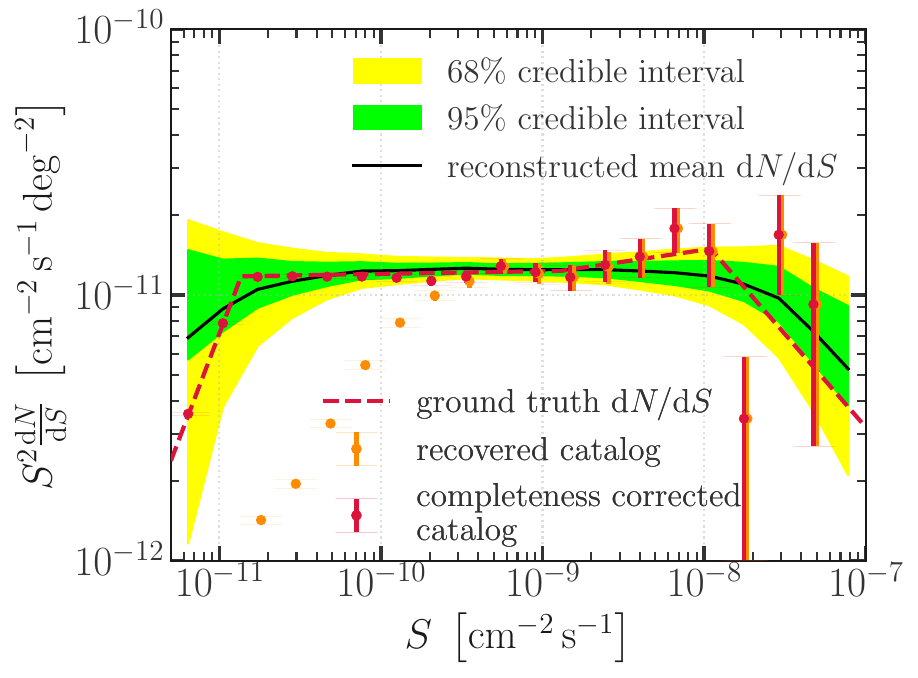}
    \caption{\textbf{Top:} Same as the left panel of Fig.~\ref{fig:dNdS-woGRF} regarding a simulated target dataset featuring a TBPL \dnds~with the best-fit parameters of the profile-likelihood 1p-PDF \cite{Zechlin:2015wdz} (dashed red line labeled as `ground truth'). The background model features all gamma-ray emission components at a normalization of 1 without additional modulations due to GRFs. \textbf{Bottom:} Same as the upper panel regarding the non-parametric \dnds~inference approach without GRFs. In both panels, we show in orange the catalog of detected point-like sources obtained with the detection network discussed in Sec.~\ref{sec:results_det}. The red points represent the completeness corrected catalog obtained by rescaling the detected sources by the false negative rate derived from simulations.
    }
    \label{fig:dNdS-1pPDF-woGRF}
\end{figure}

We can directly test this hypothesis on simulated data that follows a TBPL with the best-fit parameters of the 1p-PDF publication \cite{Zechlin:2015wdz}. We prepare, as a test scenario, a simulation with the profile-likelihood TBPL best-fit \dnds~parameters and all background template normalizations set to 1 without including GRFs. In the first sanity check, we apply the detection network (trained on simulations without GRFs) on the generated target sky map and evaluate the obtained true positive point-like sources. This catalog is shown as orange points in Fig.~\ref{fig:dNdS-1pPDF-woGRF}. Based on the intuition we get from applying the detection network on simulated data of the same underlying \dnds~(1p-PDF best fit, see the right panel of Fig.~\ref{fig:roc}), we can also assess the false negative rate of the network and correct the derived catalog accordingly. The result is displayed with red points in the same figure. Similar to the performance on real data, the source detection has an efficiency of almost 100\% above a flux of $S=3\times10^{-10}\;\mathrm{cm}^{-2}\,\mathrm{s}^{-1}$. Around this value, the credible bands of the inferred \dnds~(see next paragraph) reach the narrowest width while they open up towards lower fluxes. Since the catalog detection loses efficiency for dimmer sources, this behavior is expected. It also implies that the direct \dnds~reconstruction has access to information from the dim-flux regime where the detection network does not.

Secondly, we apply the trained and amortized parametric and non-parametric inference networks on this dataset to recover the injected parameters. The obtained mean \dnds~with 68\% and 95\% credible intervals are shown in Fig.~\ref{fig:dNdS-1pPDF-woGRF}; the upper panel illustrates the findings with the parametric while the lower panel is dedicated to the non-parametrically obtained findings. We recover the injected TBPL satisfyingly well, which encompasses, in particular, the dim flux range below the third break at around $10^{-11}$ cm$^{-2}$s$^{-1}$. We also made sure that this statement holds for different realizations of point-like source populations drawn from the same underlying \dnds. Therefore, the obtained result suggests that our source-count distributions for the real high-latitude gamma-ray sky are faithful representations of the source population. In addition, and as we will demonstrate in the following appendix sections (cf.~App.~\ref{app:svdd} and App.~\ref{app:mismodeling-GDE}), the impact of background mis-modeling on our SBI sensitivity and performance seems not to be a critical factor that biases the findings either.

\begin{figure*}[t]
    \includegraphics[width=0.95\textwidth]{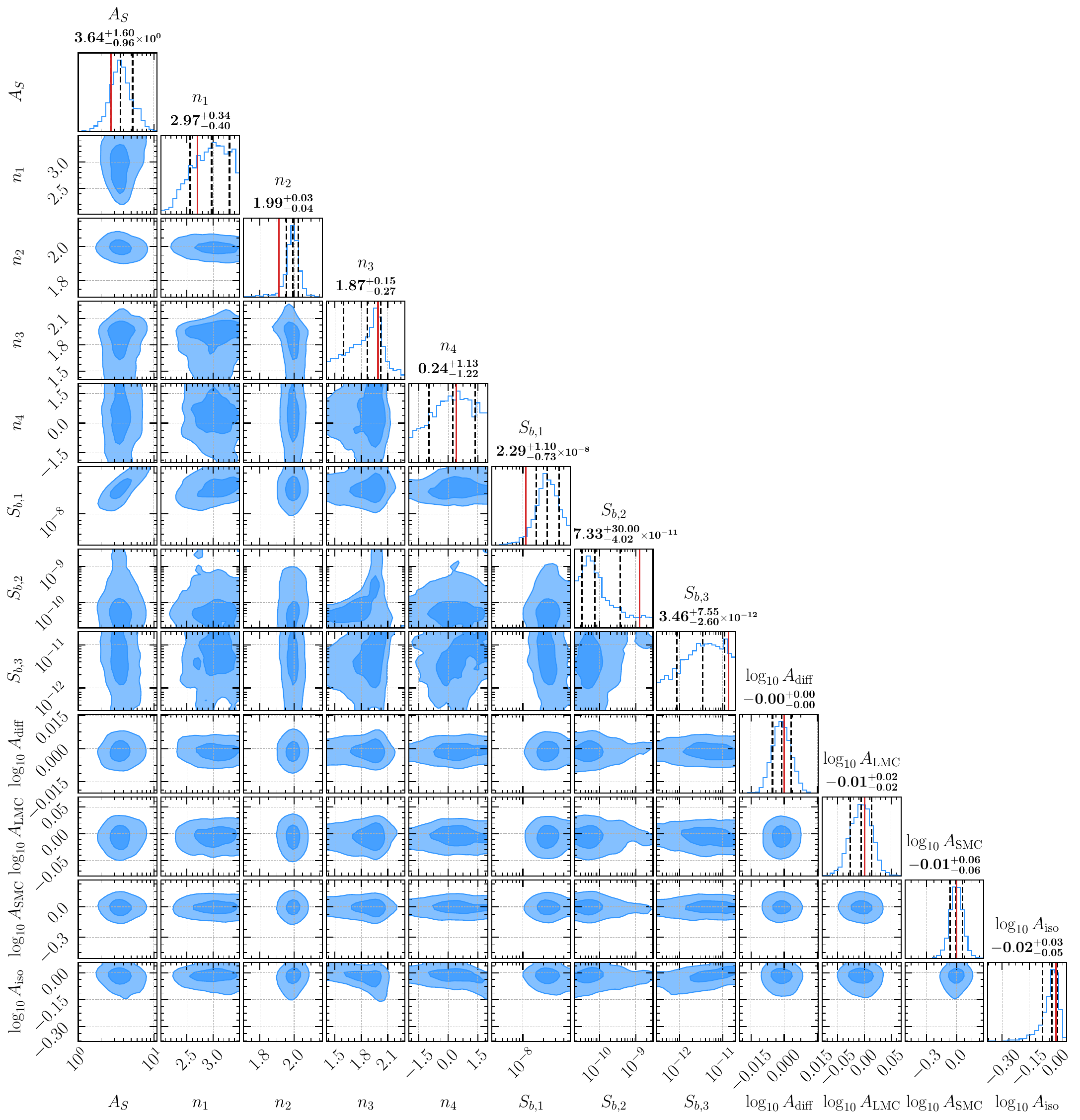}
    \caption{Same as Fig.~\ref{fig:2Dmarginal-LAT-combined} regarding a simulated target dataset featuring a TBPL \dnds~with the best-fit parameters of the profile-likelihood 1p-PDF \cite{Zechlin:2015wdz} and the astrophysical background normalizations set to 1. Here, the vertical dashed red lines mark the true injected values.}
    \label{fig:2Dmarginal-1pPDF-woGRF}
\end{figure*}

As for the real LAT data, we display in Fig.~\ref{fig:2Dmarginal-1pPDF-woGRF} the two-dimensional marginal posterior distributions regarding this simulated target dataset. The ground truth for each parameter is denoted with a vertical dashed red line inside the one-dimensional marginal posterior panels on the diagonal. The ground truth is within the band defined by the vertical dashed blue lines, the 68\% credible intervals, for eight cases. This is fully in line with the expected coverage of this credible band, which we more closely investigate in App.~\ref{app:coverage}. Besides, we observe also here that all marginal posterior distributions are well-defined with varying predictiveness; in particular, as concerns $n_4$ and $S_{b,3}$. In general, the correlations and degeneracies we observed and discussed in the case of the real LAT data are also present in this simulated dataset. 

\subsection{Performance on simulated sky maps}
\label{app:simulation-woGRF}

In the previous section, we focused on a single \dnds~parameter realization and explored the performance of our parametric and non-parametric SBI approach to recover these injected values. In this section, we showcase the obtained inference results for the large variety of \dnds~profiles allowed by the prior distributions defined in Tab.~\ref{tab:params}. Besides the source-count distribution, the quality of the recovery of the astrophysical background parameters is also an interesting aspect of the inference. What we would need to demonstrate is that the Bayesian $X\%$ credible intervals displayed in, e.g., Fig.~\ref{fig:bkg-woGRF-wGRF} contain the ground truth in $X\%$ of all tested simulated sky maps. In other words, we need to verify that the obtained posteriors for the background (and likewise \dnds~parameters) are well-calibrated. The question of calibration is explicitly answered in App.~\ref{app:coverage} so that we postpone an inspection of the quality of the background parameter recovery and the posterior calibration in general to this section.

\begin{figure*}[t]
    \includegraphics[width=0.95\textwidth]{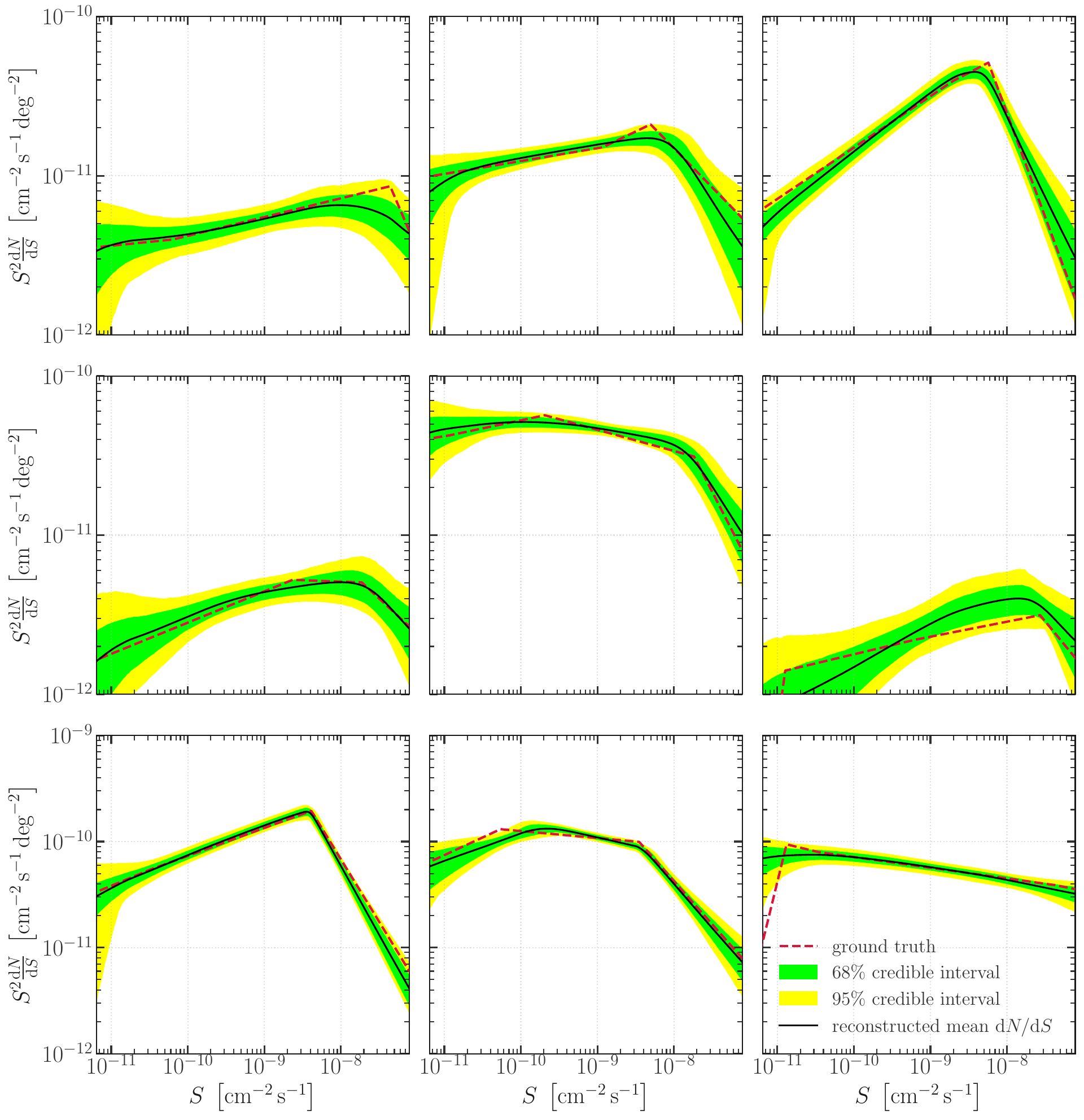}
    \caption{Examples of \textbf{parametric} inference results on test simulated data generated from the prior distributions listed in Tab.~\ref{tab:params} without GRFs. The applied neural network is the same that was used to generate the right panel of Fig.~\ref{fig:dNdS-woGRF}, i.e.~a network trained on TBPL \dnds~(without GRFs). The ground truth for each inferred source-count distribution is shown as a dashed red line, while the mean reconstructed \dnds~is denoted by a solid black line. The green and yellow bands illustrate the 68\% and 95\% credible intervals of the inferred profile, respectively.}
    \label{fig:9sim-data-parametric}
\end{figure*}

\begin{figure*}[t]
    \includegraphics[width=0.95\textwidth]{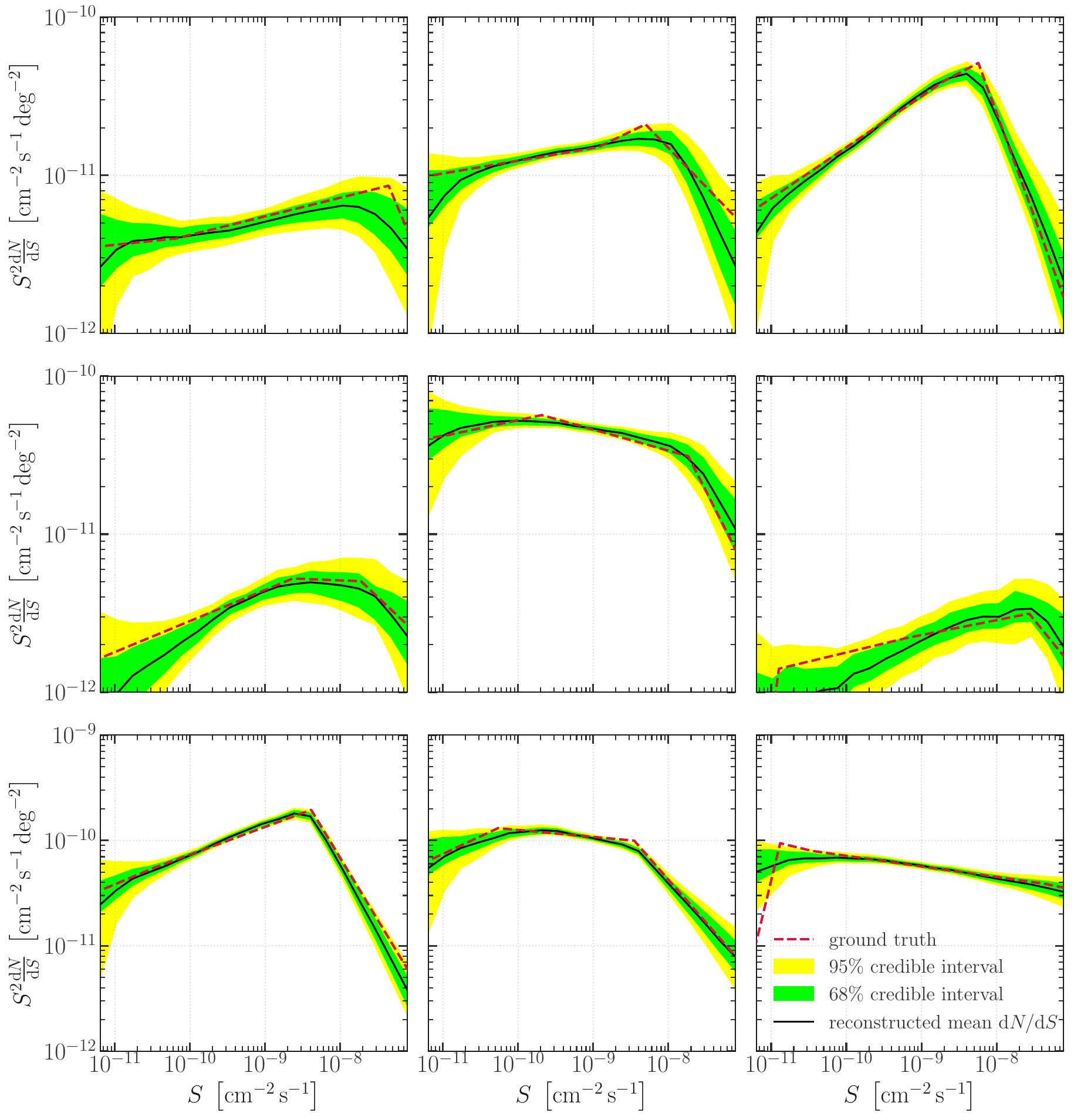}
    \caption{Same as Fig.~\ref{fig:9sim-data-parametric} but using the \textbf{non-parametric} \dnds~inference approach.}
    \label{fig:9sim-data-nonparametric}
\end{figure*}

We generated nine test gamma-ray sky maps in the simulator setup without GRFs and employed the trained networks for parametric and non-parametric inference to them. The obtained recovered mean source-count distribution profiles with 68\% and 95\% credible intervals are shown in Fig.~\ref{fig:9sim-data-parametric} (parametric) and Fig.~\ref{fig:9sim-data-nonparametric} (non-parametric). There are a couple of common features among all panels that are worth noting:
\begin{itemize}[itemindent=0.1in, leftmargin=0.2in]
    \item The credible bands of each \dnds~profile open towards the bright-source regime. This observation is caused by shot noise regarding the number of realized bright point-like sources in the map. Intuitively, when there are only a few, it is harder to accurately determine break position $S_{b,1}$ and slope $n_1$ in this flux range.
    \item The credible bands of each \dnds~profile also open towards the dim-source regime. While the sky maps might feature a large number of such dim point-like sources, they cannot be resolved individually so that their gamma-ray emission successively blends in with the diffuse isotropic background component. This generates the mild correlation we mentioned in the two-dimensional marginals plots in Figs.~\ref{fig:2Dmarginal-LAT-combined} and~\ref{fig:2Dmarginal-1pPDF-woGRF}. In addition, our SBI approach (but also traditional methods like the 1p-PDF \cite{Zechlin:2015wdz} and other neural-network-based analysis \cite{Amerio:2023uet}) loses sensitivity in this flux range, causing a larger uncertainty. 
    \item The larger the \dnds~amplitude $A_S$, the narrower the inferred credible intervals. Especially those source-count distribution realizations where the maximum is below $S^2\mathrm{d}N/\mathrm{d}S \sim10^{-11}\;\mathrm{cm}^{-2}\,\mathrm{s}^{-1}\,\mathrm{deg}^{-2}$ exhibit large uncertainties due to the rather small number of bright and intermediate point-like sources. 
    \item The ground truth is reconstructed within the 95\% credible interval in most of the cases. The bottom right panel seems to be a scenario in which both methods do not capture the pronounced drop in the dim-source regime. Here, it is most likely the loss of sensitivity to this regime that causes the non-optimal inference result. However, using more training data can easily improve the sensitivity of the method.
\end{itemize}
Based on these examples, the quality of the inferred source-count distributions and the characterization of the associated uncertainties are not very different between the parametric and non-parametric approaches. They never fail dramatically -- we checked this statement for many more examples than are shown here -- as the ground truth is reliably recovered within the obtained credible bands.

\section{Calibration and goodness of fit}
\label{app:gof}

In App.~\ref{app:simulations}, we examined the performance of our point-like source detection and parameter inference framework on simulated data on
specific examples. We demonstrated that the devised algorithms are capable of retrieving the injected values to a satisfactory degree. What is still missing is a more general assessment of the calibration of the obtained posterior distributions in the various scenarios we considered in the main text. In addition, we ought to verify that our gamma-ray simulator generates samples of gamma-ray emission at high latitudes compatible with the actual \textit{Fermi}-LAT sky. In this appendix section, we address both questions. In App.~\ref{app:coverage}, we study the calibration of our posteriors via Bayesian coverage plots, whereas App.~\ref{app:svdd} contains a quantitative analysis of the realism of our simulator through anomaly detection.

\subsection{Bayesian coverage of the trained networks}
\label{app:coverage}

In Bayesian statistics, it is possible to check the calibration of the method and of the posterior distributions it yields via so-called \textit{coverage} tests. With coverage tests, we aim to verify that the obtained $\alpha\%$ credible intervals $\Theta_{\hat{p}(\boldsymbol{\zeta}|\boldsymbol{d})}(\alpha)$ of the posteriors $\hat{p}(\boldsymbol{\zeta}|\boldsymbol{d})$~\footnote{In this appendix, we denote the estimated neural posterior with $\hat{p}(\boldsymbol{\zeta}|\boldsymbol{d})$ to differentiate it from the true target posterior ${p}(\boldsymbol{\zeta}|\boldsymbol{d})$.} cover the true value of the respective parameter $\alpha\%$ of the time. Here, $\boldsymbol{\zeta}$ refers to the set of parameters of our gamma-ray simulator, e.g.~ TBPL parameters $\vec{\theta}$, bins  $\vec{\theta}_S$, or background parameters or a combination of them depending on the performed analysis. More formally, we want to assess \cite{Cole:2021gwr}
\begin{equation}
\label{eq:coverage}
    \hat{\alpha} = \mathbb{E}_{p(\boldsymbol{{d}},\boldsymbol{\zeta)}}\!\left[\mathbb{I}\!\left(\boldsymbol{\zeta}\in\Theta_{\hat{p}(\boldsymbol{\zeta}|\boldsymbol{d})}(\alpha)\right)\right]\mathrm{,}
\end{equation}
where $\hat{\alpha}$ is the estimator for $\alpha$, $\mathbb{E}_{p(\boldsymbol{{d}},\boldsymbol{\zeta})}(\cdot)$ denotes the expectation value with respect to joint samples $(\boldsymbol{d}, \zeta)$ from the joint probability density $p(\boldsymbol{{d}},\boldsymbol{\zeta})$ and $\mathbb{I}$ refers to the indicator function introduced in Sec.~\ref{sec:method}. In the case of a perfectly calibrated posterior, $\hat{p}(\boldsymbol{\zeta}|\boldsymbol{d}) = p(\boldsymbol{\zeta}|\boldsymbol{d})$, we have that $\hat{\alpha}(\alpha) = \alpha$. Therefore, tracking how $\hat{\alpha}$ deviates from this relation as a function of $\alpha$ given the obtained posteriors, allows us to quantify how close these posteriors are to perfect calibration. We note, however, that perfect calibration does not imply optimality of the posteriors in terms of precision. 

Evaluating the coverage for likelihood-based Bayesian inference is computationally intensive because it requires repeated sampling and likelihood evaluations for many joint draws $\boldsymbol{d},\boldsymbol{\zeta}\sim p(\boldsymbol{d},\boldsymbol{\zeta})=p(\boldsymbol{d}|\boldsymbol{\zeta})p(\boldsymbol{\zeta})$. In contrast, SBI frameworks yield amortized posteriors, allowing one to sample from $\hat{p}(\boldsymbol{\zeta}|\boldsymbol{d})$ for any data $\boldsymbol{d}$ without retraining. This makes it feasible to compute posteriors for numerous mock data samples rapidly, rendering the coverage test a practical means of assessing calibration. The \texttt{swyft} package provides a routine to conduct such tests for all simulator parameters.

As discussed in the main text, we aim for the best possible calibration of our posteriors by selecting the trained network state -- per inference task -- that provides results closest to perfect calibration. Although early-stopping and learning rate schedules are based on the validation loss, the coverage test is conducted \textit{a posteriori} using 1000 samples generated by the simulator. This technical choice makes calibration our primary stopping criterion during training. In all cases examined, the best calibration did not coincide with the network state yielding the minimal validation loss. Consequently, it is possible that our training was terminated before achieving the optimal calibration. The coverage tests presented below represent the best approximation to optimal calibration achievable within our framework.

\begin{figure*}
    \includegraphics[width=0.95\linewidth]{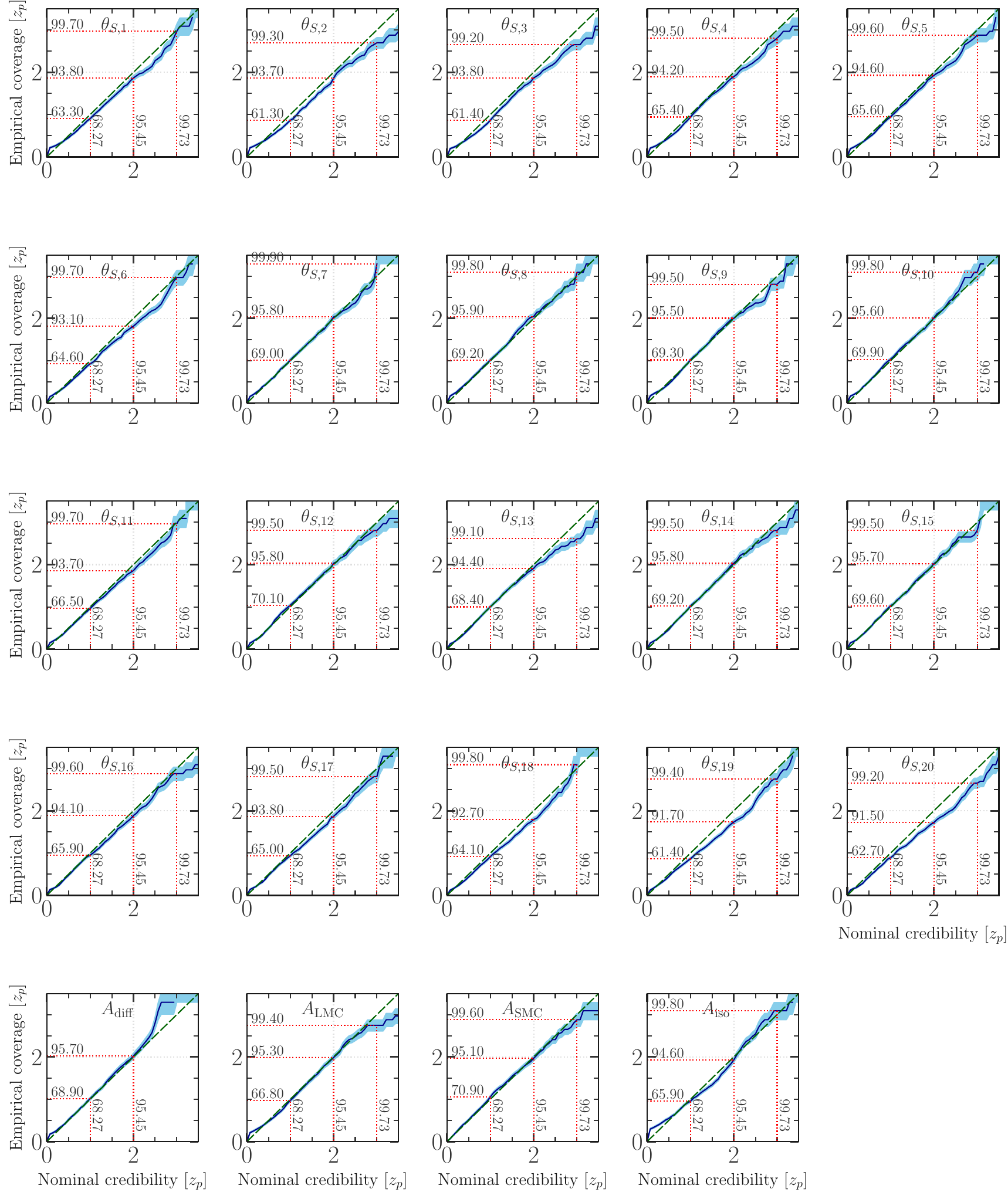}
    \caption{\texttt{swyft} coverage results for the network state with the smallest deviation from perfect calibration based on the non-parametric description of the source-count distribution without GRFs. The first four rows contain the results for each component of the flux bin parameters $\vec{\theta}_S$. The last row reports the coverage results for the astrophysical background parameters. The horizontal axis shows the nominal credibility interval (in significance), whilst the vertical axis shows the empirically determined coverage. The dashed green line indicates perfect coverage while the solid blue line is the obtained average coverage; the light-blue band denotes the 68\% containment band derived from the 1000 samples generated by our gamma-ray simulator.}
    \label{fig:coverage-woGRF-nonparametric}
\end{figure*}

\vspace{10pt} 
\noindent\textbf{Calibration of posteriors for non-parametric \dnds~without GRFs.} In Fig.~\ref{fig:coverage-woGRF-nonparametric}, we show that the posterior distributions for the non-parametric description of the \dnds~without including GRFs are satisfyingly close to an optimal calibration. As a reminder, these posteriors are based on one-dimensional log-ratio estimators not incorporating correlations between parameters. Moreover, the \dnds~parameters are not the parameters of the TBPL itself, but they rather yield the source-count distribution's value in 20 logarithmically spaced flux bins $\vec{\theta}_S$. The recovered source-count distribution of the real LAT sky regarding this network state is shown in the right panel of Fig.~\ref{fig:dNdS-woGRF}.

The horizontal axis in these panels reports the nominal credibility $\alpha$ in terms of significance ($\alpha=68.27\%$ corresponds to 1, $\alpha=95.45\%$ corresponds to 2, etc.), while the vertical axis shows the empirical coverage $\hat{\alpha}$ (solid blue line) estimated from the 1000 samples (light blue band, 68\% containment interval). Perfect calibration is referenced as a diagonal dashed green line. When the empirical coverage lies above the green line, the posteriors are considered conservative, meaning that they are wider than could optimally be. In contrast, an empirical coverage below this diagonal signifies an over-confident posterior which is too narrow. 

In detail, we find that the calibration of the astrophysical background parameters is very close to perfect calibration. In the case of the SMC, however, this is expected because the posterior is simply the prior as reported in Fig.~\ref{fig:bkg-woGRF-wGRF}. As the network was not able to extract enough information from the training dataset to constrain the SMC's norm, predicting the prior as the estimated posterior must result in optimal calibration. The \dnds~paramerers $\vec{\theta}_S$ are also satisfyingly well calibrated, with exceptions being the brightest flux bins in the fourth row (18, 19, 20; with $S\in\left[4\times10^{-8}, 10^{-7}\right]\;\mathrm{cm}^{-2}\,\mathrm{s}^{-1}$). They are slightly overconfident. 

\begin{figure*}
    \includegraphics[width=0.95\linewidth]{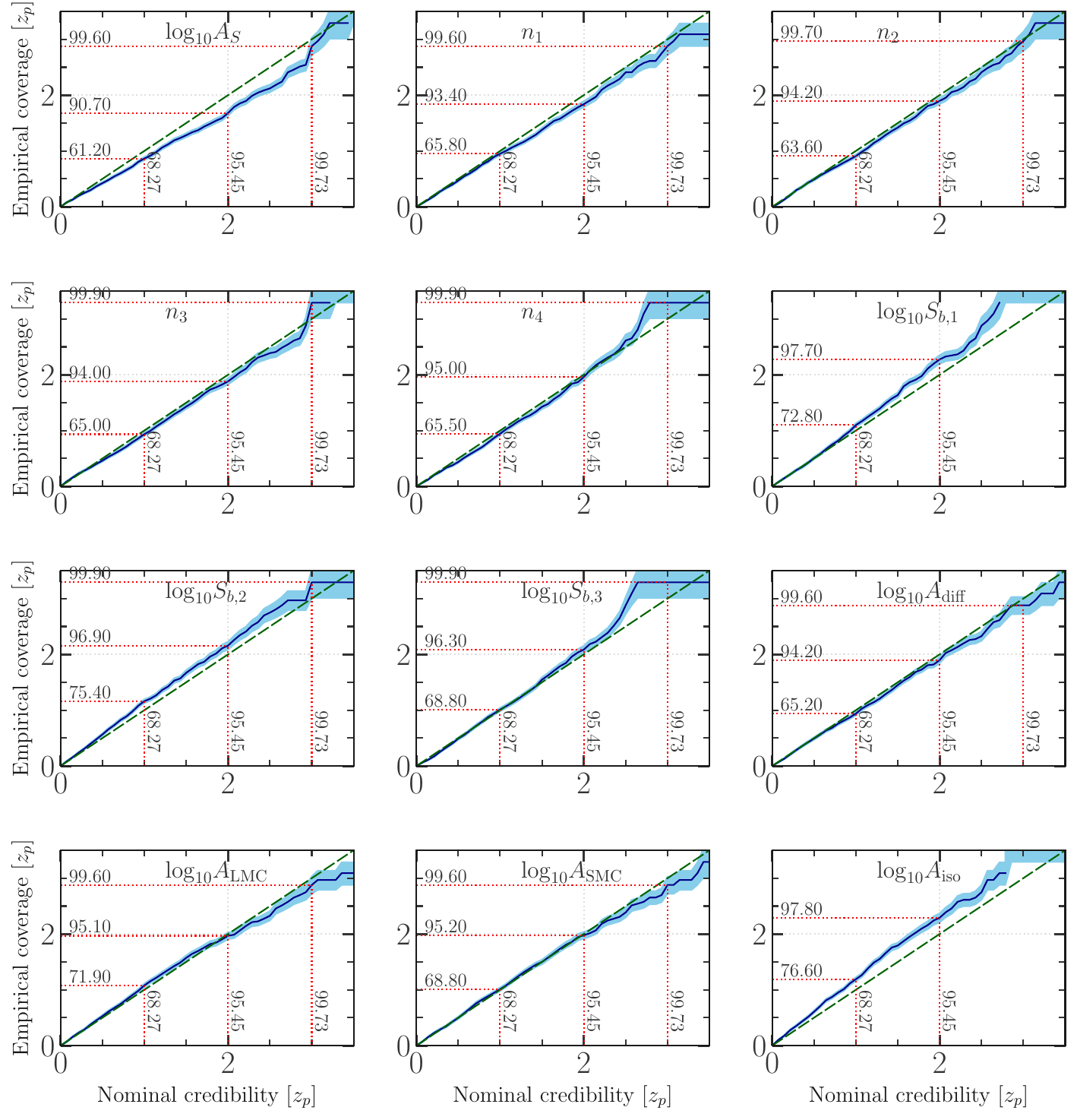}
    \caption{Same as Fig.~\ref{fig:coverage-woGRF-nonparametric} regarding the parametric \dnds~inference task using a simulator without GRFs.}
    \label{fig:coverage-woGRF-parametric}
\end{figure*}

\vspace{10pt}
\noindent\textbf{Calibration of posteriors for parametric \dnds~without GRFs.} In Sec.~\ref{sec:stat_parametric}, we explained how we derive the final one-dimensional marginal posterior distributions in the case of a parametric description of the source-count distribution (i.e., inferring the parameters $\vec{\theta}$). Employing autoregressive NRE, we obtain an estimate of the joint posterior distribution, from which we can sample with a nested sampling algorithm. However, how this NRE approach is practically implemented forbids us to directly access the coverage of the obtained joint posterior; we run into the same problem as likelihood-based techniques. As a proxy, we first perform a one-dimensional NRE like in the non-parametric case above. This first step yields the network state characterized by the best-fit coverage/calibration regarding the marginal one-dimensional posteriors of the parameters. Then, we apply the full autoregressive NRE to the same problem and check the projected marginal one-dimensional posteriors at each training epoch for the same synthetic test data sample. The projected one-dimensional posteriors that most closely match the well-calibrated posteriors of the first step are accepted as the final network state whose \dnds-predictions are shown in Figs.~\ref{fig:dNdS-woGRF} and~\ref{fig:bkg-woGRF-wGRF} with respect to the real LAT data. The coverage test findings regarding the one-dimensional NRE of the first step are given in Fig.~\ref{fig:coverage-woGRF-parametric}.

The summary of this coverage test implies that the parametric approach can achieve a satisfying degree of calibration. Some parameters like $A_{\mathrm{iso}}$ or $\log_{10}\!S_{b,2}$ are slightly conservative while the one-dimensional marginal posterior of $\log_{10}\!A_{S}$ tends to be slightly overconfident throughout the range of nominal credibility. For others, the 68\% credible interval is not exactly at its nominal value, whereas the full profile does align with the optimal expectations. In conclusion, neither the inferred parametric nor non-parametric \dnds~of the real LAT sky are strongly biased by poorly calibrated methods and, thus, reliably predict the uncertainties across the point-like-source flux range in the scenario without GRFs.

\begin{figure*}
    \includegraphics[width=0.9\linewidth]{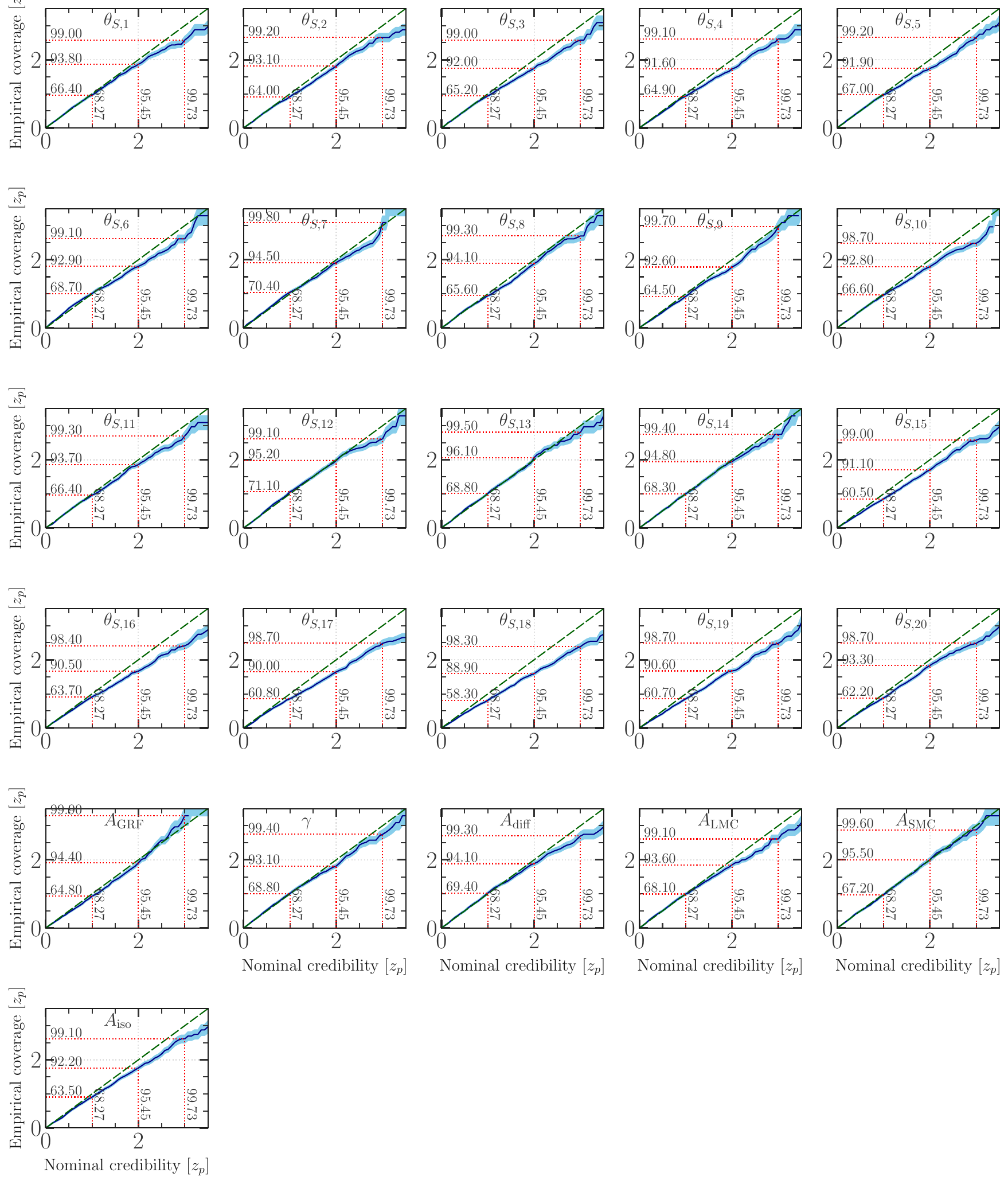}
    \caption{Same as Fig.~\ref{fig:coverage-woGRF-nonparametric} using a simulator with GRFs and inference of the corresponding parameters $A_{\mathrm{GRF}}$ and $\gamma$.}
    \label{fig:coverage-wGRF-nonparametric}
\end{figure*}

\vspace{10pt}
\noindent\textbf{Calibration of posteriors for non-parametric \dnds~with GRFs.} The achieved coverage in the case of the non-parametric \dnds~approach using a simulator with GRFs is shown in Fig.~\ref{fig:coverage-wGRF-nonparametric}. For this simulator setup, we also find the flux bins corresponding to the brightest part (and less pronounced at the dim flux end) of the point-like source population to produce slightly overconfident one-dimensional marginal posteriors. In contrast, the intermediate flux bin posteriors are well-calibrated. The background parameters, including the added GRF parameters $A_{\mathrm{GRF}}$ and $\gamma$ are also satisfyingly calibrated. Only the diffuse isotropic background posterior is somewhat overconfident. 

When comparing the inference results on real data regarding the non-parametric approach with and without GRFs in Figs.~\ref{fig:dNdS-woGRF} and \ref{fig:dNdS-wGRF}, there is no stark difference between the obtained 68\% and 95\% credible bands. They are compatible with each other. We, therefore, do not believe that the overconfidence of some of the flux bins is negatively impacting the result on LAT data. To improve the calibration, it is most likely already sufficient to increase the size of the training sample beyond the $2\times10^5$ maps currently used. The additional variance introduced by the GRFs may require a broader sample space to calibrate the posteriors optimally. However, since the addition of GRFs is meant to illustrate the capabilities of our SBI method and the real LAT data can already be described by the simulator without them (see the next App.~\ref{app:svdd}), we leave the improvement of the calibration for the next applications in future work. 

\begin{figure*}
    \includegraphics[width=0.95\linewidth]{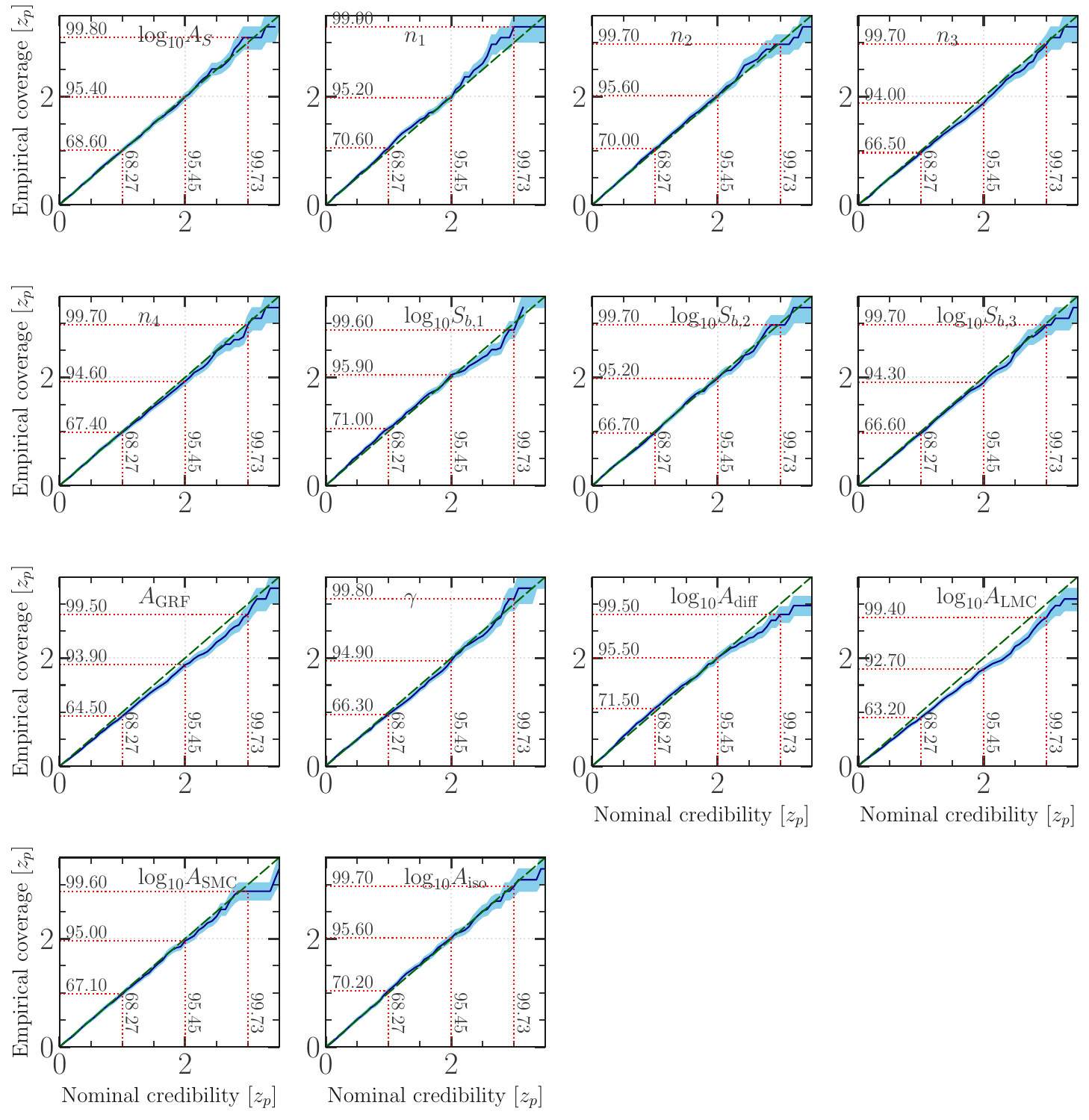}
    \caption{Same as Fig.~\ref{fig:coverage-woGRF-parametric} regarding the parametric \dnds~inference task using a simulator with GRFs.}
    \label{fig:coverage-wGRF-parametric}
\end{figure*}

\vspace{10pt}
\noindent\textbf{Calibration of posteriors for parametric \dnds~with GRFs.} Finally, we present in Fig.~\ref{fig:coverage-wGRF-parametric} the coverage test results regarding parametric \dnds inference with autoregressive NRE based on training samples with GRFs. As in the scenario without GRFs, these one-dimensional marginal posteriors are derived with an NRE scheme estimating only one-dimensional marginals. We checked that the shape of the posteriors fits the eventually utilized joint samples from the nested sampler of the autoregressive NRE approach. Again, these posteriors are satisfyingly well-calibrated with only minor overconfidence present in the panel of $A_{\mathrm{GRF}}$ and $A_{\mathrm{LMC}}$.

\subsection{Assessing the simulator's realism with anomaly detection}
\label{app:svdd}

We stressed in Sec.~\ref{sec:model} that our gamma-ray simulator was designed in the spirit of generating realistic samples of gamma-ray emission at high latitudes. While the results reported in Sec.~\ref{sec:results} align well with previous literature results -- hence, they do not cast striking doubts on the accuracy of the simulator -- we aim to quantitatively assess the realism of the generated gamma-ray maps regarding the selected \textit{Fermi}-LAT data. 

To this end, we exploit the ability of neural networks to detect anomalies in large datasets. In particular, we employ the so-called `One-Class Deep Support Vector Data Description' (Deep SVDD) \cite{pmlr-v80-ruff18a, Caron:2021wmq} method that was already utilized to characterize the reality gap in gamma-ray emission models of the MW's Galactic center region \cite{Caron:2022akb}. In simple terms, Deep SVDD maps high-dimensional data onto a predefined lower-dimensional manifold, and new data points that significantly deviate from this manifold are identified as anomalies. More precisely, the idea of the Deep SVDD method is to train a neural network that maps samples from the gamma-ray emission simulator onto a \textit{target manifold} with specific characteristics. The target manifold is defined as a $D$-dimensional vector $v^D_X\in\mathbb{R}^D$ with repeated value $X$ in all elements of the vector. Each input to the neural network is assigned an anomaly score based on the Euclidean distance between the network's output and the target manifold. As the network is trained to transform the full ensemble of training data into the target vector, samples resembling the training data should systematically produce a lower anomaly score than samples that are out-of-distribution, i.e., exhibiting features inconsistent with the simulator. To fully harness the features and correlations in the training samples during the compression step, we use an ensemble of Deep SVDD scores to create a final anomaly score. This ensemble consists of eight target manifolds with varying dimensionality $D\in\{5, 89, 144, 233\}$ and assumed constant values $X\in\{\pi,256.6\}$. The final score is the average of the eight individual SVDD scores regarding a specific input. 

Concerning the Deep SVDD neural network architecture $\mathcal{O}$, we adopt the layout employed in the context of $\mathrm{d}N/\mathrm{d}S$ parameter inference outlined in Sec.~\ref{sec:nn_architecture} and App.~\ref{app:NN-architectures}. Hence, the network features an encoder part implemented as a \texttt{DeepSphere} convolutional neural network, followed by a final fully connected linear layer that maps the \texttt{DeepSphere} output to the dimensionality of $v^D_X$. The loss function $\mathcal{L}$ during training is a simple square loss between the target manifold and the network output:
\begin{equation}
    \mathcal{L}(\boldsymbol{d}) = \left(v_X^D - \mathcal{O}(\boldsymbol{d})\right)^2\mathrm{.}
\end{equation}
For each combination of $D$ and $X$, we train the network for a maximum of 250 epochs on $100,000$ samples (80\%/20\%, training/validation), monitoring the achieved validation loss. The training is initiated with a learning rate of $10^{-3}$ and decreased by a factor of $0.1$ after eight consecutive epochs without improved validation loss. We set an early-stopping patience of eleven epochs. All other hyper-parameters not mentioned here explicitly coincide with the definitions in Tab.~\ref{tab:hyperparams} concerning the \dnds~inference.

\begin{figure*}
    \includegraphics[width=0.49\textwidth]{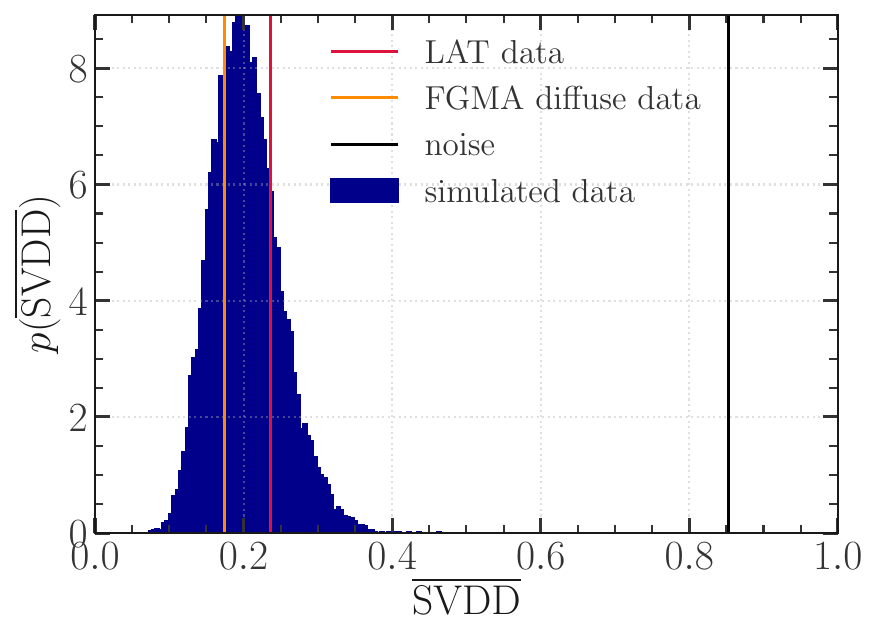}
    \includegraphics[width=0.49\textwidth]{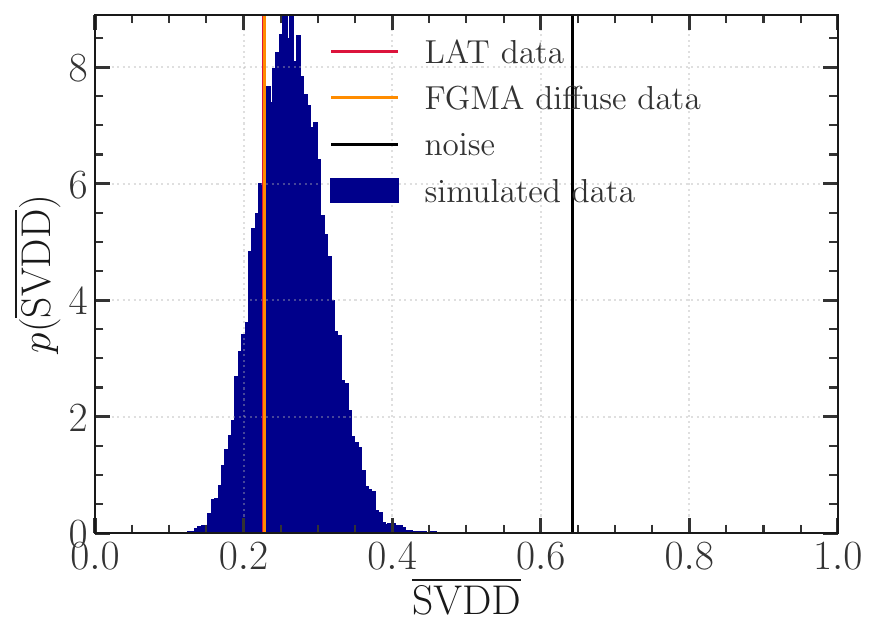}
    \caption{Averaged SVDD score prediction $\overline{\mathrm{SVDD}}$ of the distance in latent space between a target vector $v_{X}^D$ and validation data (blue, without (\textbf{Left}) and with (\textbf{Right}) GRFs to add variability to the MW's diffuse gamma-ray foreground), the real 14-year LAT data (red line), simulated data using an alternative diffuse MW foreground template (orange, also cf.~App.~\ref{app:mismodeling}) and an all-sky map filled only with a Poisson draw from exponentiated Gaussian noise (black line). The displayed scores were derived from averaging the results obtained for different target vector dimensions $D$ and target values $X$ as stated in the text. The adopted range of the SVDD score is a re-normalized distance in latent space to visually enhance the proximity of the real data to the validation dataset in relation to the noise.}
    \label{fig:svdd_plots}
\end{figure*}

\noindent\textbf{Anomaly detection results.} We apply the Deep SVDD approach for the eight combinations of $D$ and $X$ to gamma-ray emission maps generated from our simulator with and without GRFs. The resulting distributions of the averaged Deep SVDD anomaly scores $\overline{\mathrm{SVDD}}$ for the validation datasets -- w/o GRFs (\emph{left}) and with GRFs (\emph{right}) -- are shown as blue histograms in Fig.~\ref{fig:svdd_plots}. The vertical red line represents the average SVDD anomaly score of the 14-year \textit{Fermi}-LAT dataset, while the vertical orange line marks the same score for a simulated sample sky-map generated with an alternative diffuse MW foreground model the network has never seen before (see App.~\ref{app:mismodeling}). To show the adequacy of the Deep SVDD method to reveal anomalous samples, we prepare a `noise' gamma-ray sample whose score is shown as a vertical black line. To generate the noise map, we draw a random number from a unit normal distribution for each pixel, exponentiate the resulting sky map, and finally draw a Poisson realization from it. To enhance the visual accessibility of the scores, we re-normalize each individual Deep SVDD scenario to yield scores between 0 and 1. The average $\overline{\mathrm{SVDD}}$ is based on these re-normalized values.

In the case w/o GRFs, we find that the real data is well within the variation expected from the simulator output, while the noise map is clearly an outlier that cannot be generated from the simulator. Interestingly, the sample featuring an alternative MW diffuse foreground model performs equally well as the real data. The latter is also true for most of the individual Deep SVDD cases; two of them indicate that the alternative model is an outlier located in the tails of what the simulator can generate. These observations suggest that the Deep SVDD approach is sensitive to features in the training sample that differentiate them from noise and that our simulator, even w/o GRFs, encompasses the real gamma-ray sky. This is perhaps not too surprising given the fact that the default Galactic diffuse gamma-ray emission model contains re-injected fit residuals from fit to LAT data. Yet, the impact of the diffuse components seems to be rather sub-dominant in light of the performance of the alternative diffuse model. The most relevant features are most likely tied to the locations and brightness of discrete point-like sources and the correct simulation of the LAT IRFs (the noise map does not respect the LAT PSF profile). 

Given these results on simulations without GRFs, it is not astonishing that a very similar picture emerges from training the SVDD networks on samples including GRFs as shown in the right panel of Fig.~\ref{fig:svdd_plots}. Both real LAT dataset and simulated data containing FGMA instead of the benchmark Galactic diffuse model are well within the model space spanned by the simulator. Only the noise sky map shifts to a position closer to the model space -- while still being out-of-distribution. This observation can be understood from the generation of the noise map. It is a Poisson draw from an exponentiated Gaussian noise map. Hence, there can be fringe cases in the training data where very few point-like sources are present and the diffuse MW foreground dominates that is modulated by a GRF with $\gamma\sim1.5$ and large amplitude $A_{\mathrm{GRF}}$. As this parameter configuration enhances small-scale fluctuations that resemble the noise map, it can conceivably lead to a shorter distance between the noise realization and the model space of the simulator. 

We conclude that there is no pronounced reality gap between our simulator and the real gamma-ray sky neither in its version w/o nor with GRFs. Therefore, our inference results in the main text are likely unaffected by severe mis-modeling biases. 

\section{Effect of mis-modeling}
\label{app:mismodeling}

In this section, we focus on the question of how well our inference framework can accommodate a certain degree of mis-modeling. We will evaluate and quantify the performance of our SBI approach on synthetic data that purposefully contain mis-modeled features. We target the component of the Galactic diffuse emission and its impact on the inference results in App.~\ref{app:mismodeling-GDE} and investigate in App.~\ref{app:mismodeling-dnds} how the parametrization of the source-count distribution influences the posterior distributions.

\subsection{Alternative description of the Galactic diffuse emission}
\label{app:mismodeling-GDE}


As stressed in different places, the Galactic diffuse emission originates in high-energy charged cosmic rays colliding with the interstellar medium, thus tracing the structure and distribution of gas and ISRFs. The uncertainties on these quantities will propagate into the predictions of the expected gamma-ray emission. Phenomenologically, we employed GRFs to account for these uncertainties on small and large scales. Now, we address whether this addition to our gamma-ray simulator allows our inference framework to generalize to different characterizations of the Galactic diffuse emission robustly. Concretely, we probe how well the SBI framework can recover the \dnds~parameters of the same point-like source map $\boldsymbol{c}^{\ast}$ component of synthetic data that additionally contains an unknown diffuse MW foreground model when it is trained on simulations with and w/o GRFs.

We choose the so-called foreground model A (FGMA) employed in \cite{Fermi-LAT:2014ryh}\footnote{The respective templates can be retrieved from the \textit{Fermi}-LAT collaboration's public data archive: \url{https://www-glast.stanford.edu/pub_data/845/}.} as an alternative description of Galactic diffuse emission. The authors of \cite{Fermi-LAT:2014ryh} exploited this model (and two similar variants) to carefully derive and analyze the IGRB and its associated statistical and systematic uncertainties. Therefore, FGMA is tuned towards the high-latitude gamma-ray sky featuring hadronic ($\pi^0$-decay, bremsstrahlung) and leptonic (IC scattering off of multiple target radiation fields) gamma-ray emission components.

\begin{figure}
    \centering
    \includegraphics[width=\columnwidth]{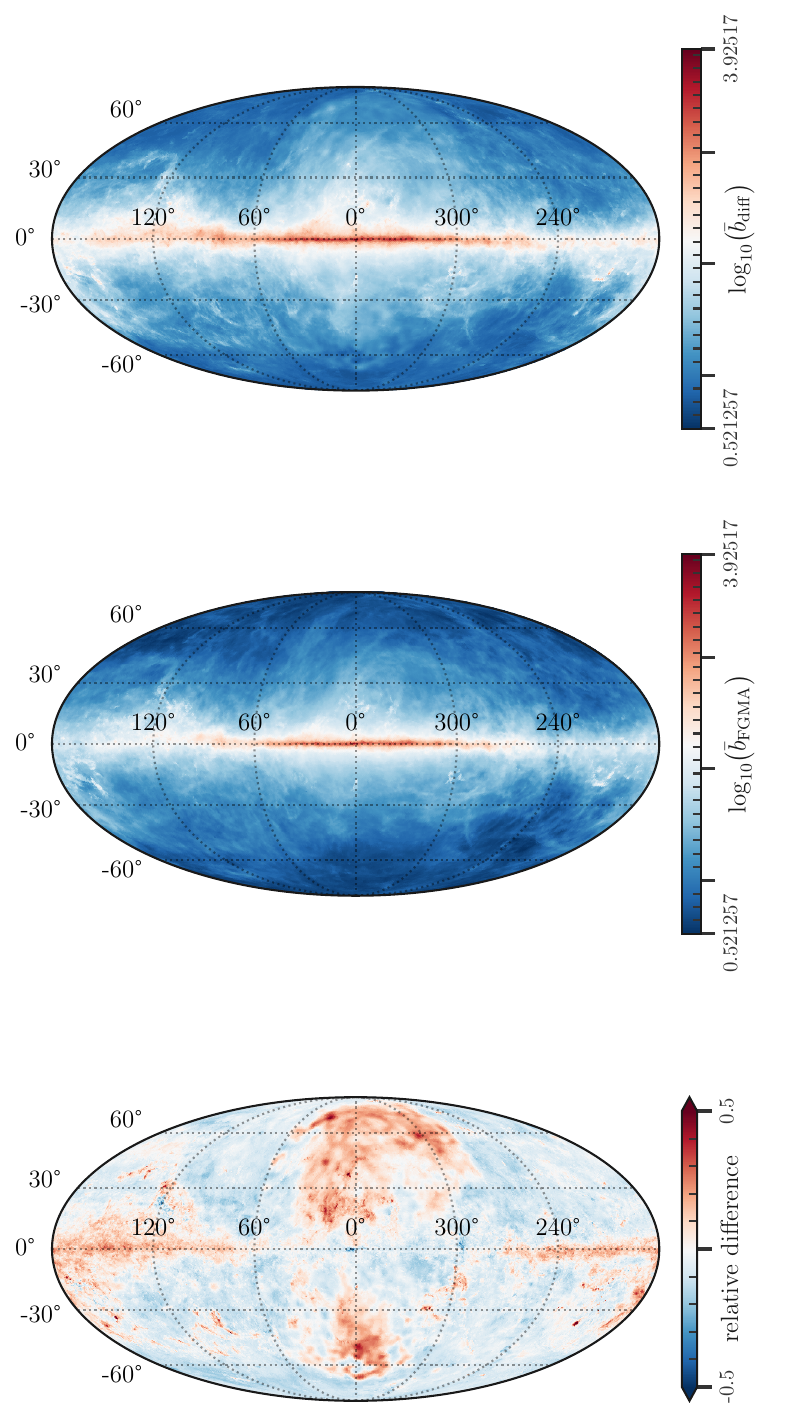}
    \caption{Summary of the counts maps of the Galactic diffuse emission models employed to investigate the impact of mis-modeling. The top panel contains the benchmark Galactic diffuse model utilized in the main body of this study (see Sec.~\ref{sec:de_fm}). The middle panel instead shows the alternative diffuse MW foreground model called FGMA adopted from \protect\cite{Fermi-LAT:2014ryh}. For visualization purposes, both panels share the same color scale range. The bottom panel highlights the relative differences of both models quoting the quantity $(\bar{b}_{\mathrm{diff}} -\bar{b}_{\mathrm{FGMA}})/(\bar{b}_{\mathrm{diff}} +\bar{b}_{\mathrm{FGMA}})$.}
    \label{fig:diff-models}
\end{figure}

It is important to note that we can expect small-scale and large-scale differences between the benchmark diffuse model and FGMA. We show both models in terms of photon counts (but before PSF convolution) in Fig.~\ref{fig:diff-models} as well as their relative differences $(\bar{b}_{\mathrm{diff}} -\bar{b}_{\mathrm{FGMA}})/(\bar{b}_{\mathrm{diff}} +\bar{b}_{\mathrm{FGMA}})$. Small-scale variations predominantly occur because the 4FGL background model is based on updated gas maps of atomic hydrogen (H \textsc{i}) adopted from the HI4PI all-sky survey \cite{2016A&A...594A.116H} while FGMA relies on outdated observations of the Leiden-Argentine-Bonn all-sky survey \cite{Kalberla:2005ts} with a lower angular resolution. Large-scale differences will also be present since FGMA does not encompass any description of the FBs or Loop I, which are present in the 4FGL background model via re-injected fit residuals. These large-scale variations are easy to recognize in the bottom panel of Fig.~\ref{fig:diff-models}. The gas map differences have their impact mainly along the Galactic plane, but a fraction of them also exists at high latitudes. Yet, there are no extreme differences between default and the alternative MW foreground model (cf.~also App.~\ref{app:svdd}).

\vspace{10pt}
\noindent\textbf{Impact on the \dnds~parameter inference.} To probe the resilience of the parametric autoregressive NRE approach against mis-modeling of the diffuse MW foreground, we work on simulated target data that is always comprised of the same point-source map realization. To this end, we choose the TBPL \dnds~with the best-fit parameters of the profile-likelihood 1p-PDF \cite{Zechlin:2015wdz} that we also utilized in App.~\ref{app:simulation-1pPDF-parametric}. In contrast, the background map $\boldsymbol{b}$ is varied in order to generate two different target datasets $\boldsymbol{d}$ with and without foreground mis-modeling, i.e., one background map features the default Galactic diffuse emission model while the other adopts FGMA. In both cases, the MW foreground component is injected with $A_{\mathrm{diff}} = 1$ as are the remaining background components (which are the same in both maps).   

\begin{figure}
    \centering
    \includegraphics[width=\columnwidth]{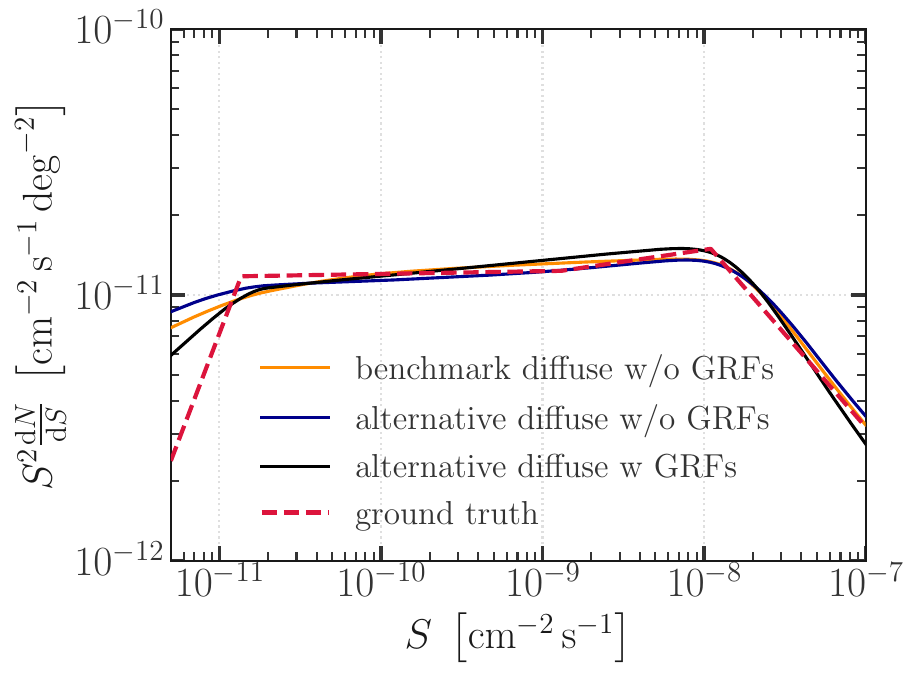}
    \caption{Comparison of inferred mean source-count distributions within the parametric autoregressive NRE approach regarding a point-like source map $\boldsymbol{c}$ following the best-fit TBPL parameters of \cite{Zechlin:2015wdz} (dashed red line; see also Fig.~\ref{fig:dNdS-1pPDF-woGRF}). The different colors denote different background maps $\boldsymbol{b}$ added to the point-source map to generate the full target data map. The orange line illustrates the benchmark case where the background is comprised of the \textit{Fermi} diffuse background model and the remaining components without GRFs. The blue line displays the results when the diffuse MW foreground is modeled via FGMA. The inference results are thus obtained for the network trained on simulated sky-maps without GRFs. Lastly, the black line states the obtained \dnds~for the latter case but uses the network trained on simulations with GRFs.}
    \label{fig:inference-diff-confusion-dnds}
\end{figure}

Then, we employ the trained parametric autoregressive NREs -- without GRFs -- on these target datasets to explore the effect that mis-modeling has on the achieved parameter reconstruction. Furthermore, we want to check to what extent the network trained with data from the simulator with GRFs succeeds in eliminating the difference between the two MW foreground models. Therefore, we also apply this network to the target dataset containing FGMA as the Galactic diffuse emission component. The obtained mean source-count distributions are compared in Fig.~\ref{fig:inference-diff-confusion-dnds}. The ground truth (dashed red profile) is well-recovered in all three scenarios: The orange line refers to the case that is also shown in Fig.~\ref{fig:dNdS-1pPDF-woGRF} while the blue line denotes the inferred source-count distribution when the background map contains FGMA. Although they are not displayed in this figure, the 68\% credible intervals of both results overlap. Lastly, adding in black the profile obtained with the network trained on simulated data with GRFs, yields a comparable reconstruction of the injected \dnds. We should not be particularly surprised that we have obtained this degree of consistency in the results. As we have shown with the anomaly tests of App.~\ref{app:svdd}, FGMA and the default diffuse MW foreground model are not so different that the sample space spanned by simulator without GRFs would not cover them. In that sense, our results -- even with respect to the real data -- should not suffer much from a certain degree of mis-modeling in the astrophysical backgrounds.   

\begin{figure}
    \centering
    \includegraphics[width=0.9\columnwidth]{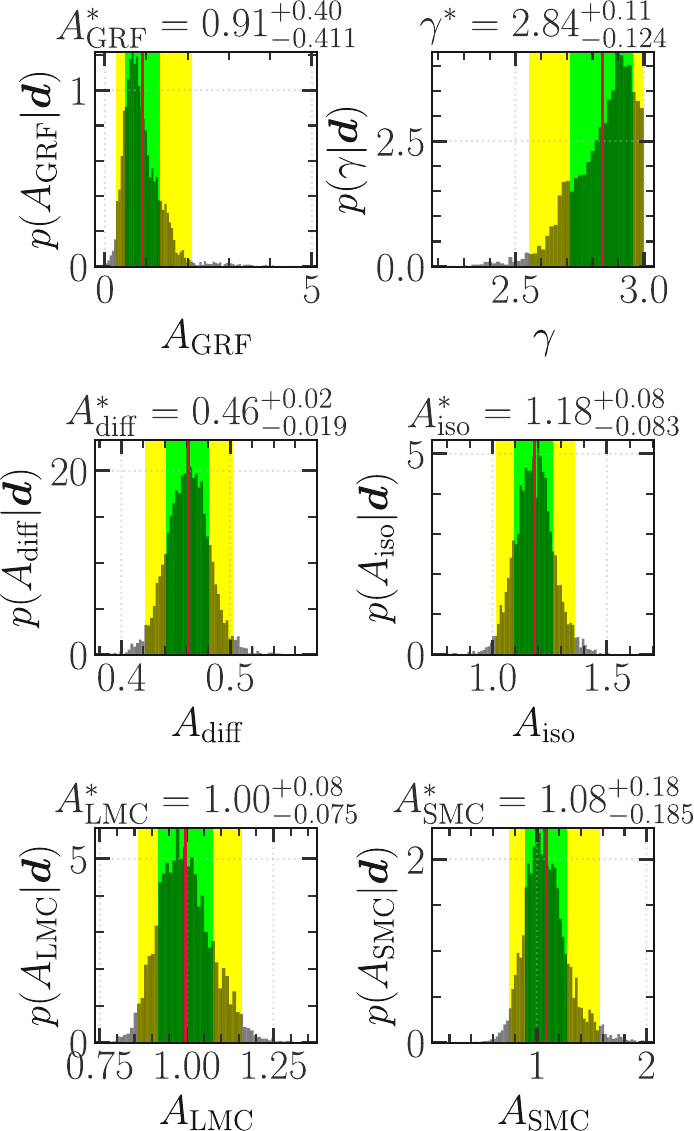}\\
    \includegraphics[width=0.9\columnwidth]{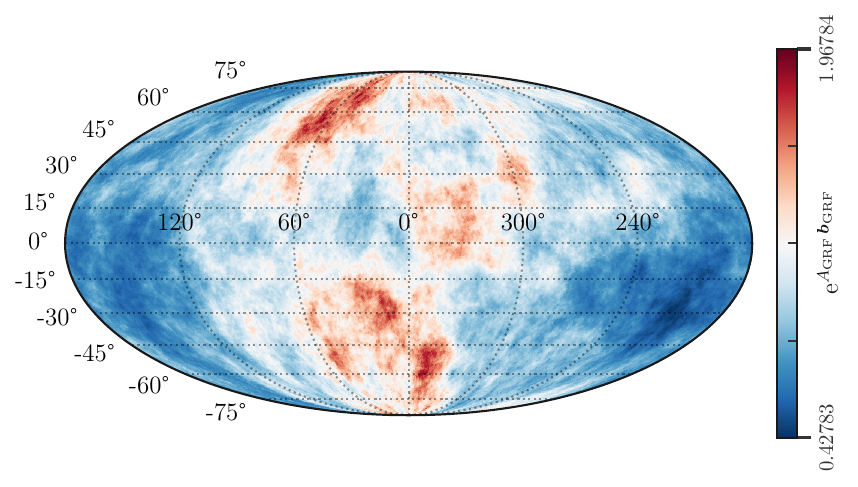}
    \caption{\textbf{Top:} Inferred posterior probability density functions for the six background components of our gamma-ray emission model with GRFs (see Tab.~\ref{tab:params}) obtained from simulated data featuring a \dnds~following the best-fit TBPL parameters of \cite{Zechlin:2015wdz} and FGMA as the diffuse MW foreground component injected with a normalization scale of 1. The neural network was trained on samples assuming our default Galactic diffuse emission model. We display as histograms the marginal one-dimensional posteriors generated by the nested sampler applied in the parametric $\mathrm{d}N/\mathrm{d}S$ approach. A vertical red line denotes the mean of these posteriors, while 68\% and 95\% credible intervals are overlaid in green and yellow, respectively. The panels' titles state the numerical values of the mean and 68\% credible region. \textbf{Bottom:} Random realization of the exponentiated GRF using the best-fit parameters for $A_{\mathrm{GRF}}$ and $\gamma$ reported in the top panel.}
    \label{fig:inference-diff-confusion-bkg}
\end{figure}

Despite this reassuring observation, we want to test whether the autoregressive NRE approach trained on samples with GRFs can detect the dominant scale at which both Galactic diffuse emission models differ. To this end, we inspect the one-dimensional marginal posteriors of the background parameters when applied to the target dataset containing FGMA. The profiles are shown in the top panel of Fig.~\ref{fig:inference-diff-confusion-bkg}. Since FGMA is predicting fewer gamma rays at high latitudes compared to the official \textit{Fermi} diffuse background model (cf.~Fig.~\ref{fig:diff-models}), its overall normalization is correctly recovered as being lower than 1. In comparison to the case without diffuse mis-modeling, the associated credible intervals are wider. The diffuse isotropic background's normalization is found to be larger than the originally injected value of 1 and outside of the 95\% credible band. This is most likely another quantity affected by the fainter Galactic diffuse emission model; the diffuse isotropic background needs to be larger to compensate for this contribution relative to what the network learned from the training sample. On the other side, LMC and SMC are correctly recovered with their injected normalizations. 

In terms of the inferred posteriors that describe the GRFs, we obtain a clear signal for the necessity of correcting the diffuse MW foreground template on large scales $\gamma\in[2.7,2.9]$. We illustrate an exponentiated GRF realization corresponding to the retrieved mean values $A^{\ast}_{\mathrm{GRF}}$ and $\gamma^{\ast}$ in the bottom panel of Fig.~\ref{fig:inference-diff-confusion-bkg}. Just by visual inspection, this figure provides a sense of the scales where correction is needed. This particular GRF realization was also chosen because it slightly resembles the structure of Loop I in the northern hemisphere. Loop I and the FBs in both hemispheres are missing in FGMA. Therefore, the obtained one-dimensional posterior for $\gamma$ appears to be fully in line with the scale of mis-modeling present in FGMA compared to the \textit{Fermi} diffuse background model. Additionally, the induced amplitude of distortions of FGMA is about a factor of 2, which is a reasonable assessment given the fact that this model was used to fit the high-latitude sky in previous work.  Consequently, we find circumstantial evidence that the gamma-ray simulator with GRFs provides a powerful extension of the base simulator capable of dealing with background mis-modeling.

\subsection{Shape of the source-count distribution}
\label{app:mismodeling-dnds}


Throughout this study, we worked with a simulator that generates high-latitude source populations following a TBPL source-count distribution. As noted in \cite{Zechlin:2015wdz}, there is no evidence for the necessity of even three breaks of the \dnds. Yet, any parametric description of the source-count distribution is a simple approximation of reality, and the true shape of this quantity can be much more complicated, whilst our limited statistics obstruct us from significantly resolving more details. The guiding idea behind our non-parametric inference approach is to devise a flexible means capable of reconstructing any \dnds~profile. Yet, all training data is generated from TBPL \dnds, which also determines the prior ranges the source-count distribution may take in each flux bin as well as their mutual correlations. Therefore, it is a fair question to ask whether the non-parametric inference approach can resolve a \dnds~profile with more breaks than it has been trained on; i.e.~how does it perform in the presence of \dnds~mis-modeling? 

\begin{figure*}
    \centering
    \includegraphics[width=0.95\textwidth]{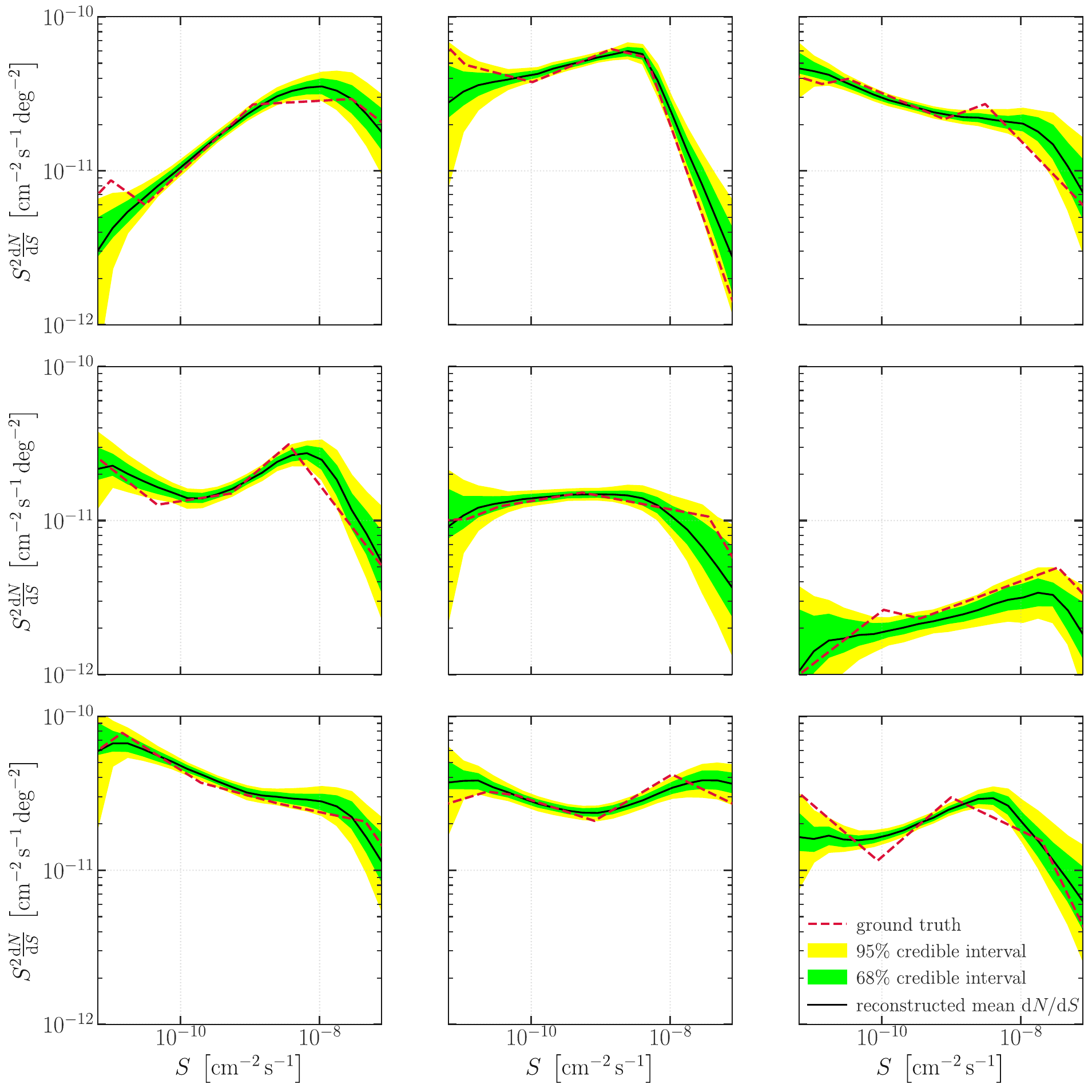}
    \caption{Examples of non-parametric inference results on test simulated data featuring a five-times broken source-count distribution. The applied neural network is the same that was used to generate the right panel of Fig.~\ref{fig:dNdS-woGRF}, i.e.~a network trained on TBPL \dnds~(without GRFs). The ground truth for each inferred source-count distribution is shown as a dashed red line, while the mean reconstructed \dnds~is denoted by a solid black line. The green and yellow bands illustrate the 68\% and 95\% credible intervals of the inferred profile, respectively.}
    \label{fig:dnds-nonparam-5BPL}
\end{figure*}

\begin{figure*}[t]
    \centering
    \includegraphics[width=0.95\textwidth]{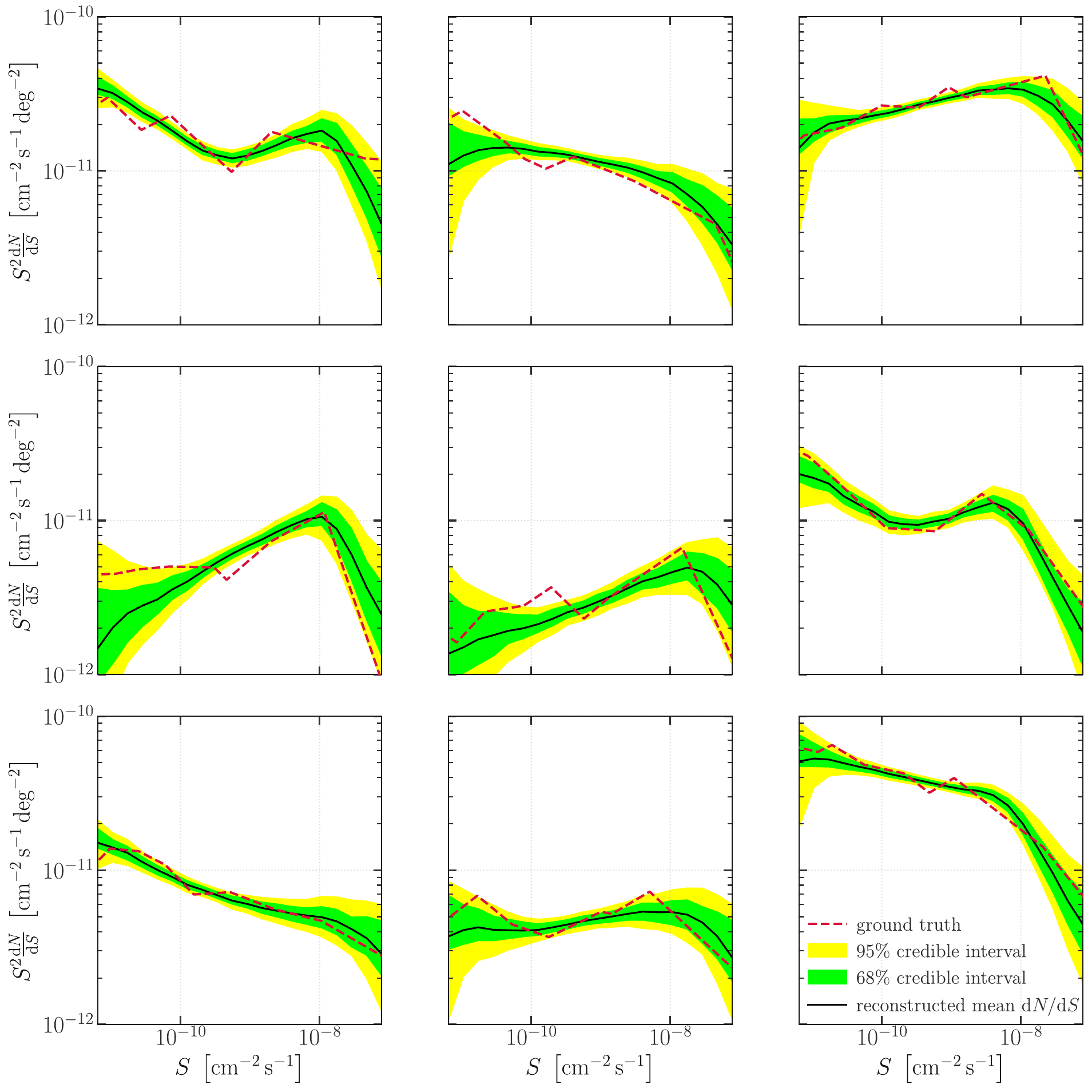}
    \caption{Same as Fig.~\ref{fig:dnds-nonparam-5BPL}, but with a ten-times broken source-count distribution.}
    \label{fig:dnds-nonparam-10BPL}
\end{figure*}

We explore this question via simulated test data featuring multiply broken source-count distributions with five (5BPL)  and ten break positions (10BPL). For both scenarios, we retain the priors for $n_1$, $n_{N_b+1}$ ($n_4$ for a TBPL) and $\log_{10}\!S_{b,1}$ stated in Tab.~\ref{tab:params}. The remaining flux range from $3\times10^{-13}\;\mathrm{cm}^{-2}\,\mathrm{s}^{-1}$ to $3\times10^{-9}\;\mathrm{cm}^{-2}\,\mathrm{s}^{-1}$ is divided into four (5BPL)/nine (10BPL) equally log-spaced bins serving as prior ranges for the additional break positions. The \dnds~slope parameters associated with each interval are uniformly sampled from 1.5 to 2.5. For both the 5BPL and 10BPL \dnds, we simulate nine examples (without GRFs but otherwise with random background normalizations) and apply the trained neural network for non-parametric source-count distribution inference on them. The results for the 5BPL are given in Fig.~\ref{fig:dnds-nonparam-5BPL}, whereas the 10BPL posteriors are displayed in Fig.~\ref{fig:dnds-nonparam-10BPL}.

Both figures indicate that the non-parametric approach does not fully capture the complexity of the injected multiply broken power laws. However, the inference remains robust. Although rapid variations in the slope parameter between break intervals are not accurately recovered -- likely due to limitations in the training data's flexibility -- the true values are almost always contained within the 95\% credible intervals. The sometimes more substantial deviations from the ground truth in the high-flux regime are likely caused by the Poisson realization of individual sources. The Poisson scatter simply has a greater impact in this flux range also because the few bright sources might, by chance, end up at low latitudes $|b|<30^{\circ}$. On the more positive side, the network consistently recovers the overall trend of the \dnds~in the faint flux regime, effectively providing an average representation of the source-count distribution despite some loss of fine detail. These results are consistent with experiences from traditional methods, such as the 1p-PDF technique, which are similarly constrained by available statistics and prior assumptions. In conclusion, source-count-distribution mis-modeling causes our non-parametric SBI approach to fail at accurately recovering the injected values, but it correctly predicts the broad shape of the profile across the different flux regimes.

\clearpage
\bibliographystyle{bib_style}
\bibliography{references} 

\end{document}